\documentclass{ws-ijmpe}
\usepackage[su
per,compress]{cite}
\usepackage[usenames,dvipsnames]{color}
\usepackage{nicefrac}
\pdfoutput=1
\usepackage{float,ulem}
\usepackage{epsfig,epstopdf,amsmath,amssymb}
\usepackage{xcolor}
\usepackage{subfigure}
\usepackage{epstopdf}

\newcommand{\be}{\begin{equation}}
	\newcommand{\ee}{\end{equation}}
\newcommand{\ba}{\begin{eqnarray}}
	\newcommand{\ea}{\end{eqnarray}}

\newcommand{\bas}{\begin{eqnarray*}}
	\newcommand{\eas}{\end{eqnarray*}}

\usepackage{scalerel}

\usepackage{soul}
\usepackage{hyperref}
\usepackage{wrapfig}

\usepackage{slashed}
\usepackage{comment}
\usepackage{float}
\usepackage{graphicx}
\begin{document}
\setcounter{tocdepth}{1} 

\markboth{S. K. Das et al.}{Dynamics of Hot QCD Matter 2024 -- Hard Probes}

\catchline{}{}{}{}{}

 \title{Dynamics of Hot QCD Matter 2024 -- Hard Probes}

\author{
Santosh K. Das$^1$\footnote{santosh@iitgoa.ac.in}~,
Prabhakar Palni$^2$\footnote{prabhakar@iitmandi.ac.in}~,
Amal Sarkar$^2$\footnote{amal@iitmandi.ac.in}~,
 Vineet Kumar Agotiya$^{3}$, Aritra Bandyopadhyay$^{4}$, Partha Pratim Bhaduri$^{5,6}$, Saumen Datta $^{7}$,Vaishnavi Desai$^{8}$, Debarshi Dey$^{9,10}$, Vincenzo Greco$^{11,12}$, Mohammad Yousuf Jamal $^{1,20}$, Gurleen Kaur $^{2}$, Manisha Kumari $^{2}$, Monideepa Maity $^{7}$,Subrata Pal $^{7}$, Binoy Krishna Patra $^{9}$,Pooja $^{13,14,1}$, Jai Prakash $^{10,21}$,  Manaswini Priyadarshini $^{2}$, Vyshakh B R $^{7}$, Marco Ruggieri $^{11,15}$, Nihar Ranjan Sahoo $^{16}$, Raghunath Sahoo $^{17}$, Om Shahi $^{19}$,Devanshu Sharma $^{18}$, Rishabh Sharma $^{3}$, Rishi Sharma $^{7}$
(authors)\footnote{
		The contributors on this author list have contributed only to those sections of the report, which they cosign with their name. Only those have collaborated together, whose names appear together in the header of a given section.}}

\address{
$^{1}$ School of Physical Sciences, Indian Institute of Technology Goa, Ponda-403401, Goa, India\\
$^{2}$ School of Physical Sciences, Indian Institute of Technology Mandi, Himachal Pradesh, India \\
$^{3}$Department of physics, Central University of Jharkhand \\
Ranchi, Jharkhand- 835222, India\\
$^{4}$Institut für Theoretische Physik, Universität Heidelberg, Philosophenweg 16, 69120 Heidelberg, Germany\\
$^{5}$Variable Energy Cyclotron Centre, 1/AF Bidhan Nagar, Kolkata 700 064, India \\
$^{6}$ Homi Bhabha National Institute, Mumbai - 400085, INDIA \\
$^{7}$Tata Institute of Fundamental Research, Colaba, Mumbai, India. 400005\\
$^{8}$P.~E.~S's R. S. N College of Arts and Science, Farmagudi Ponda, Goa, India\\
$^{9}$ Department of Physics, Indian Institute of Technology Roorkee, Roorkee-247667, Uttarakhand, India  \\
$^{10}$Department of Physics, Indian Institute of Technology Bombay, Mumbai 400076, India\\
$^{11}$Department of Physics and Astronomy “Ettore Majorana”, University of Catania, Via S. Sofia
64, I-95123 Catania, Italy\\
$^{12}$ INFN-Laboratori Nazionali del Sud, Via S. Sofia 62, I-95123 Catania, Italy\\
$^{13}$Department of Physics, University of Jyv\"askyl\"a, P.O. Box 35, 40014 University of Jyv\"askyl\"a, Finland\\
$^{14}$ Helsinki Institute of Physics, P.O. Box 64, 00014 University of Helsinki, Finland  \\
$^{15}$ INFN-Sezione di Catania, Via S. Sofia 64, I-95123 Catania, Italy  \\
$^{16}$Indian Institute of Science Education and Research (IISER) Tirupati\\
$^{17}$Department of Physics, Indian Institute of Technology Indore, Simrol, Khandwa Road\\
$^{18}$Department of Physics and Photonics Science, National Institute of Technology Hamirpur, Hamirpur-177005, Himachal Pradesh, India\\
$^{19}$Department of Physics, BITS Pilani K. K. Birla Goa Campus, Goa, India\\
$^{20}$Institute of Particle Physics, Central China Normal University, Wuhan
430079, China\\
$^{21}$School of Physics and Astronomy, Shanghai Key Laboratory for Particle Physics and Cosmology,
and Key Laboratory for Particle Astrophysics and Cosmology (MOE),
Shanghai Jiao Tong University, Shanghai 200240, China\\
}%
\maketitle
\begin{abstract}
The hot and dense QCD matter, known as the Quark-Gluon Plasma (QGP), is explored through heavy-ion collision experiments at the LHC and RHIC. Jets and heavy flavors, produced from the initial hard scattering, are used as hard probes to study the properties of the QGP. Recent experimental observations on jet quenching and heavy-flavor suppression have strengthened our understanding, allowing for fine-tuning of theoretical models in hard probes. Additionally, experimental techniques, such as machine learning algorithms, provide important tools for performing intricate data analysis in collision experiments. The second conference, “HOT QCD Matter 2024,” was organized to bring the community together for discussions on key topics in the field. This article comprises 15 sections, each addressing various aspects of hard probes in relativistic heavy-ion collisions, offering a snapshot of current experimental observations and theoretical advancements. The article begins with a discussion on memory effects in the quantum evolution of quarkonia in the quark-gluon plasma, followed by an experimental review, new insights on jet quenching at RHIC and LHC, and concludes with a machine learning approach to heavy flavor production at the Large Hadron Collider.
\end{abstract}


\ccode{PACS numbers:12.38.-t, 12.38.Aw}

\tableofcontents 

%
%

\newcommand{\tr}{{\rm{Tr}}}
\newcommand{\tot}{{\rm{tot}}}
\newcommand{\Sys}{{\rm{S}}}
\newcommand{\Env}{{\rm{E}}}
\newcommand{\Int}{{\rm{Int}}}
\newcommand{\bQ}{\bar{Q}}
\newcommand{\bfE}{{\bf{E}}}
\newcommand{\bfV}{{\bf{V}}}
\newcommand{\bfr}{{\bf{r}}}

\section{Memory effects in the quantum evolution of quarkonia in the quark gluon plasma}

\author{Vyshakh B R, Rishi Sharma}

\bigskip

\begin{abstract}
Quantum evolution in the quark gluon plasma (QGP) can be described rigorously using the framework of open quantum systems (OQS). The resulting evolution equation for the density matrix simplifies greatly if the system time scales are much longer than the environment time scales. In this hierarchy, the evolution can be described by a Lindblad equation which is memoryless. However, during a significant part of the evolution, the hierarchy is not satisfied, calling for solving general master equations with memory. We illustrate that this leads to smaller suppression than the corresponding Lindblad equation with identical zero-frequency transport coefficients.
\end{abstract}


\ccode{PACS numbers:14.40.Pq,12.38.Mh}


\subsection{Introduction~\label{sec:introducion}}
The original vision of Matsui and Satz~\cite{matsui19861} envisaged that since screening in the quark-gluon plasma (QGP) created in heavy ion ($AA$) collisions could lead to dissociation of quarkonia (bound states of heavy quarks $Q$ and anti-quarks $\bar{Q}$), it would manifest as a suppression in their measured normalized yields relative to $pp$ collisions ($R_{AA}$).  One hope was that the observation of this suppression can tell us about the medium, for example, its temperature ($T$) and the Debye screening mass~\cite{karsch19901} ($m_D$) (see~\cite{mocsy20131} for a review). Experimentally, this vision has been realized, with beautiful measurements of $R_{AA}$ of Charmonia [in particular $J/\psi$ and $\psi(2s)$] and Bottomonia [in particular $\Upsilon(1s)$, $\Upsilon(2s)$, $\Upsilon(3s)$] from STAR and PHENIX at RHIC, and ALICE, ATLAS, and CMS at LHC (see ~\cite{andronic20151} for a review). Meanwhile, theoretical developments have clarified that while quarkonium measurements can tell us about the properties of the QGP, the connection is subtle. 

Two key reasons for this are as follows. First, in addition to screening, dissociation due to gluon absorption [gluo-dissociation (GD)~\cite{peskin19791,bhanot19791}] or scattering of in-medium gluons [Landau Damping (LD)~\cite{grandchamp20041,laine20071}] contributes significantly to quarkonium suppression. In particular, since $m_D$ is smaller than the inverse size of the bound states, dissociation is the dominant effect~\cite{brambilla20081,brambilla20101}, especially for the ground states. (For Charmonia, uncorrelated recombination~\cite{grandchamp20021} also needs to be included.) Second, bound state propagation requires a quantum~\cite{strickland20111,blaizot20171,Boyd:2019arx} description. Since the state exchanges colour, energy and momentum with the environment (QGP) it is best treated in the OQS framework~\cite{akamatsu20121,borghini20121} (see ~\cite{Akamatsu:2020ypb,Yao:2021lus,Sharma:2021vvu,Andronic:2024oxz} for reviews).

A theoretical tool that has proven useful for studying quarkonium-medium interactions is an effective theory known as potential-Non Relativistic Quantum Chromodynamics (pNRQCD~\cite{brambilla20001}). It is a systematic expansion in the relative velocity, $v$, between the $Q$ and the $\bQ$ and relies on the hierarchy $M\gg 1/r\sim Mv \gg E_b\sim Mv^2$, where $M$ is the $Q$ mass, $r$ the typical relative separation between the $Q$ and $\bQ$, and $E_b$ the binding energy of the state. 

Medium interaction effects, like LD~\cite{brambilla20081,brambilla20101}, GD~\cite{brambilla20111}, and screening~\cite{brambilla20081} can be calculated in the theory by calculating the loop corrections to the quarkonium self-energy. The formalism also clarifies which processes dominate depending on the hierarchies between the pertinent scales, $1/r, E_b, T, m_D$. For example, for $T\gg E_b$ LD is more important than GD ~\cite{brambilla20101,brambilla20131}. OQS master equations have also been derived using pNRQCD~\cite{brambilla20171}. Furthermore, one can show that if $E_b\ll T$, then the evolution for the quarkonium density matrix is memoryless and Markovian~\cite{akamatsu20151} and the equation can be written as a Lindblad equation~\cite{brambilla20171,brambilla20181}. (See also~\cite{yao20183,Yao:2020eqy,Delorme:2024rdo}.)

Another nice feature of this approach is that medium dynamics and quarkonium dynamics factorize. The medium properties that appear in the Lindblad equation are characterized by two transport coefficients ($\kappa,\gamma$~\cite{Brambilla:2019tpt}, Eq.~\ref{eq:tildegamma}) which are related to a $0$ frequency limit of a response function, and can in principle be obtained from Lattice QCD~\cite{Banerjee:2022gen} calculations. The Lindblad equation has been solved~\cite{Brambilla:2020qwo,Brambilla:2022ynh,Brambilla:2023hkw,Brambilla:2024tqg} for bottomonium states and gives a good description of the results from CMS~\cite{chatrchyan20111,chatrchyan20121,CMS:2023lfu}.

However, simplification of the master equations to obtain a Lindblad form requires care. The hierarchy $E_b\ll T$ is not satisfied\cite{Sharma:2023dhj} for all the states [especially for $\Upsilon(1s)$ at late times] during the medium evolution. The master equations are therefore not memoryless and depend on the finite frequency response of the QGP. 
In these proceedings, we will investigate the effect of including memory in the evolution equation. Our main conclusion is shown in Fig.~\ref{fig:LOvsMemory} which shows that the evolution with memory ($\tau_E=1/(1.5T),1/T,1.5/T$) shows a smaller suppression compared to the leading order (LO) Lindblad equation. While the NLO Lindblad equation captures part of the effect, it is inaccurate if $\tau_{\Env}$ is ($\gtrsim 1/T$). We will review the formalism in Sec.~\ref{sec:formalism} and discuss the results in Sec.~\ref{sec:results}.

\subsection{Formalism~\label{sec:formalism}}
\subsubsection{The OQS framework~\cite{breuer20021}}
The system ($\Sys$) and the environment ($\Env$) are characterized by a density~\cite{breuer20021}matrix satisfying the Heisenberg equation of motion
\begin{equation}
i\frac{d\rho _{\tot}}{dt}
=[H_{\tot},\rho_{\tot}],\;\;
{\rm{where}}\;\; 
H_{\tot} = H_{\Sys} + H_{\Env} 
+H_{\Int}
\end{equation}
The reduced density matrix for the system, $\rho_{\Sys}$ can be obtained by tracing the environment $\rho_{\Sys}=\tr_{\Env} \{\rho_{\tot}\}$.

If the interaction between the system and the environment, $H_{\Int}$, is small~\footnote{The smallness of $H_{\Int}$ compared to $H_{\Sys}$ for quarkonia stems from the fact that it is higher order in the multipole expansion.} then it is convenient to work in the interaction picture. Further, assuming that initially (at $t=0$) $\rho_{\tot}(0)=\rho_{\Sys}(0)\otimes\rho_{\Env}$ and $\Env$ is large, the total density matrix can be approximated as $\rho_{\tot}(t)=\rho_{\Sys}(t)\otimes\rho_{\Env}$. The master equation, accurate up to ${\cal{O}}(H_{\Int}^3)$ has the following form~\footnote{We have assumed here that $\tr[\rho_{\Env} H_{\Int}]=0$, which follows from symmetry under colour transformations in the case of interest.}
\begin{equation}
\frac{\partial \rho^{I}_{\Sys}(t)}{\partial t} = - \int\limits_{0}^{t} ds \tr_{\Env}\Bigl\{\left[H^{I}_{\Int}(t),\left[H^{I}_{\Int}(t-s),\rho^{I}_{\Sys}(t)\otimes\rho_{\Env}\right]\right]\Bigr\}\;.
~\label{eq:mastergen}
\end{equation}

\subsubsection{The quarkonium Hamiltonian and the master equation}
In the center of mass of quarkonium, the system Hilbert space features the singlet states $|s\rangle$ (the separation $r$ is suppressed for clarity) and octet states $|a\rangle$ ($a=1,.8$).
\begin{equation}
\begin{split}
  H_{\Sys} = &\;   
  h_s |s\rangle\langle{s}| +  
  h_o |a\rangle\langle{a}|,\;\;{\rm{where}},\;\;
 h_{s,o}(r)=\frac{p^2}{M}+v_{s,o}(r),\;\\ 
 H_{\Int} = &- g{\bf{E}}^{a}\cdot\;{\bf{r}}\left[\frac{1}{\sqrt{2N_{c}}} |s\rangle \langle{a}| +\frac{1}{\sqrt{2N_{c}}} |a\rangle \langle{s}| +\frac{d_{abc}}{2} |b\rangle \langle{c}| \right]        
 \end{split}
\end{equation}


Eq.~\ref{eq:mastergen} can then be written in Schr\"{o}dinger picture as a master equation

\begin{equation}
\begin{split}
\frac{d\rho_{\Sys}}{dt} &= -i\left(H_{\rm{eff}}\rho_{\Sys}-\rho_{\Sys}H_{\rm{eff}}^\dagger \right)
    +  \int\limits_0^t du\; \left\{ \Gamma(u) \sum\limits_{n=1}^{3} 
\bfV_{n}(u) \rho_{\Sys}(t)\bfV_{n}^\dagger (0) + \rm{H.C}\right \}\;.
\end{split}
\label{eq:master}
\end{equation}
where,
\begin{align}
    H_{\rm{eff}}= H_{\Sys}-i\int\limits_0^t du\; \Gamma(u) \sum\limits_{n=1}^{3} \bfV_{n}^\dagger(0) \bfV_{n} (u)\;.
\end{align}
$\{\bfV_{n}(t)\}$ are time dependent jump operators corresponding to $s\rightarrow o\; (n=1)$, $o\rightarrow s\; (n=2)$, $o\rightarrow o\; (n=3)$ transitions.
The spatial indices of $\bfV$ are summed.

$\Gamma$ comes from the $\Env$ response function (${\cal{W}}_{ab}(t,{\vec{0}})$ is the adjoint Wilson line~\cite{Brambilla:2024tqg})
\begin{equation}
\Gamma(t)=\frac{g^2}{6N_c}\tr_{\Env}
\Bigl({\bfE}_i^a(t,{\vec{0}}) {\cal{W}}_{ab}(t,{\vec{0}})
{\bfE}_i^b(0,{\vec{0}})\rho_{\Env}\Bigr)\;.
\end{equation}              

The explicit form of $\bf{V}$ in $s$, $o$ space is as follows
\begin{equation}
\begin{split}
\bfV_{1}(t) = e^{-ih_ot}&\bfr e^{ih_st} 
    \left(\begin{array}{cc} 0 & 0\\1 & 0\end{array}\right),\;\;
\bfV_{2}(t) = e^{-ih_st}\bfr e^{ih_ot} 
    \sqrt{\frac{1}{(N_c^2-1)}}\left(\begin{array}{cc} 0 & 1\\0 & 0\end{array}\right) \\
\bfV_{3}(t) &= e^{-ih_ot}\bfr e^{ih_ot} 
    \sqrt{\frac{N_c^2-4}{2(N_c^2-1)}}
    \left(\begin{array}{cc} 0 & 0\\0 & 1\end{array}\right) \;.
\end{split}
\end{equation}

\subsubsection{Expansion in small $\tau_{\rm{E}}/\tau_{\rm{S}}$}
%
%

The correlator $\Gamma(t)$ is substantial for $t\lesssim\tau_{\rm{E}}\sim \frac{1}{T}$. ${\bf{V}}_n(t)$ evolves on a time scale of $\tau_{\rm{S}}\sim \frac{1}{E_b}$, where $E_b\sim h_o, h_s$. 
If one assumes $\tau_{\rm{E}}\ll\tau_{\rm{S}}$, known as the quantum Brownian approximation, one can make the following expansion:
\begin{align}
    {\bf{V}}_{n}(t) \sim e^{-ih_\alpha t} {\bf{r}} e^{ih_\beta t} 
    \approx {\bf{r}} - it (h_{\alpha}{\bf{r}} - {\bf{r}} h_{\beta}) 
    + {\cal{O}}\Bigl[\Bigl(\frac{\tau_{\Env}}{\tau_{\Sys}}\Bigr)^2\Bigr]\;.
\end{align}
Truncation at the first term is called leading order (LO)~\cite{Brambilla:2020qwo}. Keeping the term proportional to $t$ gives the next-to-leading order (NLO)~\cite{Brambilla:2022ynh} master equation. Further, making a Markovian approximation, the LO and NLO equations can be cast into the Lindbladian form, accurate up to their respective orders in ${\tau_{\Env}}/{\tau_{\Sys}}$.
The NLO equations give a good description of the LHC data for the suppression of $\Upsilon$'s \cite{Brambilla:2022ynh}. In LO and NLO only the zero frequency part of the correlator $\tilde{\Gamma}(\omega=0)$ is needed,
\begin{align}
	\tilde{\Gamma}(\omega)= \int\limits_{0}^{\infty} ds \; e^{i\omega s}\;\Gamma(s) \;,\;\; \lim_{\omega\rightarrow 0}\tilde{\Gamma}(\omega)=\frac{1}{2}(\kappa+i\gamma)\;.
 \label{eq:tildegamma}
\end{align}
However, $E_b \sim 500$ MeV for $\Upsilon(1s)$, slightly smaller for $\Upsilon(2s)$. We expect $\tau_{\Env}\sim 1/T$. For $T \lesssim 500$ MeV the hierarchy ${\tau_{\Env}}\ll {\tau_{\Sys}}$ is not satisfied and the approximation becomes worse as the medium cools as it expands. So it is crucial to go beyond the Brownian regime, particularly for $\Upsilon(1s)$. 


\subsection{Results and discussion}\label{sec:results}
The master equation (Eq.~\ref{eq:master}) can be solved using stochastic unraveling \cite{breuer20021}. The equation rewritten as an ensemble average of a stochastic evolution equation for wavefunctions. The stochastic evolution equation is comprised of deterministic evolution with a non-Hermitian effective Hamiltonian $H_{\textrm{eff}}$, interspaced with random transitions with probabilities determined by the Hamiltonian. 

The full unraveling of the non-Markovian equation is computationally expensive and hence we neglect jumps in this study. The NLO results from \cite{Brambilla:2022ynh} show that for $\Upsilon(1s)$ deterministic evolution using $H_{\textrm{eff}}$ gives a good approximation for $\Upsilon(1s)$ evolution as it is tightly bound. It is expected that for $\Upsilon(2s)$ jumps will give a significant correction but we leave this investigation for the future.  

In the study we show below, we take~\cite{Brambilla:2022ynh}
\begin{equation}
v_s(r)=-\frac{C_{F}\alpha_{s}}{r},\;\;
v_o(r)=\frac{\alpha_{s}}{2N_{c}r}\;.
\end{equation}

Numerically we solve for the radial wavefunction, initialized as singlet Coulombic eigenstates. For illustration, we take the background to be a Bjorken expanding medium with initial time $t_0=1.0$ fm (with $T_0=330$ MeV, appropriate for RHIC)~\cite{alberico20131}.  Quarkonium interactions with the medium are turned on at $t_0$ and turned off when the medium reaches the freezeout temperature $T_{\rm{f}}$. The correlator $\Gamma(t)$ is taken to be 
\begin{align}
	\Gamma(t)= \frac{\kappa}{2\tau_{\Env}}e^{{-|t|}/{\tau_{\Env}}}\;.
\end{align}
Following Ref.~\refcite{Brambilla:2023hkw}, we take ${\kappa}=4T^3$ and ${\gamma}=0$. 

\begin{figure}[ht]
     \centering
     \begin{minipage}{0.49\textwidth}
         \centering
         \includegraphics[scale=0.8]{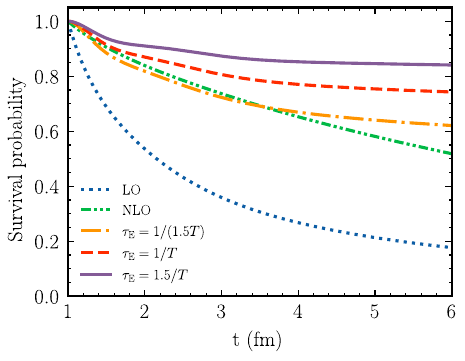}
         \label{fig:1s}
     \end{minipage}
     \hfill
     \begin{minipage}{0.49\textwidth}
         \centering
         \includegraphics[scale=0.8]{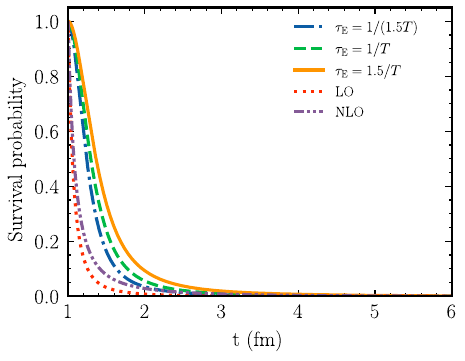}
         \label{fig:2s}
     \end{minipage}
        \caption{Survival probability for $\Upsilon(1s)$ (left) and $\Upsilon(2s)$ as a function of time. We stop the plot at $6$ fm for clearer presentation.}
        \label{fig:LOvsMemory}
\end{figure}


In Fig.~\ref{fig:LOvsMemory}, we compare survival probabilities of $\Upsilon(1s)$ and $\Upsilon(2s)$ for three different $\tau_{\Env}$, with LO and NLO values. We note that the suppression is sensitive to the hierarchy $\tau_{\Env}/\tau_{\Sys}$. Evolution with memory gives lesser suppression than LO. Furthermore, increasing $\tau_{\Env}$ leads to a smaller suppression as a larger $\tau_{\Env}$ means it takes longer to reach the Markovian limit. While NLO equations capture some memory effects at early times, at late times, higher-order corrections are important. 

In subsequent works, we will explore the effect of including jumps and a realistic hydrodynamic background. It will also be interesting to investigate to what extent the relative yields of the different quarkonium states thermalize~\cite{gupta20141,Kumar:2023acr}.


\newcommand{\IAA}{\ensuremath{{I}_\mathrm{AA}}}
\newcommand{\RAA}{\ensuremath{{R}_\mathrm{AA}}}

\newcommand{\gammadir}{$\gamma_{\rm dir}$}
\newcommand{\pizero}{$\pi^{0}$}
\newcommand{\sqrtSNN}{$\sqrt{s_{\rm NN}}$}

\newcommand{\pTjet}{$p_{\rm T,jet}^{\rm ch}$}

\section{Jet quenching: new results at RHIC and LHC}
\author{Nihar Ranjan Sahoo}

\bigskip

\begin{abstract}
We discuss recent results on jet quenching in heavy-ion collisions from experiments at the LHC and RHIC. These results include various manifestations of jet quenching, such as modifications in jet yield and jet substructure, intra-jet broadening,and jet acoplanarity. These observations provide insights into the finite-temperature QCD medium created in heavy-ion collisions.   
\end{abstract}


\ccode{PACS numbers:}


\subsection{Introduction}
The hot and dense QCD medium—known as the Quark-Gluon Plasma (QGP)~\cite{STAR:2005gfr}—filled the universe just microseconds after the Big Bang. The QGP produced in heavy-ion collisions offers a unique opportunity to study the early universe, strong interactions at finite temperature, and the QCD phase transition. In the context of strong interactions, the QCD running coupling ($\alpha_{s}$) exhibits two regimes: (i) at larger distance scales (comparable to or greater than the size of a pion), hadronization and fragmentation happen ({\it confinement}); and (ii) at shorter length scales, quarks and gluons become effectively free ({\it deconfinement}). Experiments at the LHC and RHIC aim to explore these two regimes of $\alpha_{s}$ through heavy-ion collisions.\\
A highly virtual quark or gluon (parton) undergoes gluon radiation in vacuum until its virtuality drops to the QCD scale ($\Lambda_{QCD}$). At this point, due to confinement, the parton fragments into sprays of hadrons—known as a jet. However, when traversing the Quark-Gluon Plasma (QGP), a highly virtual parton emits not only gluons but also medium-induced radiation ({\it Bremsstrahlung}). This interaction between the jet and the medium is known as jet quenching~\cite{Gyulassy:1990ye} in heavy-ion collisions. The first observations of jet quenching were made in experiments at RHIC~\cite{STAR:2005gfr}, where suppression of inclusive charged and neutral hadrons ($\pi^{0}, \eta$)\cite{PHENIX:2006avo} yields, and away-side jet suppression~\cite{STAR:2003pjh} in central Au+Au collisions, relative to $p+p$ collisions, were observed. Additionally, no suppression of direct photon yields---since photons do not interact strongly with the medium---indicates that the medium created in heavy-ion collisions is indeed QCD matter. This also validates the use of binary collision scaling using $p+p$ collisions as a baseline for heavy-ion collisions. At the LHC, these observations were confirmed at even higher collision energies in the TeV range~\cite{ALICE:2019qyj,CMS:2016uxf}. Consequently, jet quenching is regarded as one of the key signatures of QGP formation in heavy-ion collisions.\\

Detailed studies on jet quenching are being conducted in experiments at both the LHC and RHIC. In this proceeding, we discuss high transverse momentum (energy) direct photon and neutral pion coincidences with recoil jets ($\gamma$+jet and $\pi^{0}$+jet), and inclusive jet at the STAR and ATLAS experiments, as well as hadron+jet coincidences at the ALICE experiment, in both heavy-ion and $p+p$ collisions. The effects of jet-medium interactions in heavy-ion collisions are highlighted, and their implications are discussed in the following sections.
\subsection{Consequences of jet quenching}
Jet medium interaction in heavy-ion collisions results four consequences: i) jet yield suppression, ii) intra-jet broadening, iii) modification of jet substructure, and iv) jet acoplanarity. In this section, all these manifestations and their experimental observations are discussed.

\subsubsection{Jet yield suppression}
The STAR experiment reports~\cite{STAR:2023ksv,STAR:2023pal} suppression of recoil jet yield (\IAA) for direct photon (\gammadir) and neutral pion (\pizero) triggers in 0-15\% central Au+Au collisions relative to $p+p$ collisions at \sqrtSNN=200 GeV. The \IAA\ as a function of \pTjet\ indicates that recoil jets are less suppressed for a jet radius of R=0.5 compared to R=0.2. The jet radius dependence of recoil jet suppression suggests that the lost energy of the jet is redistributed within the medium. 

\begin{figure}[bth]
\includegraphics[width=6.3cm]{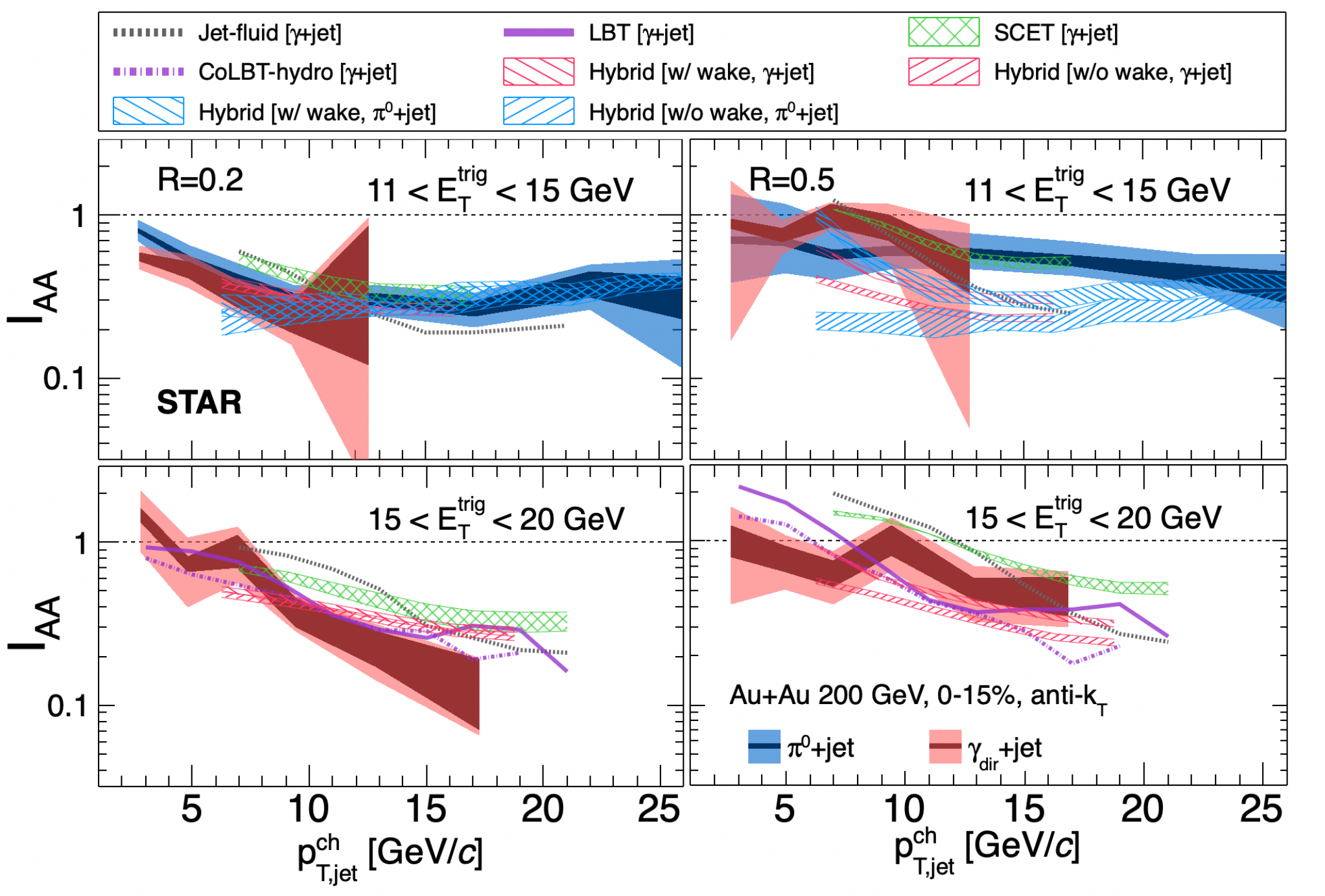}
\includegraphics[width=5.3cm]{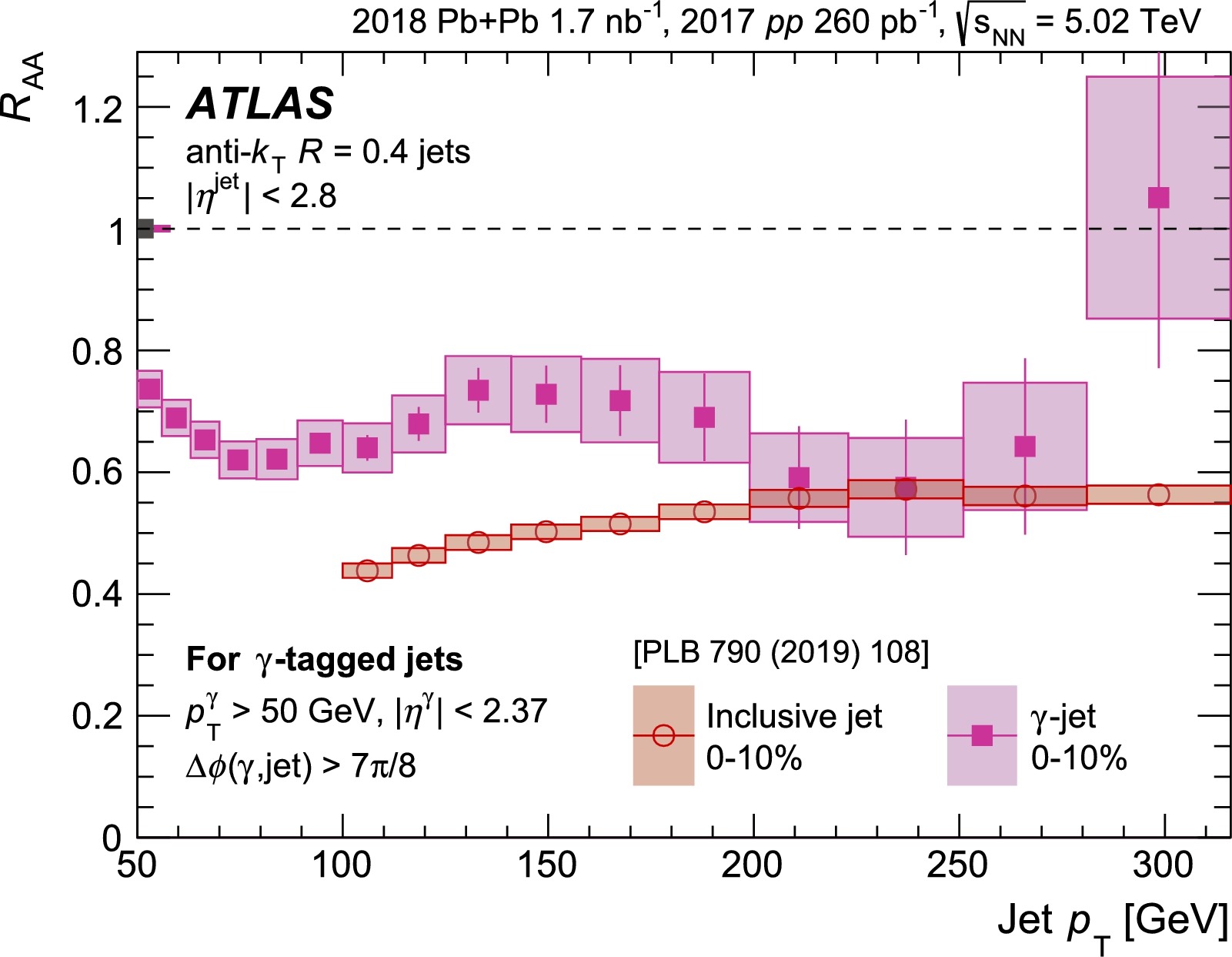}
\caption{(color online) Left: Semi-inclusive \gammadir+jet and \pizero+jet \IAA\ results from the STAR experiment~\cite{STAR:2023ksv,STAR:2023pal}. Right: \gammadir+jet and inclusive jet measurements from the ATLAS experiment~\cite{ATLAS:2023iad}.}
\label{fig:IAA}
\end{figure}

To investigate the effects of path length and color factor differences in jet energy loss, comparisons are made among \gammadir+jet, \pizero+jet, and inclusive jet measurements at RHIC and LHC energies. However, different measurement techniques, such as semi-inclusive versus inclusive methods, introduce trigger biases that complicate data interpretation. For instance, no significant difference in suppression between \gammadir+jet and \pizero+jet is observed within uncertainties in the STAR experiment, while a marked difference in \RAA\ between \gammadir+jet and inclusive jet is observed in ATLAS measurement~\cite{ATLAS:2023iad} at \sqrtSNN=5.02 TeV in Pb+Pb collisions as shown in Fig.~\ref{fig:IAA}. These contrasting results at RHIC and the LHC may be due to limited precision in RHIC measurements, differences in quark and gluon jet fractions between RHIC and the LHC, and the trigger biases discussed above. Further studies in this direction are ongoing at both facilities.

\subsubsection{Intra-jet broadening}
\label{Sec:Intrajet}
\begin{figure}[bth]
\centerline{\includegraphics[width=6.3cm]{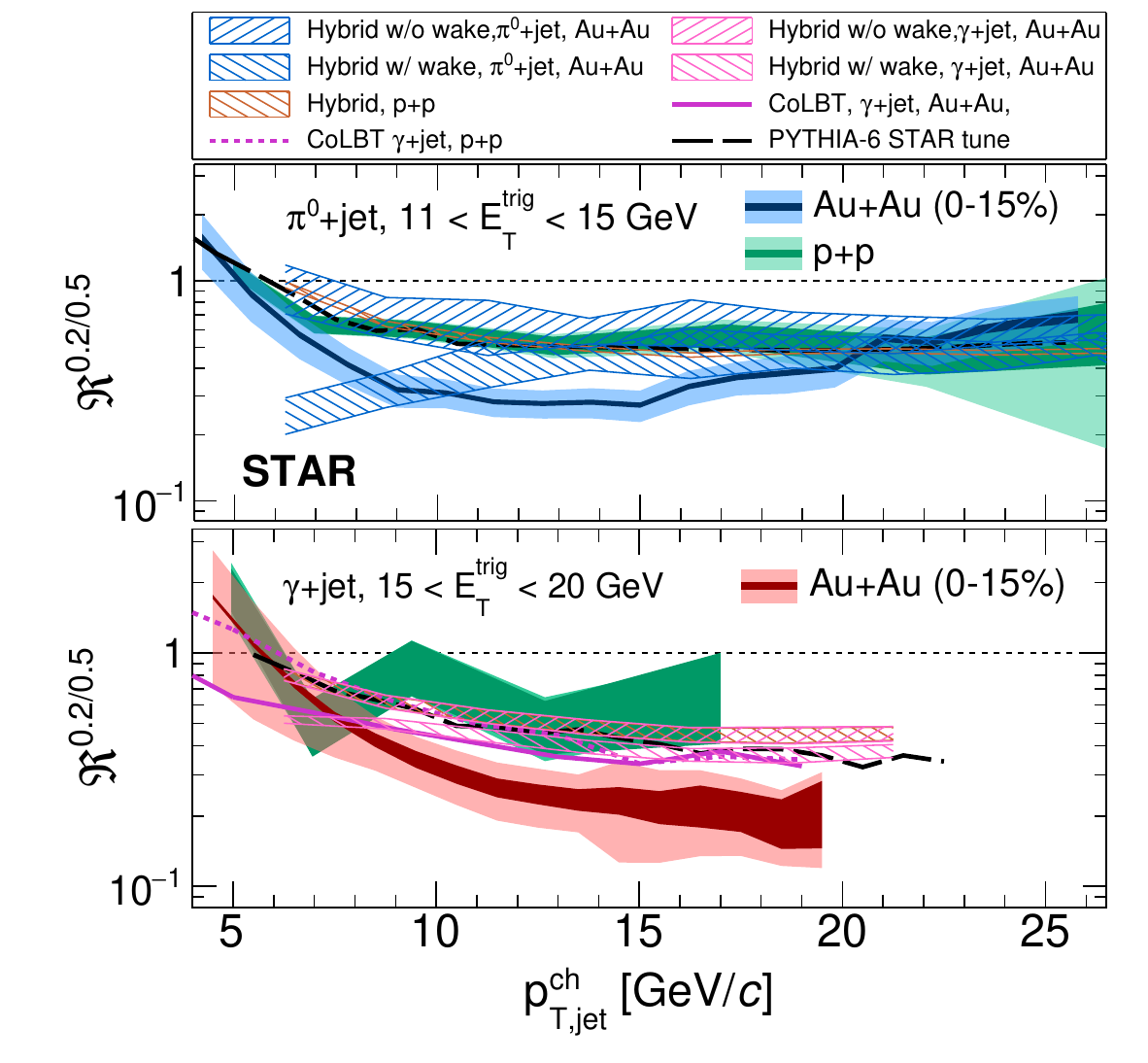}}
\caption{(color online) Yield ratio between recoil jet yield of radius R=0.2/R=0.5 for \pizero+jet (top) and \gammadir+jet (bottom). Green Bands are $p+p$; Blue and Red bands are \pizero+jet and \gammadir+jet in AuAu collisions, respectively. Curves and hatched bands represent different models.}
\label{fig:YeildRatio}
\end{figure}

Jet quenching in heavy-ion collisions arises from both vacuum radiation and in-medium gluon radiation. To investigate the medium-induced radiation in heavy-ion collisions relative to vacuum (p+p collisions), the STAR experiment reports the ratio of recoil jet yields for jet radii R=0.5 and R=0.2. This ratio is compared between $p+p$ and central Au+Au collisions, as shown in Fig~\ref{fig:YeildRatio}. At intermediate \pTjet\ values, a clear difference between $p+p$ and central Au+Au collisions is observed for both \gammadir+jet and \pizero+jet events, suggesting intra-jet broadening in heavy-ion collisions due to jet-medium interactions. Notably, various models have been unable to accurately predict the data for Au+Au collisions.

\subsubsection{Modification of jet substructure} 
\begin{figure}[bth]
\centerline{\includegraphics[width=7.3cm]{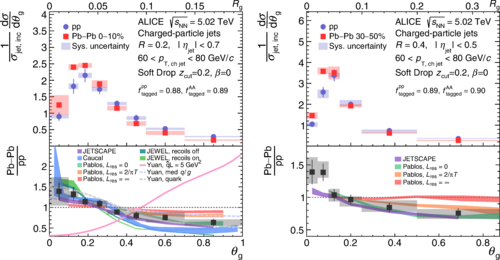}}
\caption{(color online) Unfolded $\theta_{g}$ distributions for charged-particle jets measured in the ALICE experiment~\cite{ALargeIonColliderExperiment:2021mqf}.}
\label{fig:Grooming}
\end{figure}

In heavy-ion collisions, in-medium gluon radiation modifies the internal structure of jets~\cite{ALargeIonColliderExperiment:2021mqf}. This effect has been observed in the ALICE experiment using the SoftDrop grooming method. The groomed jet momentum fraction, $z_{\rm g}$, and the groomed jet radius, $\theta_{\rm g}$, of charged-particle jets were measured in $p+p$ and Pb+Pb collisions at \sqrtSNN=5.02 TeV. A narrowing of the $\theta_{\rm g}$ distribution was observed in Pb+Pb collisions compared to $p+p$ collisions as shown in Fig.~\ref{fig:Grooming}, providing the first direct experimental evidence for the modification of the angular scale of groomed jets in heavy-ion collisions. However, no modification of the $z_{\rm g}$ distribution was observed in Pb+Pb collisions relative to $p+p$.

\subsubsection{Jet acoplanarity}
\begin{figure}[th]
\includegraphics[width=4.3cm]{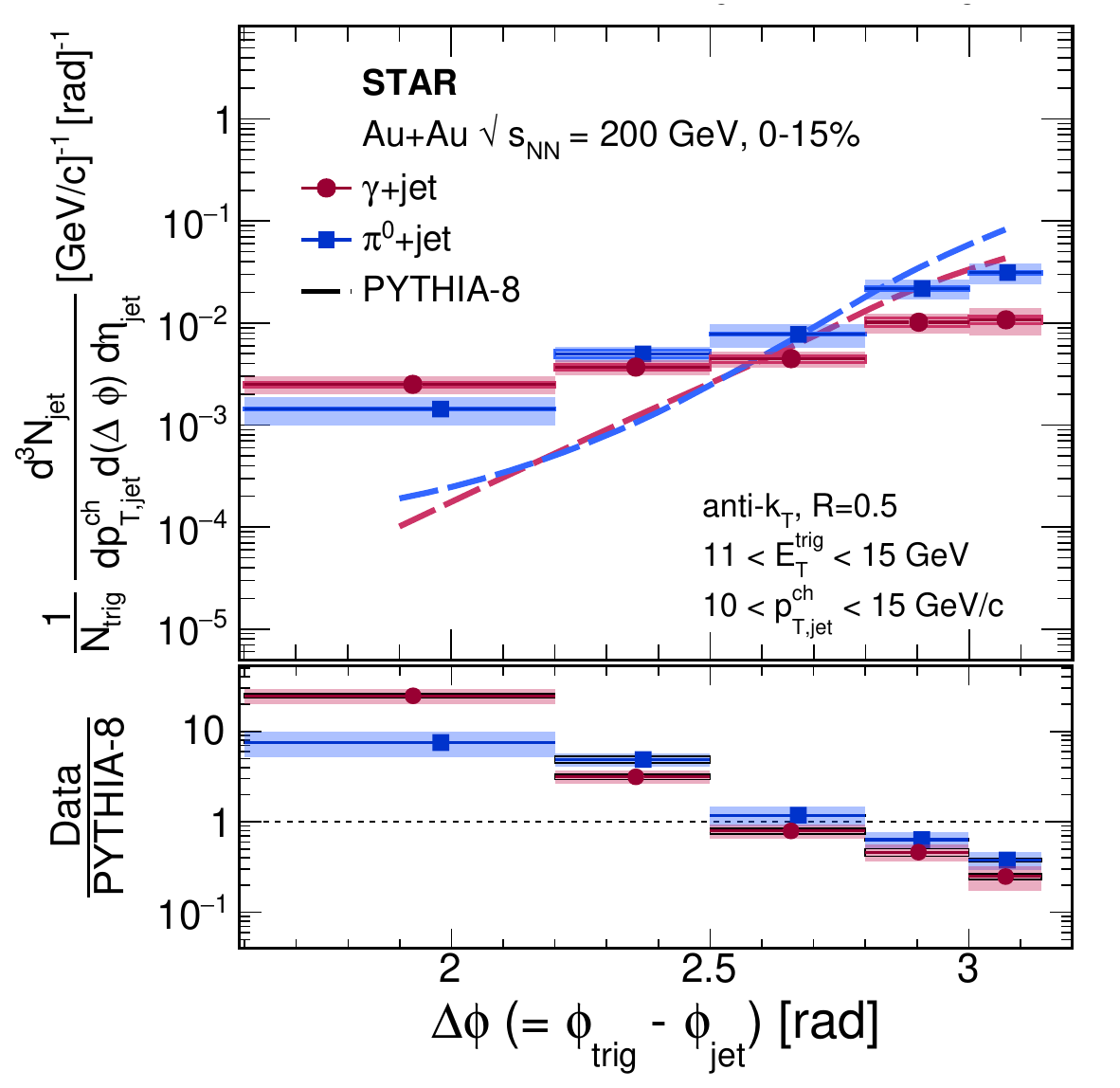}
\includegraphics[width=7.6cm]{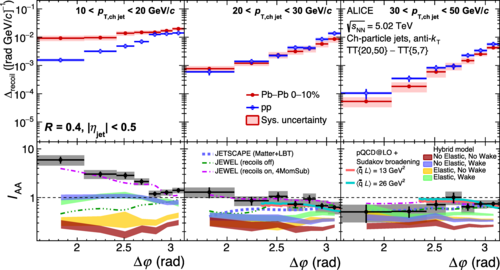}
\caption{(color online) Right: The STAR experiment's \gammadir+jet and \pizero+jet for R=0.2; with vacuum expectation as PYTHIA-6. Right: the ALICE experiment's h+jet with R=0.4.}
\label{fig:JetAcopl}
\end{figure}

The azimuthal angle correlation between the recoil jet axis and the trigger particle (here \gammadir, \pizero, or $h^{\pm}$) reveals another manifestation of jet-medium interactions in heavy-ion collisions. Large angular de-correlation, or broadening—known as jet acoplanarity—is predicted as a result of Molière scattering~\cite{DEramo:2012uzl} or a diffusion wake in the QGP~\cite{ALICE:2023jye,ALICE:2023qve}. Initial measurements from both the STAR and ALICE experiments indicate an excess of recoil jet yield away from $\pi$ in azimuthal angle relative to the trigger particle. The STAR experiment reports~\cite{Anderson:2022nxb} medium-induced jet acoplanarity for jets with R=0.5 in both \gammadir+jet and \pizero+jet events within 10 $<$ \pTjet $<$ 15 GeV/c, as shown in Fig.\ref{fig:JetAcopl}. 
A similar observation has been reported for $h$+jet events in the ALICE experiment with R=0.4. However, in the ALICE measurements, medium-induced jet acoplanarity diminishes at higher \pTjet (20 to 50 GeV/c), as shown in Fig~\ref{fig:JetAcopl}. This feature has not been measured in the STAR experiment for \gammadir+jet and \pizero+jet due to limited kinematic reach. It is worth noting that these measurements used different techniques to mitigate or remove uncorrelated background jet contributions in heavy-ion collisions. For instance, the STAR and ALICE experiments use the mixed event method and a different triggered recoil jet population subtraction method, respectively. Nonetheless, both experiments observe similar jet acoplanarity in heavy-ion collisions, despite differences in heavy-ion background subtraction methods.

Significant azimuthal broadening has also been observed in high-multiplicity $p+p$ collisions at $\sqrt{s}$=13 TeV using forward detectors in the ALICE experiment, which may suggest jet quenching effects. However, similar broadening is also observed in PYTHIA 8 event generator simulations, which do not include jet quenching effects, as noted in~\cite{ALICE:2023plt}. Notably, high-multiplicity $p+p$ events also show a significant increase in multi-jet production. This observation in $p+p$ collisions at the LHC is intriguing and warrants further study for a deeper understanding.

\subsection{Summary and outlook}
The four manifestations of jet quenching discussed—jet yield suppression, intra-jet broadening, modification of jet substructure, and jet acoplanarity—enable a deeper understanding of the interactions of high-energy partons as they traverse the QGP. Efforts toward a coherent theoretical understanding and model calculations are ongoing. Additionally, experiments at the LHC and RHIC are collecting data to further explore these phenomena with precision measurements.


\newcommand{\qqb}{\overline{Q} Q}
\newcommand{\qqs}{\left( \overline{Q} Q \right)_s}
\newcommand{\qqo}{\left( \overline{Q} Q \right)_o}  
\newcommand{\qbar}{\overline{Q}}
\newcommand{\qqg}{\overline{Q} G Q}
\newcommand{\jpsi}{J/\Psi}
\newcommand{\mq}{M_{\scriptscriptstyle Q}}
\newcommand{\vr}{\vec{r}}
\newcommand{\om}{\omega}
\newcommand{\omin}{\omega_{\rm min}}
\newcommand{\sig}{\sigma_T(\om, r)}
\newcommand{\vt}{V_{\scriptscriptstyle T}(\vec{r})}
\newcommand{\vtt}{V_{\scriptscriptstyle T}(\vec{r}, t)}
\newcommand{\vo}{V^{\rm re}_8(r)}
\newcommand{\fre}{F(\vec{r}; T)}
\newcommand{\cadj}{C_{\scriptscriptstyle EE}^{\rm adj}}
\newcommand{\cfund}{C_{\scriptscriptstyle EE}^{\rm fund}}
\newcommand{\vre}{V^{\rm re}_{\scriptscriptstyle T}(\vec{r})}
\newcommand{\vim}{V^{\rm im}_{\scriptscriptstyle T}(\vec{r})}
\newcommand{\md}{m_{\scriptscriptstyle D}}
\newcommand{\rwt}{\rho(\omega, \vec{r}; T)}
\newcommand{\rlow}{\rho_{{\rm low} \ \omega}(\omega, \vec{r}; T)}
\newcommand{\wrt}{W_{\scriptscriptstyle T}(\tau, \vec{r})}
\newcommand{\wrbt}{W_{\scriptscriptstyle T}\left(\frac{1}{2T}-\tau, \vec{r}\right)}
\newcommand{\wrm}{W_{\scriptscriptstyle T}(t, \vec{r})}
\newcommand{\wil}{\mathbb{U}}
\newcommand{\calo}{\mathcal{O}}
\newcommand{\intom}{\int_{- \infty}^\infty \, d\omega}
\newcommand{\delt}{\frac{\partial}{\partial t}}
\newcommand{\mlt}{m_{\rm loc}(t)}
\newcommand{\mlta}{m_{\rm loc}^a(t)}

\section{Non-perturbative Input to Study of Quarkonia in QGP}

\author{Saumen Datta}

\bigskip

\begin{abstract}
  Quarkonia, $\qqb$ mesons made of heavy quarks $Q$ and
  antiquarks $\qbar$, are some of
  the most interesting probes of the quark-gluon plasma formed in the
  relativistic heavy ion collision experiments. A theoretical
  study of the behavior of the quarkonia in the deconfined medium can
  be based on an open quantum system framework. Due to the highly
  nonperturbative nature of the plasma, the various
  interactions of the quarkonia and the plasma, required as an input
  for the framework, are diffcult to estimate and require numerical
  nonperturbative calculations.

  In this contribution we discuss some of these inputs that have been
  calculated on the lattice. We discuss an effective thermal
  potential that explains the evolution of the $\qqb$ in a certain regime.
  Since the interaction with the medium can convert the $\qqb$ to a color
  octet pair, we need information of this interaction potential in
  both color singlet and color octet configurations of the $\qqb$ pair.
  We also report on some studies related to the propagation of the $\qqb$
  in the plasma in other regimes, including the singlet-to-octet transition.
\end{abstract}


\ccode{PACS numbers:}


\subsection{Introduction}
\label{sec.intro}
Quarkonia have been among the most studied probes of quark-gluon
plasma in heavy ion collision experiments. The intuitive picture is
very simple \cite{Matsui:1986dk}: in the deconfined medium the attractive
force between the $Q$ and the $\qbar$ is reduced, and if a $\qqb$ pair
is formed in such a medium, it may not form a quarkonium. This will
lead to a clear experimental signature:
e.g., the depletion of the $\jpsi$ yield will show up in the dimuon
spectra.

The highly nonperturbative nature of the quark-gluon plasma (QGP) at
temperatures attainable in the relativistic heavy ion collision
experiments makes a precise theoretical calculation of the
modification of the quarkonia yield far more complicated. The
interaction of the $\qqb$ with the plasma depends on the relative
ordering of the various scales in the theory \cite{Laine:2006ns,
Brambilla:2008cx}.  Theoretically the most satisfactory framework
for the evolution of the $\qqb$ in the plasma is an open quantum
system based one \cite{Akamatsu:2014qsa, Akamatsu:2020ypb}.  The 
formalism applicable depends on the relative ordering of the quarkonia scales
($\mq, \; \mq v, \; \mq v^2$) and the thermal scales ($T, \; \md, ...$).

A successful effective field theory (EFT) for the study of quarkonia
in vacuum is pNRQCD (potential nonrelativistic QCD) \cite{Brambilla:2004jw}.
Due to the nonrelativistic nature of the $Q$ and the $\qbar$ in quarkonia,
a potential description is applicable. pNRQCD provides a way of
systematically adding terms suppressed as $1/\mq$. The open quantum system
formalism in this EFT has been worked out \cite{Brambilla:2016wgg}, and
has been used for phenomenological calculations for bottomonia
\cite{Brambilla:2021wkt} (see review \cite{Sharma:2021vvu} for more details).

The nonperturbative nature of the plasma enters such calculations through
a few parameters. pNRQCD is an EFT for length scales larger than $r$, the
size of the quarkonia. If $T > 1/r$ then the main effect of the medium
will be a modification of the $\qqb$ potential. On the other hand, for
$1/r \gg T$, which is probably the applicable hierarchy for $\Upsilon(1S)$,
the medium will see the $\qqb$ as a color singlet ``point'' object, which
may turn into a color octet on interacting with the medium. The calculation
of the medium effect then is the calculation of the effect of this
singlet-octet transition. 

In this report, I will discuss the nonperturbative calculation of a
thermal potential for the study of the system evolution of quarkonia,
and briefly report on a calculation of the singlet-octet transition effect.
These are nonperturbative inputs for calculation of dynamical quantities
related to quarkonia yield. Another major effort on the lattice is a direct
extraction of dynamical quantities like the spectral function from lattice
correlators. I will not report on these calculations  
(see Ref. \cite{Asakawa:2003re, Datta:2003ww} for the first studies, and
the reviews \cite{Datta:2014wga, Rothkopf:2019ipj} for a
discussion of the later studies, and the various systematics).

\subsection{Thermal potential for quarkonia}
\label{sec.pot}
The original quarkonia paper \cite{Matsui:1986dk} talked about screening
of the $\qqb$ potential. But a proper, formal definition of the screened
potential came much later\cite{Laine:2006ns, Brambilla:2008cx}, where it
was realized that the potential will include an imaginary part, related
to Landau damping, i.e., break-up of the quarkonia due to
absorption of space-like gluons. 

\begin{figure}[th]
  \centerline{\includegraphics[width=4cm]{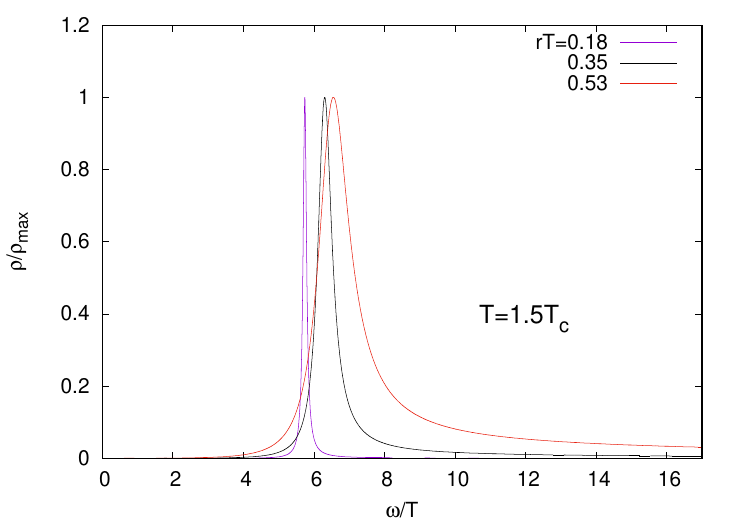}
    \includegraphics[width=4cm]{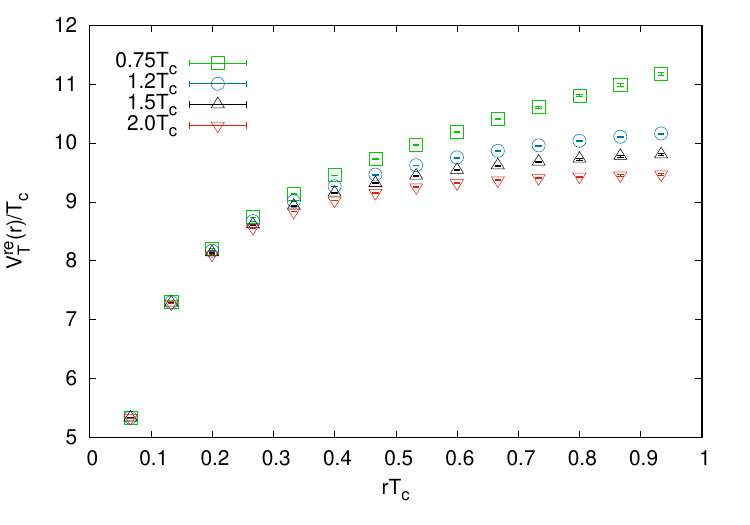}
    \includegraphics[width=4cm]{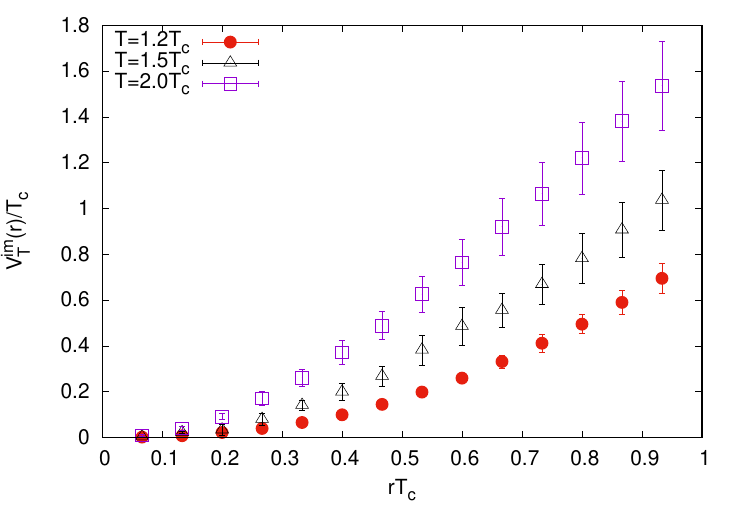}}
\caption{(Left) The low $\omega$ peak of $\rho$, as obtained from
  thermal Wilson loop at 1.5 $T_c$, in the quenched theory.
  (Middle) $\vre$ and (right) $\vim$ at various
  temperatures, for quenched QCD. From Ref. \cite{Bala:2019cqu}.}
\label{Vsqu}\end{figure}

The effective thermal potential, $V_T(r)$, can be
defined from the long time behavior of a (real time) Wilson loop
\cite{Laine:2006ns}:
\begin{equation}
  i \delt \, \wrm \xrightarrow{t \to \infty} \vt \, \wrm, \qquad
  \vt \ = \ \vre \; - \; i \, \vim .
\label{evol} \end{equation}
Nonperturbatively, we can calculate the Euclidean time Wilson loop. We
can do a spectral decomposition
\begin{equation}
  W(R, \tau) \ = \ \mathcal{N} \ \int d\omega \ e^{-\omega \tau} \
  \rho(R, \omega).
  \label{spectral} \end{equation}
The potential then depends on the low $\omega$ structure of $\rho(R,
\omega)$ \cite{Rothkopf:2011db}. Based on an analysis of the struture
of $W(R, \tau)$ and the requirement of a thermal potential as in Eq.
\eqref{evol}, it was argued\cite{Bala:2019cqu} that
at low $\omega$, $\rho(R, \omega)$ needs to have a specific peak
structure; the peak will look like a Lorentzian centered around
$\vre$ and with a width $\vim$, but will have an exponential fall-off
towards -ve $\omega$ side and a power law fall-off in the other side:
see the left panel of Figure \ref{Vsqu}.

The potential, as calculated from this spectral function, is also shown
in Fig. \ref{Vsqu}. \cite{Bala:2019cqu} At short distances, one sees
the attractive Coulomb potential. As one goes to longer distances, the
linearly rising confining potential seen below $T_c$ gets screened above
$T_c$, though the potential is different
from the perturbative, screened Coulomb form. Also the imaginary part
$\vim$ was found to be considerably larger than the
perturbative result. The potential has been calculated recently also for
QCD with 2+1 flavors of thermal quarks\cite{dibyendu}, with
qualitatively similar features. 

Note that the extraction of $\rho(\omega)$ from Eq. \eqref{spectral} is an
ill-posed problem, and the result depends to some extent on the
conditions put in on the structure of $\rho(\omega)$.
The results in Fig. \ref{Vsqu} are based on
physically well-motivated hypotheses \cite{Bala:2019cqu}. Other
structures have led to different results for the effective
thermal potential \cite{Bazavov:2023dci}. In my opinion the
assumptions put in there are not physically justified.

\begin{figure}[th]
  \centerline{\includegraphics[width=4cm]{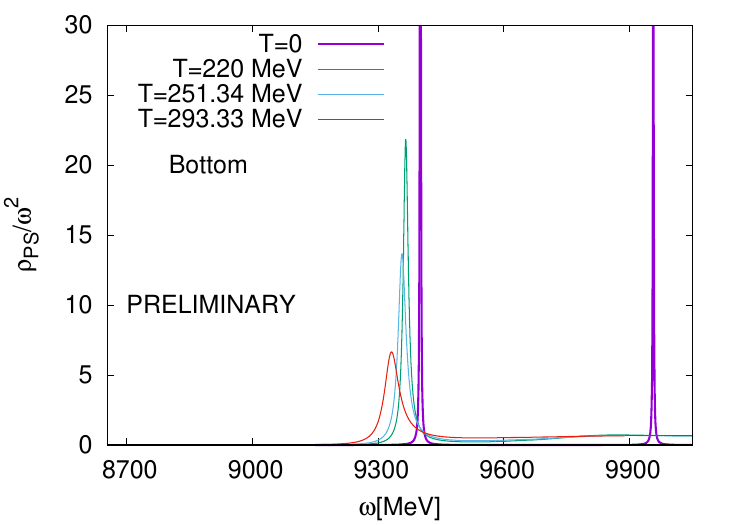}
    \includegraphics[width=4cm]{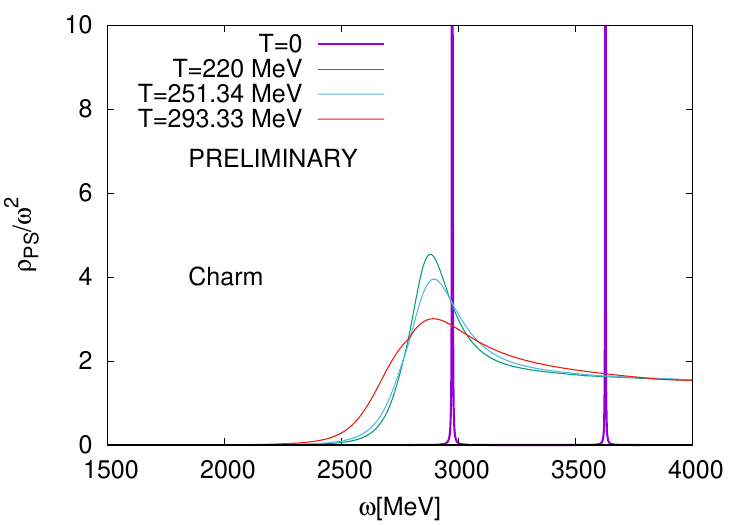}}
  \caption{The spectral function for the vector current
    calculated from the potentials shown in Figure \ref{Vsqu},
    showing the $\bar{b} \gamma_i b$ (left) and the $\bar{c} \gamma_i c$
    (right) peaks in the dilepton spectra. From ref. \cite{dibyendu}.}
\label{Spect2p1}\end{figure}

From the thermal potential, one can calculate the spectral function
for the vector current $\qbar \gamma_i Q$, \cite{Burnier:2007qm,Bala:2019cqu}
which is directly related to the quarkonia peak in the dilepton current.
The results for such a calculation \cite{dibyendu} for the charm and
the bottom, using the
potentials for 2+1 flavor QCD, are shown in Figure \ref{Spect2p1}.
Note that here the light quarks are somewhat heavier than the physical $u, d$
quarks, and $T_c \approx$ 180 MeV.
The spectral function calculated from the thermal potential can be compared
to the direct calculations of the spectral
function from the lattice \cite{Datta:2003ww,Asakawa:2003re}. The potential
has also been connected to observables related to open heavy quarks
\cite{Du:2019tjf}.

\subsubsection{A potential for $\qqb$ in a color octet configuration}
\label{sec.octet}

For a study of quarkonia in QGP, one also needs information about the
evolution of $\qqb$ in a color octet configuration. This is because
interaction with the medium will convert the $\qqs$ to $\qqo$ and vice
versa.

While one can formulate a potential description for the $\qqo$ similar
to that for $\qqs$, a nonperturbative calculation of the potential turns
out to be nontrivial. A straightforward adaptation of Eq. \eqref{evol}
will lead to Wilson loops with insertions of color matrices $T^a$, which
are not gauge invariant objects. One may think of doing a gauge fixed
calculation. But it turns out that the physical interpretation of the
``potential'' extracted from such gauge fixed Wilson loops depends on
the gauge and, in particular, it does not correspond to the potential
between a color octet $\qqb$ pair \cite{Philipsen:2013ysa}. Note that
this is not a problem with numerical analysis: even at zero temperature,
where the extraction of $V_s$ from Wilson loops is a numerically well-posed
problem, the same problem with $V_o$ persists.

In order to get an insight into the interaction between $\qqo$ in medium,
we therefore followed a different strategy \cite{Bala:2020tdt}. We started
with a gauge invariant,
extended $\qqg$ state, i.e., a hybrid of $Q, \qbar$ and a gluonic object $G$
chosen such that the $\qqb$ are in a color octet configuration. Now we can
construct the potential for this hybrid state in a way similar to
Eq. \eqref{evol}. If, as happens in perturbation theory, the $r$ dependence
of the potential is largely blind to the details of $G$, then we can extract
the $\qqo$ interaction potential from it.

\begin{figure}[th]
  \centerline{\includegraphics[width=4cm]{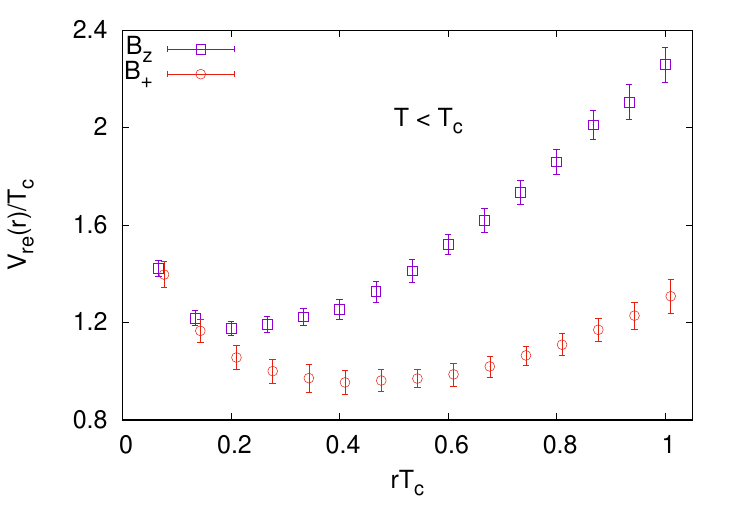}
    \includegraphics[width=4cm]{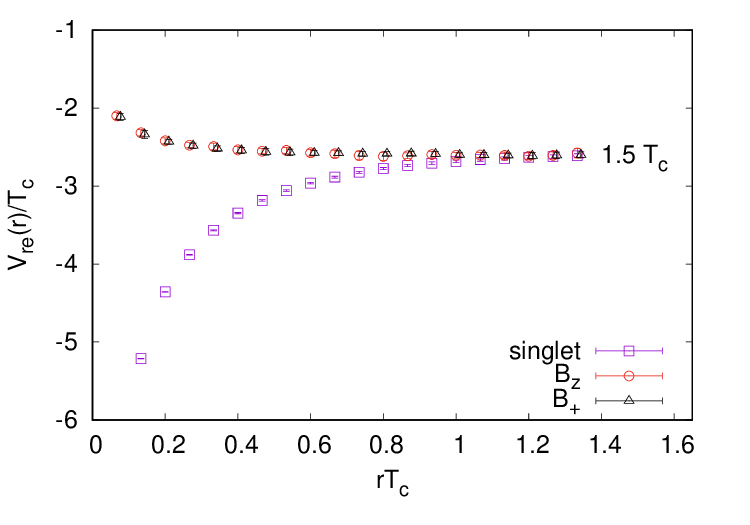}
    \includegraphics[width=4cm]{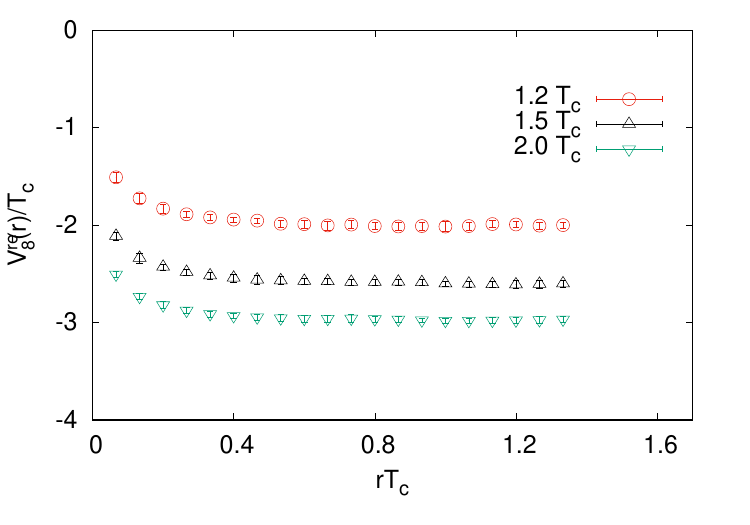}}
  \caption{(Left) The potential obtained from the hybrid Wilson loop
    at 0.63 $T_c$ in quenched theory, with $G=B_z$ and $B_+$. (Middle)
    $V^{\rm re}(r, T=1.5 T_c)$ from the two hybrid and the singlet Wilson
    loops, and (right) $\vo$ at different temperatures
    above $T_c$, for the quenched theory. From Ref. \cite{Bala:2020tdt}.}
\label{Vo}\end{figure}

In the confined phase, the potential turns out to depend sensitively on $G$
: only at very short distances we can find an
``octet potential'' separate from the details of the gluonic insertion
\cite{Bali:2003jq}. In the left panel of Fig. \ref{Vo} we show the potential
obtained in quenched QCD \cite{Bala:2020tdt} for two gluonic insertions:
magnetic field operators $B_z$ and
$B_+=B_x+i B_y$, where $z$ is the direction of separation between the $Q$ and
the $\qbar$. These two choices correspond to the $L=0$ hybrid state
$\Sigma_u^-$ and the $L=1$ state $\Pi_u$, respectively, in the standard atomic
physics notation. While at very short distances the potentials for the two
states agree, showing the repulsive octet potential, soon they
separate from each other. At long distances both of them show an
attractive potential, as expected for getting a hybrid state, but with
very different rising behaviors. As one crosses $T_c$, however, things
change dramatically. In the middle panel the potentials for the same
operators are shown at 1.5 $T_c$. They agree at all distances, and
can therefore be interpreted as the potential for $\qqo$. We note that
there is no sign of an attractive part in the octet potential at the
distance scale studied by us. As the figure
shows, at large distances the octet and the singlet potential approach
each other.  

In the right panel of Figure \ref{Vo} we have shown the results for
$\vo$ at different temperatures above $T_c$. At each temperature we
compared, and found agreement of, results with $B_z$ and $B_+$
insertions. As in the singlet case, the octet potential above $T_c$ is
complex in general. See ref. \cite{Bala:2020tdt} for the results of
the imaginary part.

\subsection{$\tfrac{1}{r} \gg T$: quarkonia diffusion}
\label{sec.kappa}
For $\tfrac{1}{r} \gg T$, which seems applicable for ground state bottomonia,
the thermal particles cannot resolve the $Q$ and
the $\qbar$. The interaction of the $\qqb$ with the medium has a color dipole
structure $\vr \cdot \vec{E}^a$ \cite{Brambilla:2008cx,Brambilla:2016wgg}.
Such an interaction changes the $\qqb$ from color singlet to octet, and
vice versa. Correlators related to the propagation of such color singlet
and octet states can be calculated nonperturbatively \cite{tumqcd:2024ip}.
Under certain assumptions\cite{Brambilla:2016wgg, Sharma:2023dhj},
the decay width of the $\Upsilon(1S)$ can be approximated as
$\Gamma \, = 3 a_0^2 \kappa$, where $a_0$ is the Bohr radius and
\begin{equation}
  \kappa \; = \; {\rm Re} \; \frac{g^2}{6 N_c} \, \int_{- \infty}^\infty ds \,
  \cadj(s), \qquad \cadj(s) =
  \langle T \,  E^{a, i}(s, \vec{0}) \,
  W^{ab}(s, 0) \, E ^{b, i}(0, \vec{0}) \rangle_T \, .
  \label{kappa} \end{equation}
$W^{ab}(s, 0)$ is adjoint Wilson line.
A nonperturbative calculation of $\kappa$ would require a spectral
decomposition of the Euclidean correlator $\cadj(\tau)$.

\begin{figure}[th]
\centerline{\includegraphics[width=5cm]{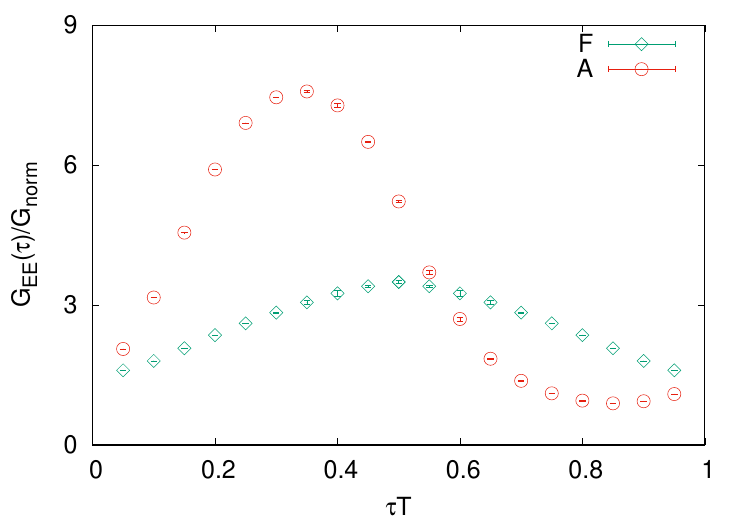}}
\caption{Comparison of $\cadj(\tau)$ and $\cfund(\tau)$ at 1.5 $T_c$
  in quenched QCD.}
\label{fadj}\end{figure}

At the moment no calculation of $\kappa$ exists, though computations
are in progress \cite{mumtum:2024mb}. The quantity $\cadj(\tau)$ is
somewhat similar to the correlator
\begin{equation}
  \cfund(\tau) = \langle W_{kl}(1/T,\tau) \, E^i_{lm}(\tau, \vec{0}) \,
  W_{mn}(\tau, 0) E^i_{nk}(0, \vec{0}) \rangle_T
  \label{kf} \end{equation}
used for the calculation of the heavy quark diffusion
coefficient $\kappa_f$ \cite{Caron-Huot:2009ncn}. $\kappa_f$ has been
calculated on the lattice by various groups (see Ref.
\cite{Banerjee:2022gen, Altenkort:2023oms} for recent
results in quenched and 2+1 flavor QCD, respectively). In the absence
of a calculation of $\kappa$, estimates of $\kappa_f$ have sometimes
been used instead \cite{Brambilla:2016wgg}. The relation between
$\cadj(\tau)$ and $\cfund(\tau)$ has been estimated in perturbation
theory \cite{Scheihing-Hitschfeld:2023tuz}: the correlators agree in
leading order, and in NLO, the spectral functions differ only by a
temperature independent term. Nonperturbatively, the correlators are
seen to differ quite substantially \cite{tumqcd:2024ip, mumtum:2024mb}
(see Figure \ref{fadj}). In
particular, $\cadj(\tau)$ is strongly asymmetric around the distance
$\tau=1/2T$, which makes the task of extraction of $\kappa$
nontrivial.

\section{Heavy Quark dynamics in a hot magnetized medium}

\author{Aritra Bandyopadhyay}



\bigskip

\begin{abstract}
In this proceeding, we examine the momentum diffusion of heavy quarks in a hot magnetized medium across a wide range of magnetic field strengths. Using effective gluon and quark propagators for hot magnetized medium, we systematically incorporate the magnetic field effects. To derive gauge independent analytic form factors valid for all Landau levels, we apply the hard thermal loop technique for the resummed gluon propagator. This allows us to analytically calculate the longitudinal and transverse momentum diffusion coefficients for charm and bottom quarks, extending beyond the static limit. We also explore an alternative method in the static limit, using a magnetized medium-modified Debye mass, and compare the results to assess the need for structural adjustments.
\end{abstract}


\ccode{PACS numbers:}


\subsection{Introduction}

Heavy quarks are widely studied in the heavy-ion community as key indicators for understanding the properties of hot, dense quark matter. Due to the non-centrality of heavy-ion collision (HIC) experiments and the resulting strong magnetic fields, recent studies have focused on heavy quarks in a magnetized medium. In contrast to most previous research, which considered limited scenarios, present study addresses the more general case of arbitrary external magnetic field strengths.

The heavy quark (HQ) momentum diffusion coefficients are crucial theoretical quantities needed to describe the evolution of HQ. We use the widely adopted Langevin equations which assumes that HQs receive random "kicks" from the thermal partons in the surrounding medium. In absence of the magnetic field $eB$ and within the nonrelativistic static limit of HQ there is no anisotropy imposed on the system resulting in a single diffusion coefficient $\kappa$. Several studies have evaluated this $\kappa$ using various techniques~\cite{Caron-Huot:2007rwy}. Moving beyond the static limit~\cite{Braaten:1991we,Moore:2004tg} and introducing an external magnetic field attributes two anisotropy directions in the system and consequently $\kappa$ splits into longitudinal and transverse components. Most current HQ studies within magnetized medium have adopted the Lowest-Landau-Level (LLL) approximation~\cite{Fukushima:2015wck,Bandyopadhyay:2021zlm}, or the weak magnetic field approximation\cite{Dey:2023lco}. However, extending calculations to arbitrary external magnetic fields liberates us from the limitations imposed by the scale of $eB$.

To calculate heavy quark (HQ) momentum diffusion coefficients, which depend on the HQ scattering rate in the presence of arbitrary magnetic fields, it is essential to compute the effective gluon propagator for a hot arbitrarily magnetized medium. This requires using hard thermal loop (HTL) approximations to derive form factors for the effective gluon propagator across all Landau levels. Recent approaches have introduced magnetic field effects in a different way, such as through the medium-dependent Debye mass in calculations of HQ potential and energy loss~\cite{Nilima:2022tmz,Jamal:2023ncn}. By comparing results within the static limit, we gain insights into the limitations of using the simplified Debye mass approximation.

\subsection{Formalism}
When the HQ moves (with momentum $P$ / velocity $\vec{v}$) in the presence of an external magnetic field ($\vec{B}$), focusing on two distinct scenarios is beneficial: 
$\vec{v} \shortparallel \vec{B}$ and $\vec{v} \perp \vec{B}$. In the former case $\vec{v} \shortparallel \vec{B}$, two distinct diffusion coefficients $\kappa_L$ and $\kappa_T$ arise, which are linked to the HQ scattering rate $\Gamma$ as
 \begin{align}
\kappa_T (P) = \frac{1}{2}\int d^3q\frac{d\Gamma(P)}{d^3q}q_\perp^2,~~ \kappa_L (P) = \int d^3q\frac{d\Gamma(P)}{d^3q}q_z^2.
\label{coeffs_case1}
\end{align}
On the contrary, $\vec{v} \perp \vec{B}$ case generates three different diffusion coefficients $\kappa_j$'s ($j=\{x,y,z\}\equiv\{1,2,3\}$), i.e.
\begin{align}
\kappa_j (P) = \int d^3q\frac{d\Gamma(P)}{d^3q}q_j^2.
\label{coeffs_case2}
\end{align}

It's clear that when we consider the static limit ($\vec{v}\rightarrow 0$) in a magnetized medium, only one anisotropy remains, dictated by the direction of $\vec{B}$, which causes the scenario $\vec{v} \perp \vec{B}$ to vanish.

From Eqs. \ref{coeffs_case1} and \ref{coeffs_case2}, it's evident that to calculate the $\kappa$'s, we first need to determine the $\Gamma$ of $2\leftrightarrow 2$ scatterings between the light quark/gluon and the heavy quark (i.e. $qH\leftrightarrow qH$ and $gH\leftrightarrow gH$). To evaluate these scattering rates, we employ an effective method initially introduced by Weldon~\cite{Weldon:1983jn} which connects $\Gamma$ to the cut/imaginary parts of the HQ self-energy as
\begin{align}
\Gamma(P) = -\frac{1}{2E}~\frac{1}{1+e^{-E/T}}~\rm{Tr}\left[(\slashed{P}+M)~\rm{Im} ~\Sigma(p_0+i\epsilon,\vec{p})\right],
\end{align}
where $\Sigma(P)$'s represent the effective HQ self energy with an HTL resummed effective gluon propagator which duly incorporates the necessary soft contributions. The detailed calculation of $\Gamma(P)$ and hence $\kappa_i$'s has been done in Ref.~\cite{Bandyopadhyay:2023hiv}.

\subsection{Results}

\begin{figure*}[ht]
\begin{center}
\includegraphics[scale=0.35]{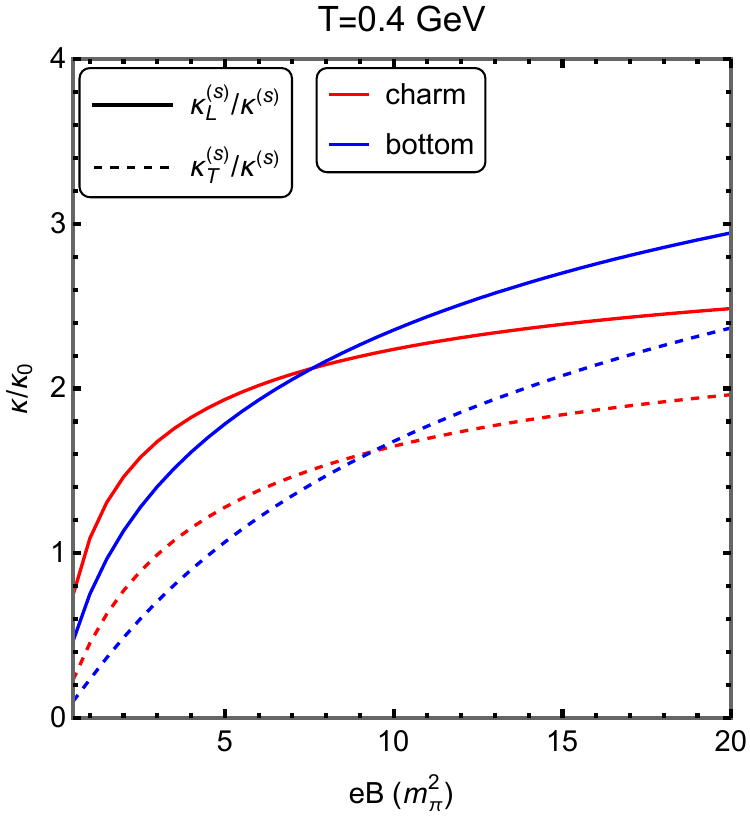}
\hspace{1cm}
\includegraphics[scale=0.35]{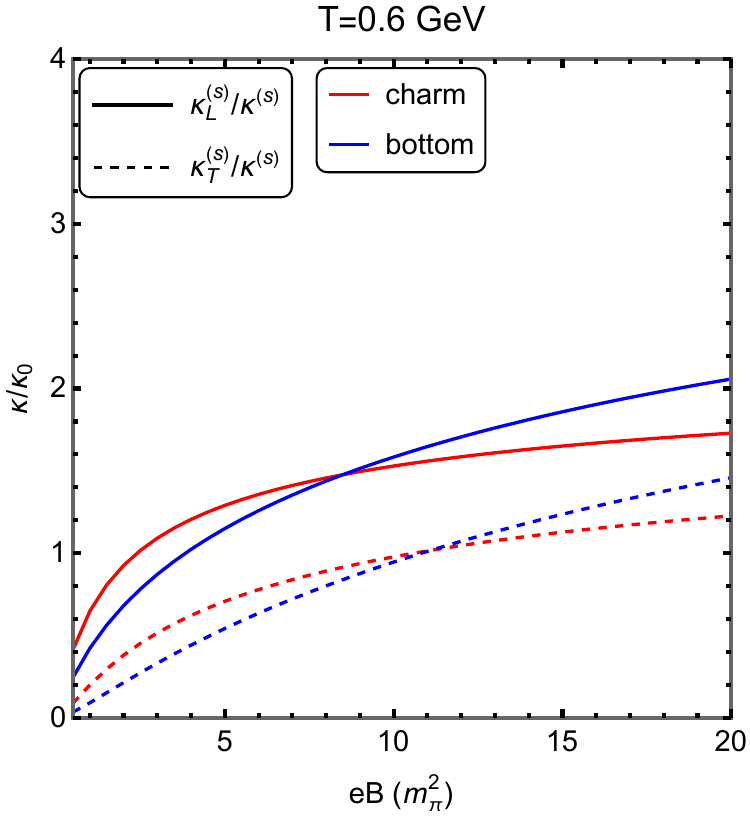}
\caption{The variation of scaled momentum diffusion coefficients with $eB$ in the static limit.} 
\label{kappa_ratio_static_wzB}
\end{center}
\end{figure*}

We first present our static limit results for a magnetized medium in Fig.~\ref{kappa_ratio_static_wzB}, showing how the ratio $\kappa_{L/T}^{(s)}/\kappa^{(s)}$ changes with the external magnetic field. Here, $\kappa^{(s)}$ represents the isotropic momentum diffusion coefficient at zero magnetic field. We focus on higher temperatures, $T=0.4$ and $T=0.6$ GeV, in line with the HTL approximation used in our study. At lower $eB$, both longitudinal (solid) and transverse (dashed) diffusion coefficients increase more sharply, especially for charm quarks (red), crossing over the bottom quark (blue) curves. Throughout, $\kappa_L$ remains higher than $\kappa_T$ for both quark types, and the overall ratio decreases with rising temperature due to the interplay between $eB$ and $T$. The dominance of $\kappa_L$ over $\kappa_T$ can be attributed to the predominant gluonic contributions in the $t$-channel scatterings studied in this context.

\begin{figure*}[ht]
\begin{center}
\includegraphics[scale=0.35]{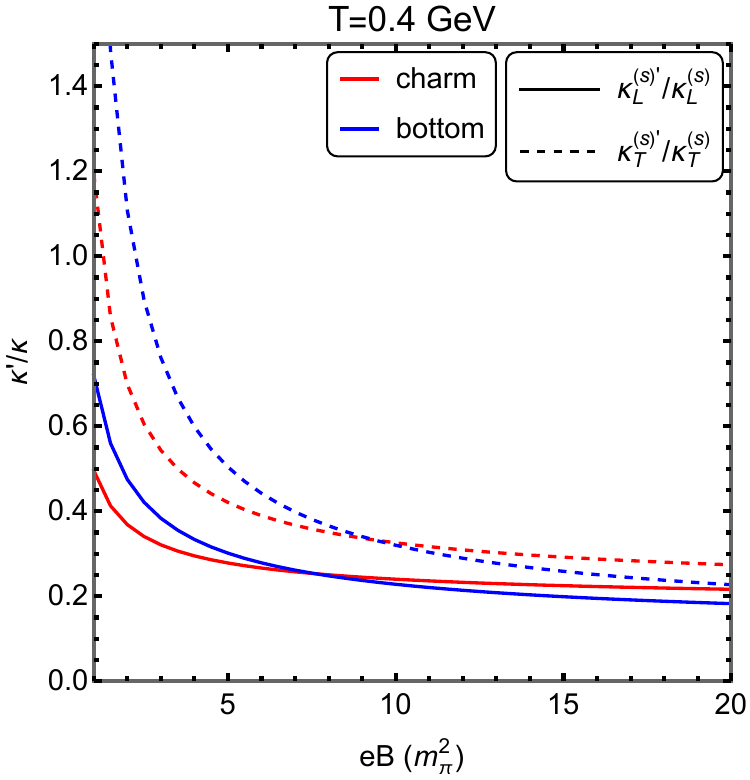}
\hspace{1cm}
\includegraphics[scale=0.35]{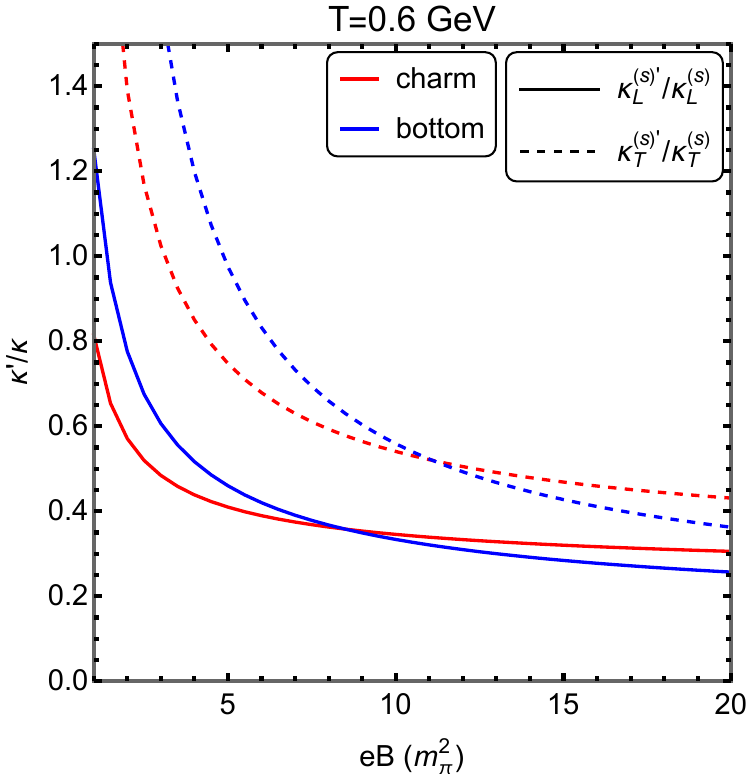}
\caption{The variation of the ratio between the Debye mass approximated results ($\kappa'$) and the exact results ($\kappa$) as a function of magnetic field strength $eB$ in the static limit.} 
\label{kappa_ratio_static}
\end{center}
\end{figure*}

In the present study, we also explore an alternative method where, all medium effects are encapsulated through the medium-modified Debye screening mass. In Fig.~\ref{kappa_ratio_static}, we present comparisons between results obtained from the exact calculation ($\kappa$) and those from the Debye mass approximation ($\kappa'$). The general trend we found is similar for all the cases: the Debye mass approximated results underestimate the exact results for larger values of $eB$ and overestimate them for smaller values of $eB$. These discrepancies are more pronounced for bottom quarks due to their heavier mass ($M_b = 4.18$ GeV) compared to charm quarks ($M_c = 1.27$ GeV). 

\begin{figure*}
\begin{center}
\includegraphics[scale=0.35]{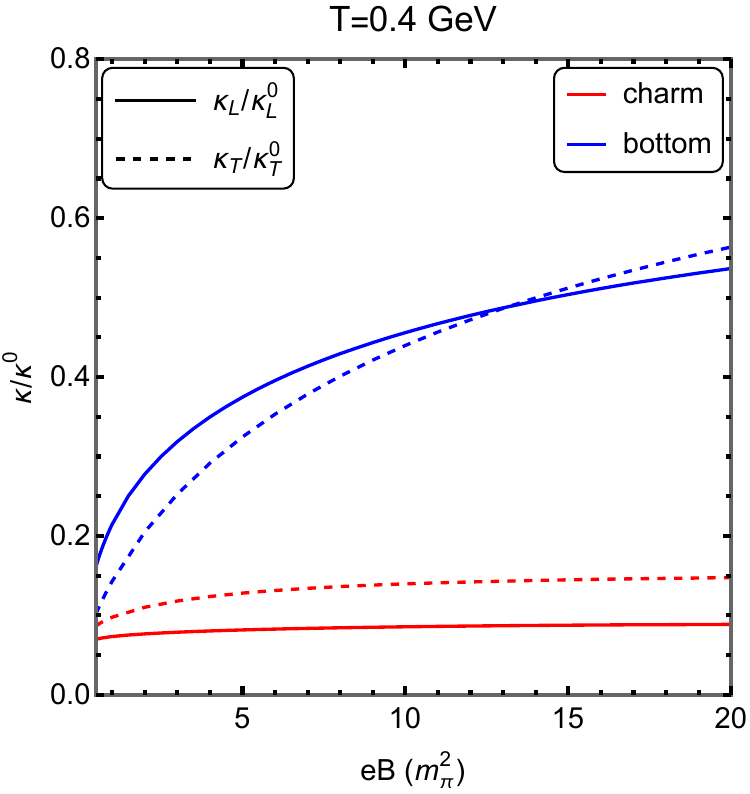}
\hspace{1cm}
\includegraphics[scale=0.35]{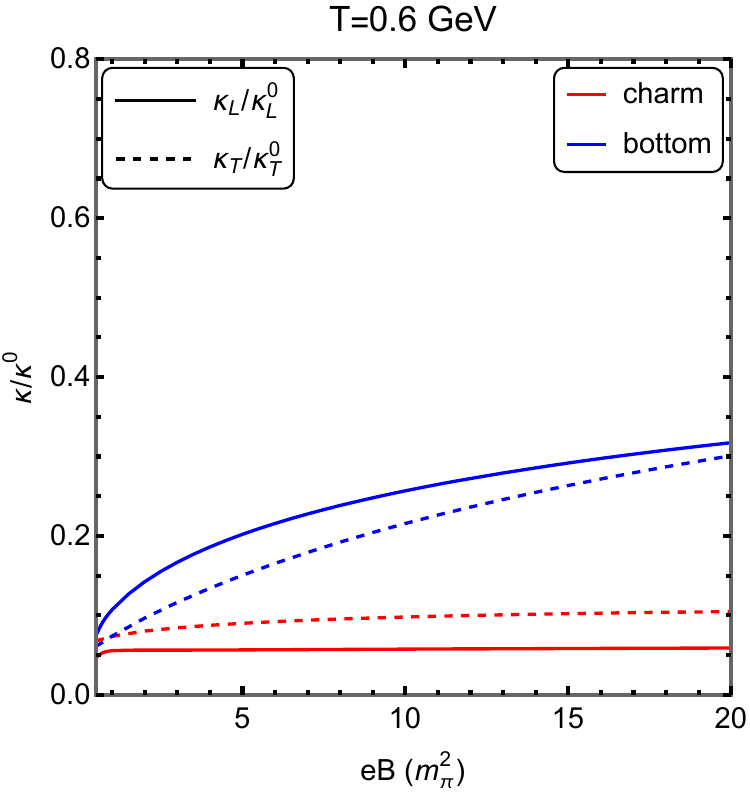}
\caption{$\vec{v}\shortparallel \vec{B}$ case : scaled (w.r.t eB=0 values) momentum diffusion coefficients as functions of $eB$.} 
\label{kappaveB_case1}
\end{center}
\end{figure*}

\begin{figure*}
\begin{center}
\includegraphics[scale=0.35]{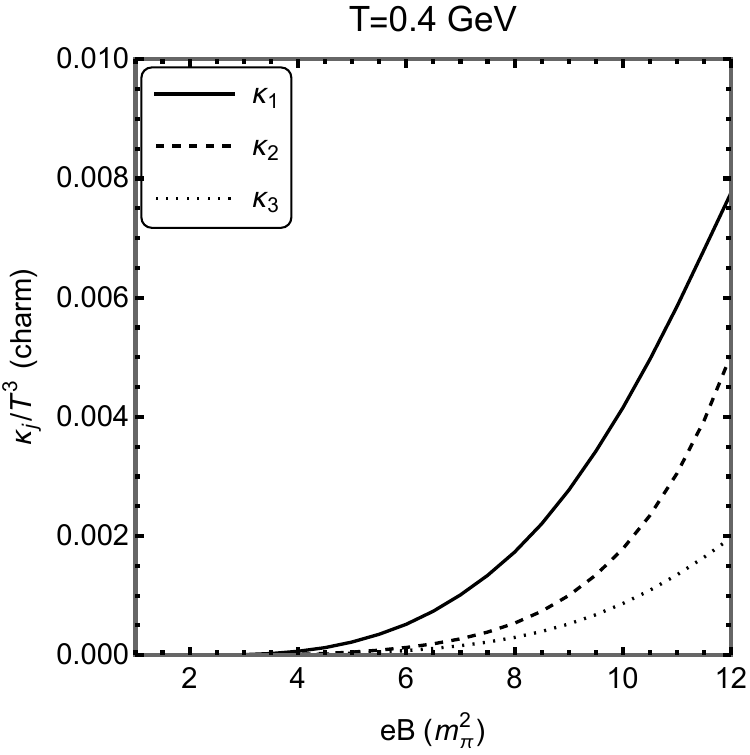}
\hspace{1cm}
\includegraphics[scale=0.35]{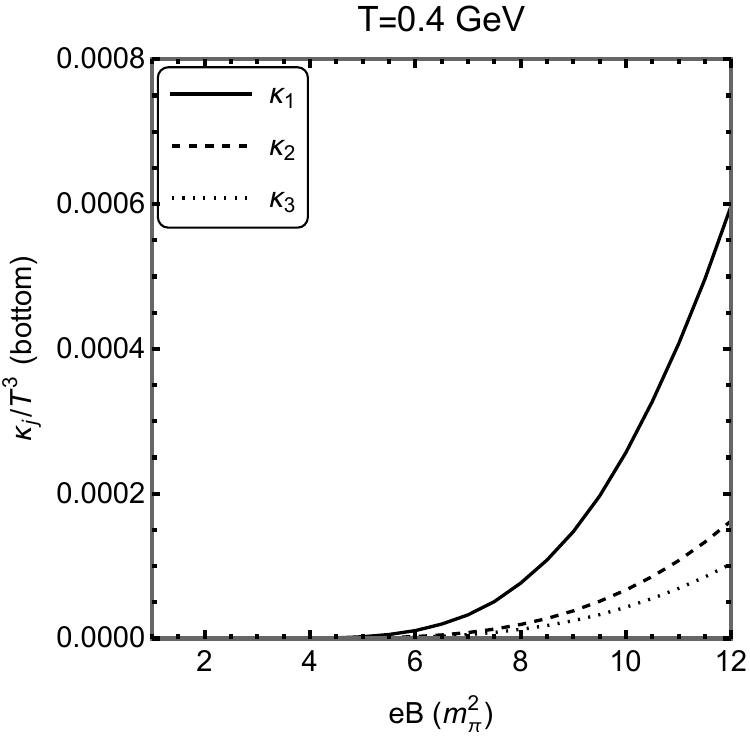}
\caption{$\vec{v}\perp \vec{B}$ case : scaled (w.r.t $T^3$) momentum diffusion coefficients as functions of $eB$. } 
\label{kappaveB_case2}
\end{center}
\end{figure*}

Next we present our estimates for longitudinal and transverse momentum diffusion coefficients beyond the static limit, fixing the HQ momentum at $p = 1$ GeV. In the $\vec{v} \parallel \vec{B}$ scenario, Fig.\ref{kappaveB_case1} illustrates that similar to the static limit, for larger $eB$, the rate of change of $\kappa_{L/T}$ flattens, particularly for charm quarks (red curves). Interestingly, $\kappa_T$ (dashed) dominates over $\kappa_L$ (solid) across the entire $eB$ range for charm quarks. For bottom quarks (blue curves), $\kappa_L$ exceeds $\kappa_T$ until a higher $eB$ threshold, where a crossover occurs. These crossovers reflect the interplay between the mass ($M$), temperature ($T$), and magnetic field ($eB$). Due to their higher mass, bottom quarks require a stronger magnetic field to display similar $\kappa_{L/T}$ behavior as charm quarks. This trend is clearer in the right panel, where no immediate crossover indicates that an even higher $eB$ is needed for bottom quarks.

Fig.~\ref{kappaveB_case2} shows results for the perpendicular case, $\vec{v} \perp \vec{B}$. Since no such configuration exists at $eB=0$, we scale $\kappa_j$ with $T^3$. Here, the transverse components $\kappa_1$ (solid) and $\kappa_2$ (dashed) dominate over the longitudinal component $\kappa_3$ (dotted) for both charm and bottom quarks. Unlike previous cases, the rates of change of $\kappa_j$'s increase with $eB$. Additionally, bottom quark coefficients (right panel) are roughly an order of magnitude lower than those for charm quarks (left panel).

\subsection{Summary}

In this proceeding, we have analyzed the momentum diffusion coefficients of heavy quarks (HQ) in a magnetized medium. By comparing exact results with a Debye mass approximation in the static HQ case, we highlighted the limitations of the latter, emphasizing the importance of including the full gluon two-point correlation functions in a hot, magnetized medium. In both static and non-static scenarios (especially when HQ moves parallel to the magnetic field), we observed a rapid increase in diffusion coefficients at low $eB$, followed by saturation, particularly for charm quarks. When HQ moves perpendicularly, the diffusion coefficients continue to grow with increasing $eB$. In the static limit, longitudinal diffusion dominates due to soft gluon scatterings, but beyond the static limit, this trend reverses. The interplay of scales—mass ($M$), momentum ($p$), magnetic field ($eB$), and temperature ($T$)—is evident in our findings. This study paves the way for future research, such as exploring HQ in-medium evolution and its impact on experimental observables like directed and elliptic flow of open heavy flavor mesons.


\section{Heavy quark momentum diffusion in weakly magnetized QGP}

\author{Debarshi Dey, Binoy Krishna Patra}

\bigskip

\begin{abstract}
The momentum diffusion coefficient $\kappa$ is evaluated for Charm and Bottom quarks in a QGP medium in the presence of a weak background magnetic field, for the cases of the heavy quark moving either parallel ($\bm{v}\parallel \bm{B}$) or perpendicular ($\bm{v}\perp \bm{B}$) to the direction of the magnetic field. For both the cases, our results show that the
momentum transfer between the HQ and the medium takes place preferentially along the direction of HQ velocity, thus leading to a significant increase in the momentum-diffusion anisotropy, compared to $\bm{B}=0$. 
\end{abstract}


\ccode{PACS numbers:}


\subsection{Heavy quarks in a thermal medium}
The heavy-quark (HQ) mass is the largest scale in the problem; $M_Q\gg T\gg \sqrt{eB}$. As such, it takes a large number of collisions with the light medium partons to change the HQ momentum $p$ by $\mathcal{O}(1)$. Thus, the evolution of HQ momentum is described via Langevin equations\cite{Fukushima:PRD'2016,Bandyopadhyay:PRD'2022}: 
\begin{align}
	\frac{dp_{z}}{dt}&=-(\eta_D)_{\parallel}p_z+\xi_z\,,\quad \langle \xi_z(t)\xi_z(t')\rangle=\kappa_{\parallel}(\bm{p})\,\delta(t-t')\\
	\frac{d\bm{p}_{\perp}}{dt}&=-(\eta_D)_{\perp}\bm{p}_{\perp}+\bm{\xi}_{\perp}\,,\quad \langle \bm{\xi}_{\perp}(t)\bm{\xi}_{\perp}(t')\rangle=\bm{\kappa}_{\perp}(\bm{p})\,\delta(t-t'),
\end{align} where, $\eta_z$ and $\bm{\eta_{\perp}}$ are the components of random forces parallel and perpendicular to $\bm{B}$., and the momentum diffusion coefficients are obtained as
\begin{align}
	\kappa_{\parallel}&=\int d^3 q \frac{d \Gamma(E)}{d^3 q} q_{\parallel}^2\label{kappal} \\[0.2em]
	\bm{\kappa_{\perp}}&=\frac{1}{2}\int d^3 q \frac{d \Gamma(E)}{d^3 q} \bm{q_{\perp}}^2\label{kappap}
\end{align}
The scattering rate $\Gamma$ is evaluated using Weldon's formula\cite{ PhysRevD.28.2007}
\begin{equation}
	\Gamma(P \equiv E, \mathbf{v})=-\frac{1}{2 E} [1-n_F(E)] \operatorname{Tr}\left[(\not P+M_Q) \operatorname{Im} \Sigma\left(p_0+i \epsilon, \bm{p}\right)\right].\label{gamma}
\end{equation}
$\operatorname{Im} \Sigma$ is related to the $t$-channel scatterings of the HQ with light thermal quarks and gluons via cutting rules\cite{Thoma} the one-loop HQ self-energy $\Sigma$ is evaluated using a HTL resummed gluon propagator $D^{\mu\nu}$ \cite{Karmakar:EPJC'2019}. Finally, $\Gamma$ evaluates to\cite{Dey:PRD'2024} 
\begin{equation}
	\Gamma(E,\bm{v})=\frac{\pi g^2}{2E}\sum_{i=1}^4\int\frac{d^3 q}{(2 \pi)^3}  \int_{-\infty}^{+\infty} d \omega\left[1+n_B(\omega)\right]\frac{\rho_i(\omega, q) A_i}{2 E}\delta(\omega-\bm{v}\cdot\bm{q}).\label{gamma_final},
\end{equation}
where, $A^i$'s are the relevant traces that appear in $\Sigma$\cite{Dey:PRD'2024}.
\subsubsection{Case 1: $v\parallel B$}
Modifying the $A^i$'s for the case $\bm{v}\parallel\bm{B}$, and using $\Gamma$ from Eq. \eqref{gamma_final} in Eq.s \eqref{kappal} and \eqref{kappap}, we arrive at
\begin{align}
	\kappa_L&=\frac{\pi g^2T}{4E^2\,v}\int dq\, q^3\int_{-vq}^{vq} d \omega\sum_{i=1}^4\frac{\rho_i(\omega, q,\frac{\omega}{vq}) A_i^{\parallel}}{\omega}\left(\frac{\omega^2}{v^2q^2}\right).\label{kappa4L}\\
	\kappa_T&=\frac{\pi g^2T}{4E^2\,v}\int dq\, q^3\int_{-vq}^{vq} d \omega\sum_{i=1}^4\frac{\rho_i(\omega, q,\frac{\omega}{vq}) A_i^{\parallel}}{\omega}\left(1-\frac{\omega^2}{v^2q^2}\right).\label{kappa4T}
\end{align}
\subsubsection{Case 2: $v\perp B$}
The direction of HQ velocity in the $x$-$y$ plane can be specified by the azimuthal angle $\phi'$ (The polar angle $\theta'=0$, as it is in the $x$-$y$ plane). The vector $\bm{q}$ is specified by $q$, $\theta$ and $\phi$ (our integration variables), where, $\theta$ is the polar angle, and $\phi$ is the azimuthal angle. Then,
\begin{equation}
	\bm{v}\cdot\bm{q}=vq\sin\theta\cos(\phi-\phi'),
\end{equation}
where, $v=|\bm{v}|$, $q=|\bm{q}|$. The interaction rate becomes
\begin{align*}
	\Gamma(E,\bm{v})=\frac{\pi g^2}{4E^2(2\pi)3}\sum_{i=1}^4&\int dq\int_0^\pi d\theta\sin\theta\int_0^{2\pi}d\phi  \int_{-\infty}^{+\infty} d \omega\left[1+n_B(\omega)\right]\\\nonumber
	&\rho_i(\omega, q) A_i^{\perp}\delta[\omega-vq\sin\theta\cos(\phi-\phi')].
\end{align*}
For the case $\bm{v}\perp \bm{B}$, 3 coefficients are defined conventionally
\begin{equation}
	\kappa_{1}=\int d^3 q \frac{d \Gamma(E,v)}{d^3 q} q_{x}^2,\quad 
	\kappa_{2}=\int d^3 q \frac{d \Gamma(E,v)}{d^3 q} q_y^2,\quad 
	\kappa_{3}=\int d^3 q \frac{d \Gamma(E,v)}{d^3 q} q_z^2.
\end{equation}	
They finally evaluate to:
\begin{align*}
	\kappa_{1}&=\frac{g^2}{16E^2v^3\pi^2}\int dq\int_{-\phi'}^{2\pi-\phi'}dy\int_0^{v q\cos y} d \omega\,
	\frac{\omega^3\cos^2(y+\phi')[1+n_B(\omega)]}{\cos^3y\sqrt{v^2q^2\cos^2y-\omega^2}}\sum_{i=1}^4A_i^{\perp}\rho_i(\omega, q).\\[0.3em]
	\kappa_{2}&=\frac{g^2}{16E^2v^3\pi^2}\int dq\int_{-\phi'}^{2\pi-\phi'}dy\int_0^{v q\cos y} d \omega\,
	\frac{\omega^3\sin^2(y+\phi')[1+n_B(\omega)]}{\cos^3y\sqrt{v^2q^2\cos^2y-\omega^2}}\sum_{i=1}^4A_i^{\perp}\rho_i(\omega, q).\\[0.3em]
	\kappa_{3}&=\frac{g^2}{16E^2v^3\pi^2}\int dq\int_{-\phi'}^{2\pi-\phi'}dy\int_0^{v q\cos y} d \omega\,
	\frac{q\omega\sqrt{v^2q^2\cos^2y-\omega^2}[1+n_B(\omega)]}{\cos^3y}\sum_{i=1}^4A_i^{\perp}\rho_i(\omega, q).
\end{align*}

\subsection{Results}
\begin{figure}[H]
    \begin{minipage}{0.48\textwidth}
        \centering
        \includegraphics[width=5.8cm,height=5.3cm]{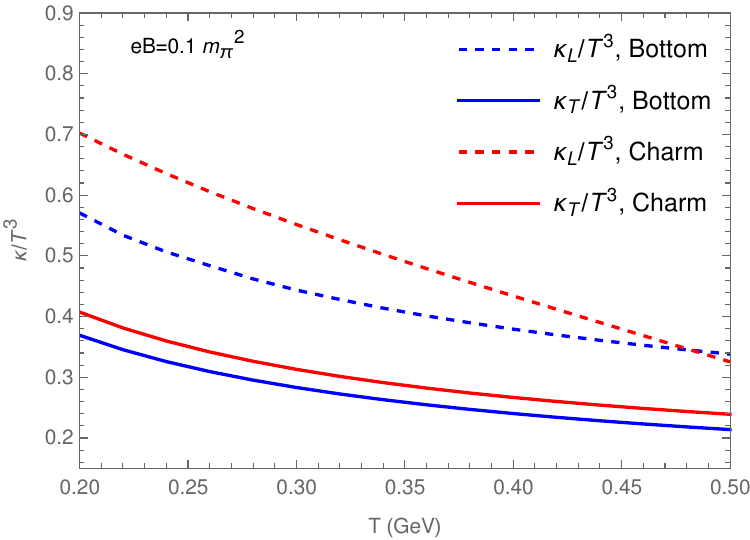}
        \label{seebeck_med}
    \end{minipage}
    \hspace*{\fill}
    \begin{minipage}{0.48\textwidth}
        \centering
        \includegraphics[width=5.8cm,height=5.3cm]{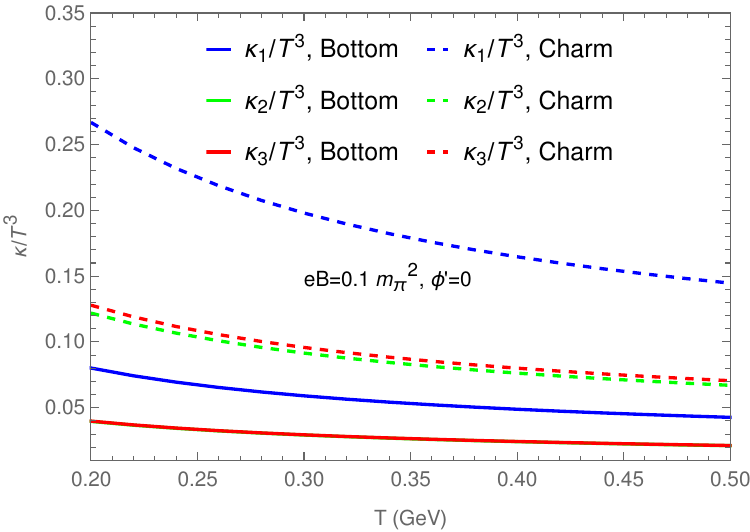}
        \label{nernst_med}
    \end{minipage}
    \caption{Variation of normalised momentum diffusion coefficients for charm and bottom quarks with temperature at a fixed magnetic field. \textbf{Left:} $\bm{v}\parallel\bm{B}$, \textbf{Right:} $\bm{v}\perp\bm{B}$.}
\end{figure}

The figures clearly show the significant anisotropy in the momentum diffusion. This is in stark contrast with the $B=0$ case\cite{Dey:PRD'2024}. Normalised $\kappa$'s are decreasing functions of temperature. The charm values are larger than the bottom values because of the lighter mass of the former. The key takeaway is that in both the cases ($\bm{v}\parallel \bm{B}$, $\bm{v}\perp \bm{B}$), the momentum transfer between the HQ and the medium happens preferentially along the direction of
HQ velocity.



\section{Jet quenching and medium response in a transport calculation}

\author{Monideepa Maity, Subrata Pal}

\bigskip

\begin{abstract}
We have incorporated medium-induced gluon radiation formalism within the AMPT transport model
in addition to the existing parton elastic parton scattering. Within the unified framework of jet
and bulk medium evolution, we make realistic predictions of jet based observables such as the nuclear 
modification of jet energy loss and full jet shapes in central Pb-Pb collisions 
relative to $pp$ collisions at 5.02 TeV.
\end{abstract}


\ccode{PACS numbers: 12.38.Mh, 24.85.+p, 25.75.-q}

\subsection{Introduction}
Jet quenching has been a key signature in the discovery and characterization of hot
and dense Quark-Gluon Plasma (QGP) formed in ultrarelativistic heavy-ion collisions
\cite{CMS:2016qnj,CMS:2018zze,ATLAS:2018gwx,ALICE:2019qyj}.
The high-$p_T$ quark and gluon jets produced in initial hard collisions suffer about a 
factor of two or more energy loss in traversing the QGP medium as compared to cold nuclear
matter envisaged in proton-proton collisions. Extensive model analysis of 
high-$p_T$ hadron suppression and dijet momentum imbalance data have established
that the quenching is dominated by medium-induced radiative parton energy loss over the 
collisional energy loss 
\cite{Schenke:2009gb,Zapp:2013vla,He:2015pra,Tachibana:2017syd,Cao:2024pxc}. 

To better characterize the QGP medium, particular interest has been in more differential jet 
observables pertaining to finding the remnants of the lost energy as well as in the jet-induced 
medium flow and medium excitations resulting in excess of soft particle production. 
The full jet structure within a jet cone radius $R$ (a distance parameter from the jet axis) 
is then sensitive to medium effects of rescattering the final parton shower from the
core of a jet to outside of the jet cone. While early attempts were mainly confined within 
QCD analytic frameworks, evidently the modifications of jet substructure 
and jet shape at large $R$ depend crucially on realistic heavy-ion model simulation 
of the space-time dynamics of medium evolution, the associated collective flow, 
and the complicated interactions between the full jet with the background medium.
This also underscores the importance to employ the same background subtraction method which becomes a part 
of jet reconstruction and jet definition as used in experimental analysis. 
Studies within jet-transport coupled to relativistic hydrodynamic model as background assumes
an instantaneous thermalization of lost energy into the medium thereby not retaining any memory
of correlation to the jet direction \cite{Tachibana:2017syd}.

In this article we explore the medium impact on the full jet structure by developing the MultiPhase Transport
(AMPT) model \cite{Lin:2004en,Pal:2012gf} to include radiative multiple gluon emission. 
The current version of the model that has parton transport solely via two-body elastic scattering, 
has achieved remarkable success in addressing the anisotropic flow and its fluctuations of 
bulk hadrons \cite{Bhalerao:2014xra}.
While collisional energy loss may to some extent mimic the jet physics at RHIC
\cite{Cao:2024pxc,Pal:2003zf}, the fat jets and highly opaque medium at 
LHC should enforce dominant medium-induced gluon radiation.

\subsection{Formalism}

We employ the string-melting version of AMPT model \cite{Lin:2004en,Pal:2012gf}
that involves four stages: Initial nucleon
distribution within colliding nuclei and subsequent soft and hard collisions, ZPC parton transport,
hadronization and ART hadron transport. Hard jets and associated parton showers are initially triggered 
via Pythia 6.4 (encoded within HIJING 2.0 \cite{Pal:2012gf}) for high statistics events. 
The full jet and the soft bulk partons
evolve in Boltzmann transport with elastic collisions. The 2-parton $ab\to cd$ collision includes $qq\to qq$, 
$qg \to qg$, $gg \to qq$ and equivalently for $\bar q$. The elastic scattering cross section 
regulated for collinear divergences by the Debye mass $\mu$ is given by \cite{Lin:2004en,Pal:2003zf}
\begin{equation}
\frac{d \sigma_{ab \to cd}}{d\hat t} = C_a 4\pi \alpha_s^2
\left(1 + \frac{\mu^2}{\hat s} \right) \frac{1}{(\hat t - \mu^2)^2} ,
\label{eq:elastic}
\end{equation} 
where we have taken the strong coupling constant $\alpha_s = 0.33$ and $\mu = 3.226$ fm$^{-1}$.
As mentioned, for jet studies it is imperative to include the inelastic scattering by medium-induced 
gluon bremsstrahlung (see Fig. \ref{fig:Elos}) which we adopt from higher-twist energy loss calculation
\cite{Guo:2000nz,Majumder:2009ge} as
\begin{equation}
\frac{dN_g^a}{dz dk_\perp^2 dt} = \frac{2\alpha_s C_a}{\pi} {\hat q}_a
\frac{k_\perp^4 P_a(z)}{(k_\perp^2 + z^2 m_a^2)^4}  \sin^2 \left( \frac{t-t_i}{2\tau_f} \right).
\label{eq:split}
\end{equation} 
Here $k_\perp$ and $z=\omega/E$ are the transverse momentum and fractional energy of the emitted gluon
from the jet parton $a$ of mass and energy ($m_a, E$), 
with $t_i$ and $\tau_f = 2Ez(1-z)/(k_\perp^2 + z^2 m_a^2)$ 
are the formation times of parent $a$ and radiated gluon, respectively. The vacuum splitting function
$P_a(z)$ is taken from the usual DGLAP kernel. Since we model gluon radiation as induced by elastic 
scattering, the gluon transverse momentum broadening rate can be expressed as 
\begin{equation}
{\hat q}_a = \int d^2{\bf q}_\perp \frac{d \sigma_{ab\to cd}}{d{\bf q}_\perp^2} \: \rho_b \: {\bf q}^2_\perp .
\label{eq:qhat}
\end{equation}
The inelastic scattering rate $\Gamma_{\rm inel}^a$ and the average multiplicity of emitted gluons 
$\langle N_g \rangle$ in the time interval $\Delta t = t - t_i$ can be obtained by integrating out 
Eq. (\ref{eq:split}) accordingly, which then provides the inelastic scattering probability
$P_{\rm inel}^a = 1 - {\rm exp}(-\Gamma_{\rm inel}^a \Delta t)$. The total probability 
$P_{\rm tot}^a = P_{\rm el}^a + P_{\rm inel}^a - P_{\rm el}^a P_{\rm inel}^a$ 
at each collision is used to determine whether
the collision is elastic or inelastic. The outgoing channel for pure elastic collision in AMPT
can be explicitly determined. In case of inelastic process, the
number $N_g$ of emitted gluons is first obtained from a Poisson distribution with $\langle N_g \rangle$.
The energy-momentum of each gluon is then sampled from Eq. (\ref{eq:split}) and finally their 
four-momenta is scaled to ensure energy-momentum conservation for the ($2\to 2 + N_g$) process
\cite{He:2015pra}. Note these gluons can interact only after the formation time $\tau_f$ via elastic scattering 
and also by gluon radiation if their $p_T > p_{T, cut} = 2$ GeV and diffuse into the QGP medium.
The surviving (final state) hard jets, shower and radiated partons at freeze-out hadronize 
by independent vacuum-like fragmentation using Pythia.

\subsection{Simulation results on single and full jets}

The model simulations are performed for (0-10)\%  central Pb-Pb collisions at ${\sqrt s_{\rm NN}} = 5.02$ TeV. 
For phenomenological jet studies, we follow the same jet construction/definition procedure
as used in experimental analysis \cite{CMS:2016qnj,CMS:2018zze}: 
Jets are constructed using anti-$k_T$ algorithm in the FastJet framework \cite{Cacciari:2011ma} with a 
jet cone radius $R$, followed by iterative noise/pedestal event subtraction 
to construct signal pair-distribution 
$S(\Delta\eta,\Delta\phi) = 1/N_{\rm jet} (d^2N^{\rm same}/d\Delta\eta d\Delta\phi)$
with $\Delta\eta = \eta-\eta_{\rm jet}$ and $\Delta\phi =\phi-\phi_{\rm jet}$.
Limited acceptance is corrected by mixed-event correlation $M(\Delta\eta, \Delta\phi)$ to obtain 
acceptance-corrected per trigger jet distribution 
$1/N_{\rm jet} (d^2N/d\Delta\eta d\Delta\phi) = S(\Delta\eta,\Delta\phi) M(0,0)/ M(\Delta\eta,\Delta\phi)$
followed by sideband subtraction \cite{CMS:2016qnj}.

\begin{figure}[ht]
\begin{minipage}{0.5\textwidth}
\includegraphics[width=6.0cm]{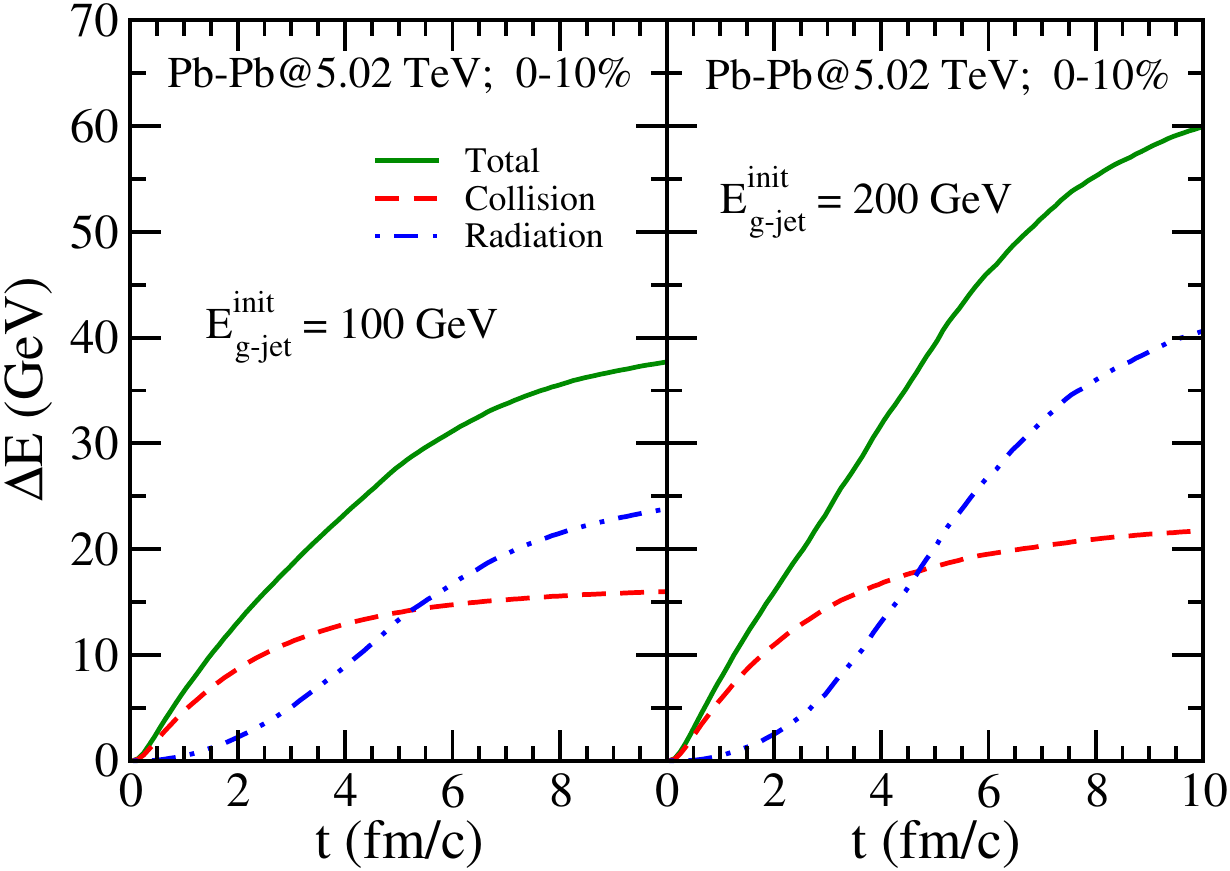}
\end{minipage}
\begin{minipage}{0.49\textwidth}
\includegraphics[width=5.4cm]{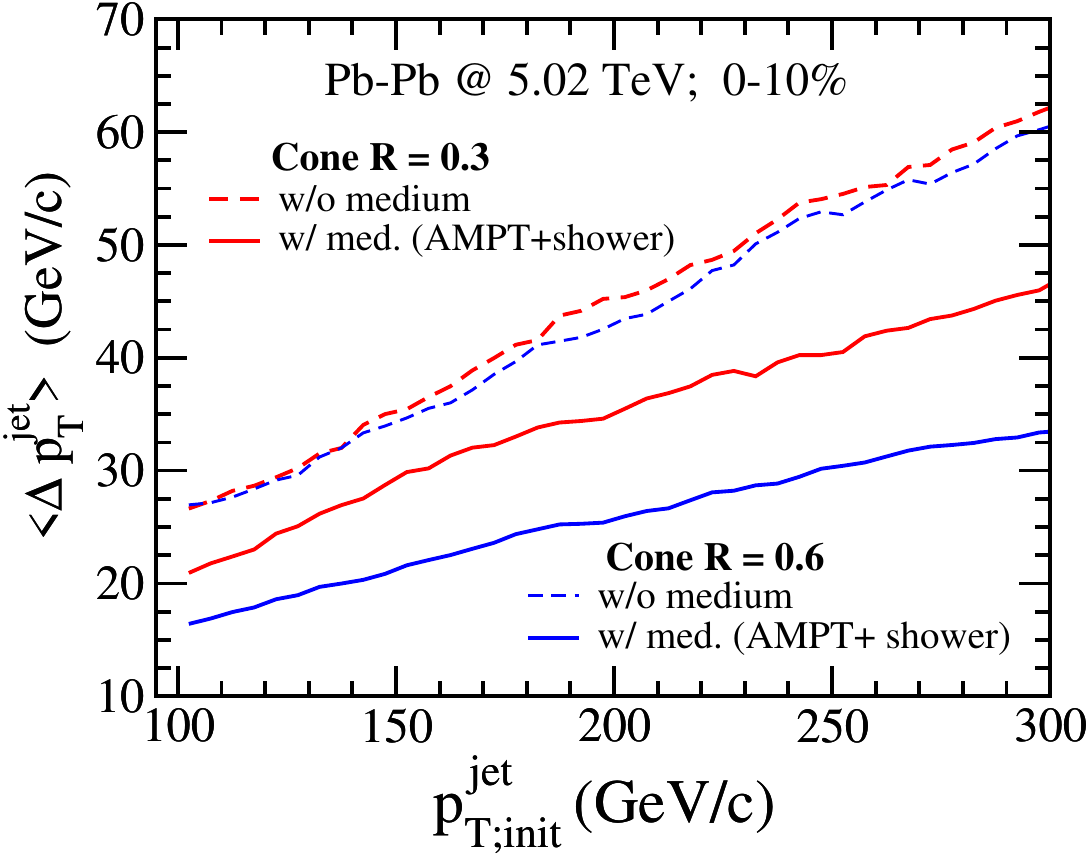}
\end{minipage}
\caption{{\it Left}: Energy loss experienced by a gluon jet with initial energy $E=100$ and 200 GeV 
in AMPT model in elastic scattering (dashed red), medium-induced radiation 
(dash-dotted blue) and total (solid green).
{\it Right}: $p_T$ loss of jets versus initial jet $p_{T, {\rm ini}}^{\rm jet}$ for cone size $R= 0.3, 0.6$ 
in the model for full shower partons (dashed lines) and with medium response (solid lines).}
\label{fig:Elos}
\end{figure}

Figure \ref{fig:Elos} (left) shows the time dependence of collisional, medium-induced gluon radiation 
and total energy loss of a gluon jet with initial energies $E =100, 200$ GeV in 
the newly developed AMPT model. 
While the collisional loss is found important at $t \lesssim 4$ fm/c, the radiation
loss becomes increasingly dominant at later times in the fully formed opaque QGP medium. For
energetic jets the radiation loss is even larger emphasizing its importance in jet studies
at LHC energies.

In Fig. \ref{fig:Elos} (right) we present the jet-cone size dependence of the full jet $p_T$ loss.
For shower partons only (without medium response) the effect of cone size $R$ is quite 
small due to collinear radiation near the core of the jet cone. On inclusion of medium 
contributions derived from radiations, the lost energy is largely recovered for larger 
cone size $R$. This is due to the fact that $p_T$ collisional broadening of the jet-induced 
partons (after their formation time $\tau_f$) diffuse and spread to larger angular distance about the jet axis. 
Further, the lost energy can be carried by the recoil medium partons as a part of jet-induced
medium response which can also contribute to the final jet energy within a given jet cone size.
This suggests that the lost energy may not thermalize instantaneously with the medium 
(as adopted in the coupled jet-hydrodynamic model) but retain some memory of the correlations.

\begin{figure}[t]
\centerline{\includegraphics[width=5.7cm]{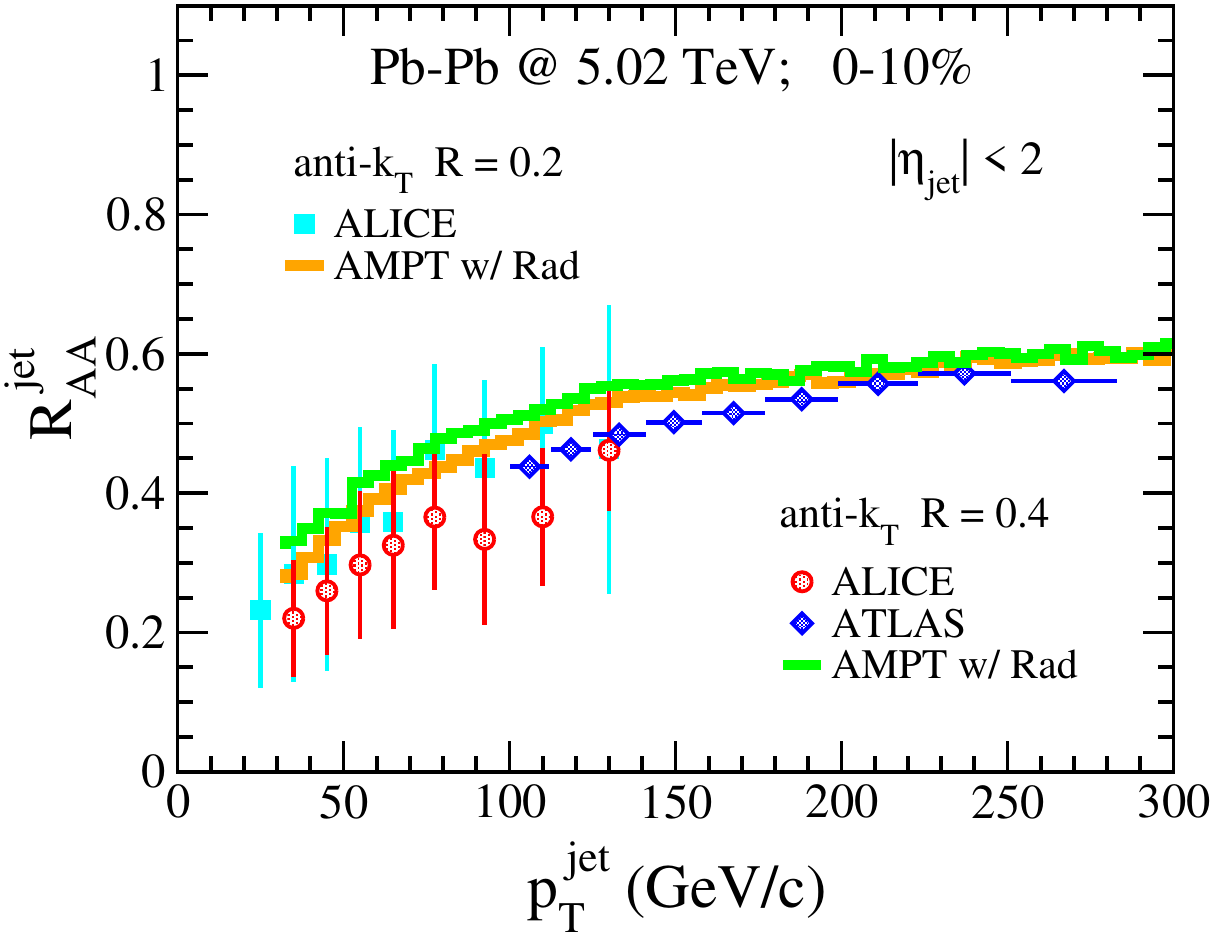}}
\vspace{-0.2cm}
\caption{Nuclear modification factor $R_{\rm AA}^{\rm jet}$ for inclusive jets at midrapidity 
for cone radii $R = 0.2, 0.4$ in AMPT simulations as compared to data from 
ATLAS \cite{ATLAS:2018gwx} and ALICE \cite{ALICE:2019qyj} Collaborations.}
\label{fig:RAAj}
\end{figure}

Figure \ref{fig:RAAj} depicts the full jet energy loss in Pb-Pb relative to p-p collisions
quantified by the jet nuclear modification factor 
$R_{AA}^{\rm jet} = dN_{\rm PbPb}^{\rm jet}/dp_T /(\langle N_{\rm bin} \rangle dN_{pp}^{\rm jet}/dp_T)$
for single inclusive jet spectrum as a function of $p_T^{\rm jet}$. The results
include the recoil medium partons which can recover some of the lost energy as the jet cone radii 
is increased from $R=0.2$ to 0.4. The jet $R_{AA}$ model predictions indicate somewhat weak
quenching for low $p_T^{\rm jet} < 150$ GeV as compared to ALICE data especially for large R=0.4  
These low-$p_T$ final-state full jets are produced either from jets near the surface,
traverse tangentially, and confront less dense medium or from much higher $p_T$ initial jets 
directed/inside the medium where the abundant radiated energy is transported to large distances.

\begin{figure}[t]
\begin{minipage}{0.5\textwidth}
\centerline{\includegraphics[width=5.2cm]{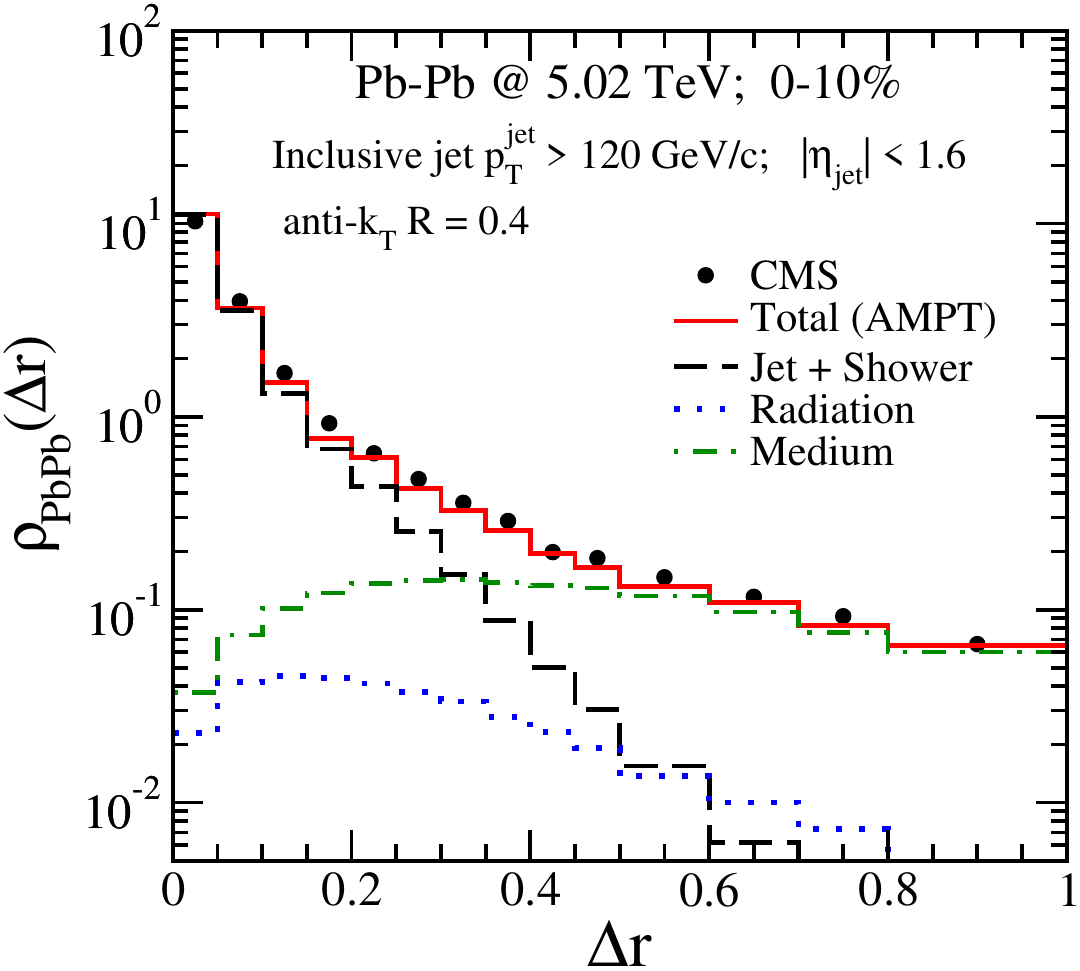}}
\end{minipage}
\begin{minipage}{0.49\textwidth}
\centerline{\includegraphics[width=5.2cm]{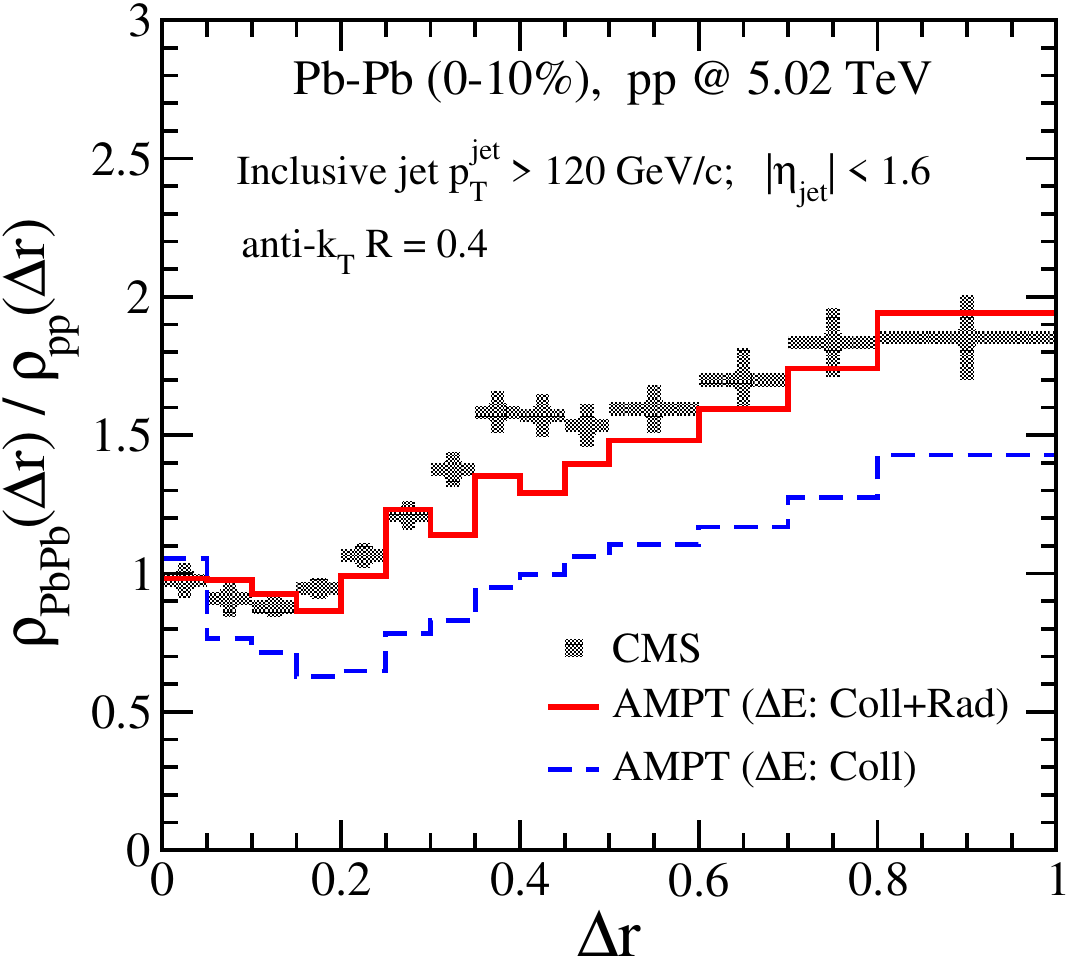}}
\end{minipage}
\caption{{\it Left}: Jet shape $\rho(\Delta r)$ for inclusive jets with $p_T^{\rm jet} > 120$ GeV at 
$|\eta^{\rm jet}| < 1.6$ from AMPT compared to CMS data \cite{CMS:2018zze} in (0-10)\% 
central Pb-Pb collisions. The contributions are from leading 
plus initial parton showers (dashed black line), semi-hard radiated gluons (dotted blue line) and 
jet-induced medium excitations (dash-dotted green line).
{\it Right}: Jet shape ratio $\rho_{\rm PbPb}/\rho_{\rm pp}$ in AMPT for collisional plus radiative
and only collisional energy loss as compared to CMS data.}
\label{fig:rhor}
\end{figure}

To examine the medium modification in detail, we present the jet shape  
$\rho(\Delta r) = \sum_{\rm jets} \sum_{{\rm trk} \in \Delta r_\mp} p_T^{\rm trk}/(\delta r \: N_{\rm jets})$ 
where $\Delta r_\mp = \Delta r \mp \delta r/2 $ are the inner/outer annular edges of 
$\Delta r = \sqrt{\Delta\eta^2 + \Delta\phi^2}$ of annulus size $\delta r = 0.05$,
and $\rho(\Delta r)$ is normalized to unity within $\Delta r < 1$. 
Figure \ref{fig:rhor} (left) compares our model prediction with CMS data for inclusive jets of 
$p_T^{\rm jet} > 120$ GeV with $R=0.4$.
The shape near the jet core $\rho(\Delta r < 0.2)$ is dictated by the fragmentation/hadronization of the 
leading jet and its associated initial parton showers at the final state after traversing the QGP medium 
(dashed black line) causing a dip in central Pb-Pb collision due to their in-medium energy loss as compared to $pp$.
While the semi-hard radiated gluons (dotted blue line) has a minor effect over the entire 
shape $\Delta r < 1$, the majority of the radiated gluons are transported by multiple collisional kicks 
(jet-induced flow) to large $\Delta r$ outside the jet cone and can further lead to medium-excitation. 
These latter particles are at much lower $p_T$ due to their thermalizing with the medium and significantly 
enhance and dominate the jet shape at $\Delta r \gtrsim 0.4$ in Pb-Pb collisions (dash-dotted green line).  
The jet shape ratio $R_{\rm PbPb}^{\rm \rho} = \rho_{\rm PbPb}/ \rho_{\rm pp}$ in Fig. \ref{fig:rhor} (right) 
clearly displays the medium modifications (solid red line). We also present $R_{AA}^{\rm \rho}(\Delta r)$ in 
the usual AMPT without radiative energy loss (dashed blue line) to underscore the importance of 
medium-induced gluon radiation for enhanced particle yield at large $\Delta r$ consistent with the CMS data.

%


\section{Charm and beauty in the early stage of high-energy nuclear collisions
}

\author{Pooja, Santosh Kumar Das, Vincenzo Greco, Marco Ruggieri} 

\bigskip

\begin{abstract}
Relativistic high-energy nuclear collisions allow us to study strong interactions at relativistic energies. These collision processes are well-characterized by the gluon saturation model, where two colored glass sheets interact, thus immediately producing strong longitudinal gluon fields known as the Glasma. Heavy quarks are generated in the very early stages of the collisions, and due to their large mass and low concentration, their motion minimally impacts the Glasma evolution, making them ideal probes for studying the Glasma directly.
In this study, we analyze the behavior of heavy quarks within the Glasma field by solving their equations of motion consistently within the evolving Glasma environment.
We calculate both the linear momentum broadening and the angular momentum fluctuations of heavy quarks in the initial Glasma stage, finding that the anisotropic nature of the background gluon fields leads to anisotropic fluctuations of the heavy quark in beam and transverse directions. We observe the non-linear behavior of the transverse momentum broadening of heavy quarks at initial times, indicating a non-Markovian diffusion pattern in the early stages. This behavior is attributed to memory effects inherent in the gluon fields. Additionally, we examine how heavy quark mass influences their momentum broadening.
\end{abstract}


\ccode{PACS numbers:}

\subsection{Introduction}
Relativistic heavy-ion collisions (HICs) present a valuable opportunity to 
experimentally replicate the conditions of the early universe within controlled laboratory settings. 
By colliding heavy nuclei at ultra-relativistic speeds, these experiments reach the extreme temperatures and energy densities thought to resemble those of the universe's first microseconds, creating an environment to study the properties of the quark-gluon plasma (QGP) phase. However, the formation of QGP in these HICs is not immediate. Prior to the thermalized QGP phase, a short, pre-equilibrium stage, known as the Glasma\cite{Lappi:2006fp}, is believed to emerge. This Glasma phase is a strongly coupled system dominated by intense color-electric and color-magnetic fields generated by the colliding nuclei. While extensive research over the past four decades has focused on understanding the properties of QGP, our main interest lies in exploring the dynamics of the Glasma, the precursor stage to QGP. To investigate this phase of matter, we plan to employ heavy quarks (HQs) as potential probes.

HQs\cite{Rapp:2018qla,Dong:2019unq, Das:2024vac, Ruggieri:2018rzi, Sun:2019fud, Liu:2019lac, Liu:2020cpj, Khowal:2021zoo, Ruggieri:2022kxv, Pooja:2022ojj, Pooja:2023gqt, Pooja:2024rnn, Avramescu:2023qvv, Boguslavski:2020tqz, Boguslavski:2023fdm, Das:2023zna, Pandey:2023dzz, Scardina:2017ipo,Das:2010tj,Chandra:2024ron,Ghosh:2011bw}, such as $c$ and $b$ quarks, are efficient probes for studying the early stages of ultra-relativistic HICs. They form very quickly after the collision and have minimal interactions with the background medium. This allows them to traverse the evolving gluon fields and QGP, providing insights into these phases of matter. While the Glasma phase is short-lived, HQs can interact with the dense gluon system during this period. Even though this interaction is brief, it can leave observable imprints on the final state. Therefore, we study the dynamics of HQs in the Glasma phase for understanding the early stages of HICs.

\subsection{Methodology}
A common approach in the heavy-ion community is to model relativistic HICs as a series of distinct stages, with each stage represented by an appropriate effective theory. In this study, we focus specifically on the initial stage, using the Color Glass Condensate (CGC)\cite{McLerran:1993ni, Lappi:2006fp, Gelis:2010nm} effective theory to characterize it. 
To this end, we use the McLerran-Venugopalan (MV) model.
In this model, the colliding nuclei $L$ and $R$ are represented by randomly distributed static color charge densities, $\rho^a$, which follow a normal distribution, namely
\begin{eqnarray} 
	\langle \rho^a_{L,R}(\Vec{x}_{\perp})\rangle &=& 0  ,\label{Eq:MV_charge_density_one_point_functionn} \\
	\langle \rho^a_{L,R}(\Vec{x}_{\perp}) \rho^b_{L,R}(\Vec{y}_{\perp})\rangle &=& (g\mu_{L,R})^2\delta^{ab}\delta^{(2)}(\Vec{x}_{\perp} - \Vec{y}_{\perp}) . \label{Eq:MV_charge_density_two_point_functionn}
\end{eqnarray}
The static color sources generate pure gauge fields. These fields combine to form the initial, boost-invariant Glasma fields within the forward light cone. These Glasma fields evolve as per the classical Yang-Mills (CYM) equations, in the static environment:
\begin{eqnarray}
	\frac{dA^a_i(x)}{dt} &=& E^a_i(x),   \label{Eq:Static_CYM_A_a_i}\\
	\frac{dE^a_i(x)}{dt} &=& \partial_jF^a_{ji}(x)+g f^{abc} A^b_j(x)F^c_{ji}(x) \label{Eq:Static_CYM_E_a_i},
\end{eqnarray} 
where $F^a_{ij}$ is the magnetic part of the field strength tensor.

Concurrently with the formation of the Glasma from the collisions of soft partons, HQs, such as $c$ and $b$, are produced through hard scatterings. These colored probes subsequently propagate through and interact with the Glasma. Approximating these probes as classical color charges, we model their dynamics using Wong equations\cite{Wong:1970fu, Heinz:1984yq, Liu:2019lac, Liu:2020cpj, Khowal:2021zoo, Ruggieri:2022kxv, Pooja:2022ojj, Pooja:2023gqt}, given by
\begin{eqnarray}
	\frac{dx^i}{dt} &=&  \frac{p^i}{E},  \label{Eq:Wong_X}\\
	\frac{dp^i}{dt} &=&    g Q_a F^{i\nu}_a  p_\nu  , \label{Eq:Wong_P} \\
	\frac{dQ_a}{dt} &=&  \frac{g}{E} f_{abc} A^{i}_b p^i Q_c ,\label{Eq:Wong_Q}
\end{eqnarray} 
where $i=x,y,z$ and $E=\sqrt{\vec{p}^2 + m^2}$ denotes the relativistic energy of the individual colored probe. 
 In the above equations, $x^i$, $p^i$, and $Q_a$ denote the position, momentum, and effective color charge of the HQs, respectively.

\subsection{Results}
We determine the parameters for the evolving Glasma fields, by fixing the saturation scale, $Q_s$, within the range of 1 to 3 GeV. 
Next, we calculate the strong coupling constant, $\alpha_s$, at the scale $Q_s$ using the one-loop QCD $\beta$-function. From $\alpha_s$, we obtain the QCD coupling, $g$\cite{Khowal:2021zoo}, using the relation $g = \sqrt{4\pi\alpha_s}$. Finally, we calculate the color charge densities, $\mu$, of the colliding nuclei using the approximation $\mu \approx Q_s/(0.6 g^2)$\cite{Lappi:2007ku}.

\begin{figure}[th]
\centering
\includegraphics[width=6cm,clip]{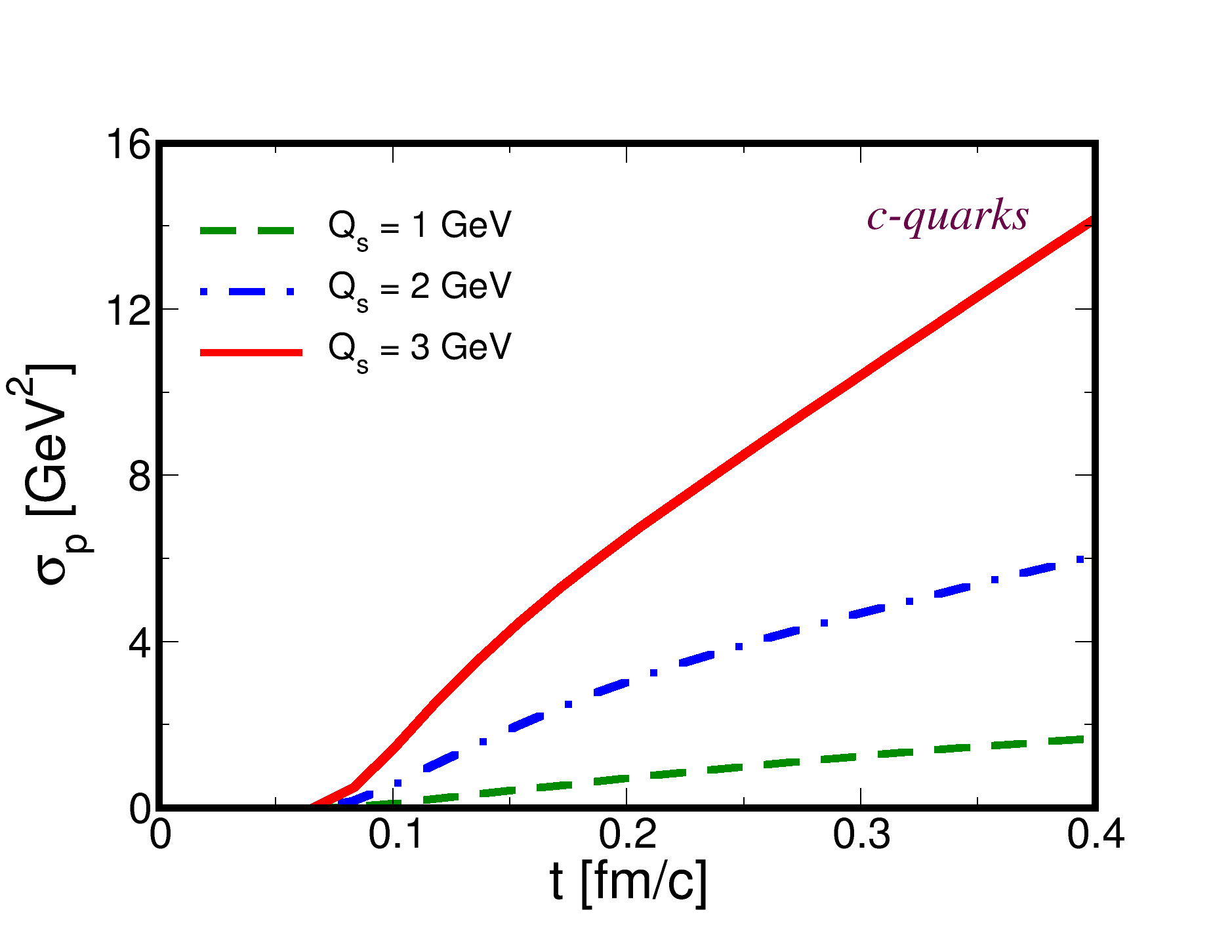}
\caption{Evolution of the transverse momentum broadening, $\sigma_p$ with proper time, $t$ for charm quarks, considering initial  $p_T=0.5~\mathrm{GeV}$. This calculation is for the evolving Glasma fields within a static box. }
\label{Fig:Charm_sigmaP} 
\end{figure}

In Fig.~\ref{Fig:Charm_sigmaP}, we illustrate the evolution of transverse momentum broadening, $\sigma_p$,
\begin{equation*}
	\sigma_p = \frac{1}{2} \big\langle (p_x(t) - p_{0x})^2 + (p_y(t)- p_{0y})^2 \big\rangle,
	\label{Eq:sigmap}
\end{equation*}
with respect to proper time $t$ for $c$ quarks propagating through the evolving Glasma fields at different initial saturation scales, \(Q_s\). Adjusting this parameter, \(Q_s\), alters the energy density, consequently impacting the effective temperature of the evolving Glasma medium in which HQs diffuse. Thus, higher $Q_s$ leads to higher rate of diffusion of HQs in the Glasma medium.
Notably,  at early times, $\sigma_p$ exhibits a non-linear growth, which deviates from the standard Brownian motion (BM) behavior. This non-linearity is due to memory effects in the Glasma. The gluon fields in the Glasma exert correlated forces on the 
c quarks, leading to this non-linear behavior. After a short transient period, $\sigma_p$ transitions to a linear growth, resembling BM without drag, as investigated in Ref.~\citen{Liu:2020cpj}. This transition occurs over a timescale of approximately $\tau_{mem}\approx 1/Q_s$\cite{Liu:2020cpj}, the memory time of the Glasma fields.

\begin{figure}[th]
\centering
\includegraphics[width=6cm,clip]{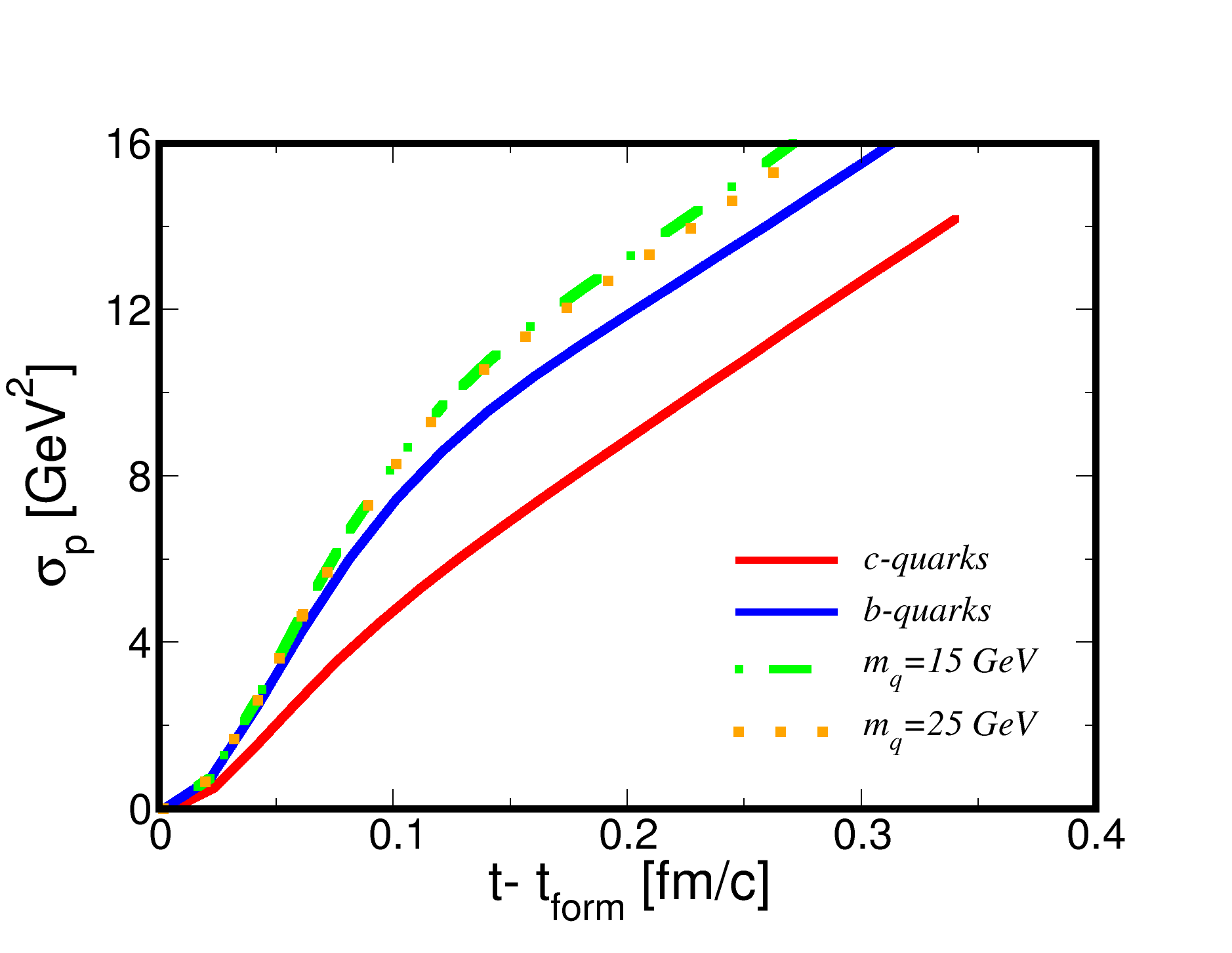}
\caption{Evolution of transverse momentum broadening, $\sigma_p$ with proper time, $t$ for HQs with different  masses, having the evolving Glasma fields within a static box. }
\label{Fig:sigmaP_c_b} 
\end{figure}

To explore the differences in transverse momentum broadening between $c$ and $b$ quarks, Fig.~\ref{Fig:sigmaP_c_b} shows \(\sigma_p\) for various HQ masses at a constant saturation scale, \(Q_s = 3~\mathrm{GeV}\). To maintain consistency, we set \(t_\mathrm{form} = 0.02~\mathrm{fm/c}\) for all quarks, representing the approximate initialization time for \(b\) quarks.
This figure clearly indicates that heavier HQs, such as b quarks, experience a more rapid growth in $\sigma_p$ compared to lighter HQs, such as $c$ quarks. 
This is because HQs with greater mass and the same transverse momentum exhibit, on average, a lower transverse velocity. Hence, they spend more time interacting with the correlated gluon fields within the Glasma. Initially, the non-linear growth of $\sigma_p$ is due to the correlated forces exerted by the gluon fields. As the HQs gain velocity over time, they begin to traverse different uncorrelated domains and experience Brownian motion, eventually leading to a linear growth in $\sigma_p$.

\begin{figure}[th]
\centering
\includegraphics[width=6cm,clip]{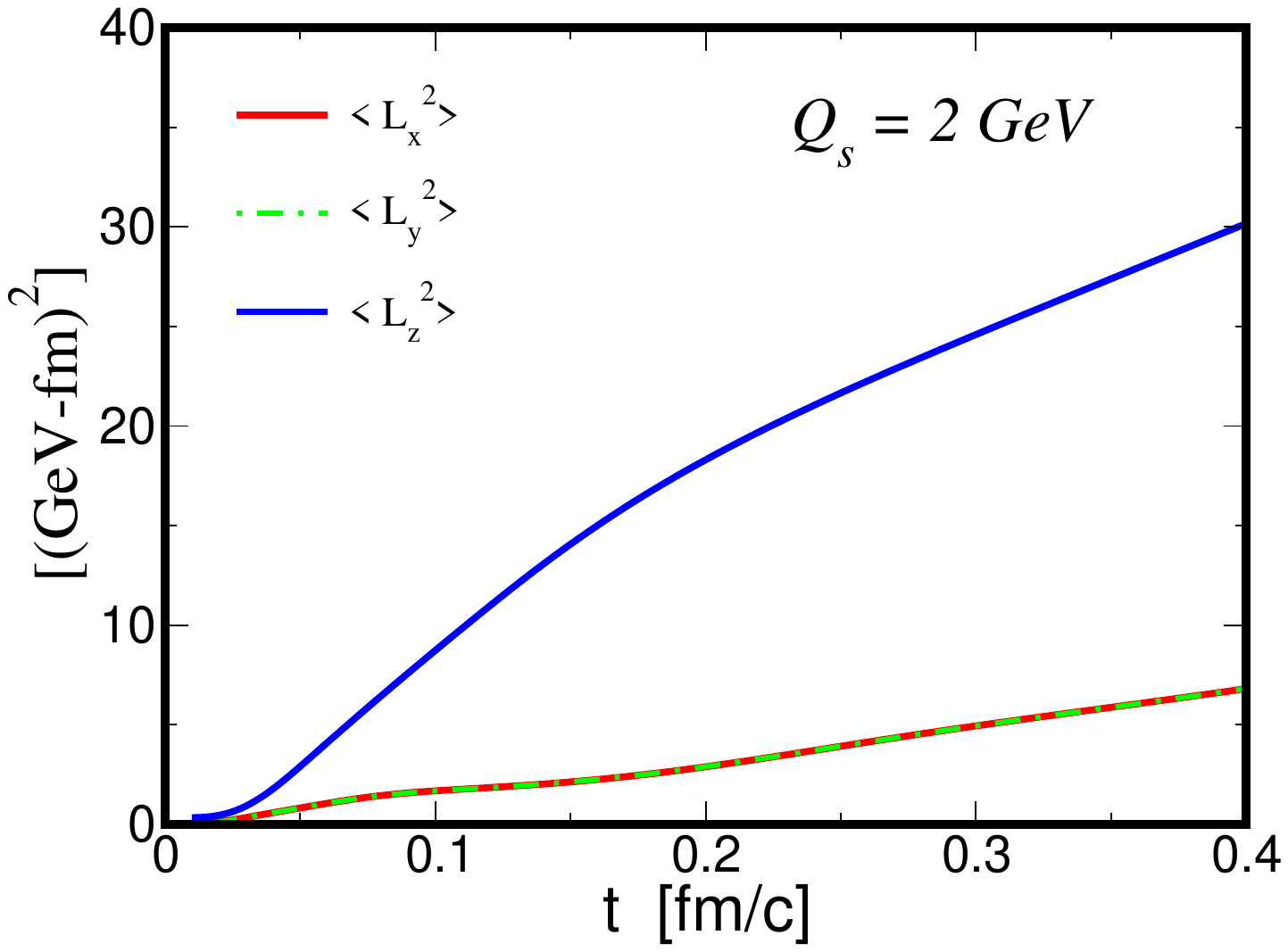}
\caption{Evolution of average of the  angular momentum components squared with proper time in the non-relativistic limit of HQs, considering initial  $p_T =  p_z = 0.5~\mathrm{GeV}$ at $Q_s = 2~\mathrm{GeV}$.}
\label{Fig:L_broadening} 
\end{figure}

Next, we analyze the fluctuations in the angular momentum of HQs by studying the HQ dynamics within a classical color field in the non-relativistic limit\cite{Pooja:2022ojj, Heinz:1984yq}. In Fig.~\ref{Fig:L_broadening}, we present the average of square of the orbital angular momentum components, \(\langle L_x^2 \rangle\), \(\langle L_y^2 \rangle\), and \(\langle L_z^2 \rangle\), for a saturation scale of \(Q_s = 2~\mathrm{GeV}\). 
 It has been verified numerically that \(\langle L_x \rangle = \langle L_y \rangle = \langle L_z \rangle = 0\) throughout the evolution, the values plotted reflect the spreading of each angular momentum component individually.

Notably, \(\langle L_z^2 \rangle\) evolves much faster than \(\langle L_x^2 \rangle\) and \(\langle L_y^2 \rangle\), indicating that HQ interactions with the gluon fields induce anisotropic fluctuations in orbital angular momentum. In other words, the fluctuations in the components of \(\vec{L}\) are more pronounced along the longitudinal direction than in the transverse directions.
In addition to examining the fluctuations of the orbital angular momentum, we have also computed the evolution of spin fluctuations.
We have observed that the fluctuations of the spin angular momentum are considerably smaller in magnitude compared to orbital angular momentum fluctuations and exhibit isotropic behavior, as demonstrated in Ref.~\citen{Pooja:2022ojj}. Additionally, our numerical simulations indicate that spin and orbital angular momentum fluctuations are uncorrelated. Consequently, we approximate the fluctuations of total angular momentum, 
\(\vec J = \vec L + \vec S\), as $
\langle J^2_i \rangle = \langle (L_i + S_i)^2 \rangle \approx \langle L_i^2 \rangle$.

\subsection{Summary}
We studied how HQs propagate through the early pre-equlibrium stage of high-energy nuclear collisions, known as the Glasma phase. We used classical Yang-Mills equations to describe the behavior of the gluon fields in the Glasma and Wong equations to model the motion of the HQs.
Numerical simulations of HQs propagating through the evolving Glasma fields were performed, focusing on linear momentum broadening and angular momentum fluctuations of the HQs. Our findings indicate that the diffusion of HQs in the pre-equilibrium phase is non-linear at initial times, influenced by strong, coherent gluon fields possessing significant memory effects. Furthermore, we observed pronounced anisotropy in the fluctuations of total HQ angular momentum within the Glasma fields.



\section{Anomalous diffusion of the heavy quarks in hot QCD matter}

\author{Jai Prakash}

\bigskip

\begin{abstract}
The study revisits heavy quark dynamics in hot QCD matter using the fractional Langevin equation with Caputo derivatives. It shows that the mean squared displacement of the heavy quark deviates from linearity relation with time, indicating anomalous diffusion. The key observables are calculated, including the mean squared momentum,  and nuclear modification factor. 
\end{abstract}


\ccode{PACS numbers:}

\subsection{Introduction}
\label{sec:intro}

The hot and dense quark matter called quark-gluon plasma (QGP) is expected to be produced in the heavy ion collision experiments at the Relativistic Heavy Ion Collider (RHIC) and Large Hadron Collider (LHC). The HQs are considered novel probes \cite{Das:2024vac,Sun:2023adv,Plumari:2017ntm,Rapp:2018qla,Cao:2018ews,Debnath:2023zet,Prakash:2023wbs,Prakash:2021lwt,Ruggieri:2022kxv} for studying the QGP properties due to their large masses and delayed thermalization. Their thermalization time is longer than the QGP's lifetime, allowing them to traverse the entire medium. The  HQs capture the whole evolution of the QGP and retain essential information about its properties. The HQ dynamics in the QGP are typically studied by tracking their position and momentum via Langevin equations (LE).
Unlike conventional LE, the fractional Langevin equation (FLE) includes fractional derivatives, allowing for the study of anomalous diffusion, which is characterized by deviations from the linear mean squared displacement (MSD) behavior with time, which is defined as fellows,
\begin{align}\label{MSD}
\langle  x(t)^2\rangle \propto t^{\nu}, \  {\nu} \neq 1, 
\end{align}
subdiffusion occurs when ${\nu}< 1 $, and superdiffusion occurs for ${\nu}> 1 $. The process describes standard diffusion, corresponding to the case where  ${\nu} = 1 $. This study focuses on superdiffusion, where the MSD of the HQs grows faster than linear with time, significantly impacting experimental observables like the $R_{AA}$. Anomalous diffusion arises from memory effects, where the particle's dynamics depend on its interaction history. This effect is crucial in understanding the energy loss mechanisms of the HQs more precisely. By using the fractional derivative approach, this study aims to address discrepancies between observed and theoretical predictions of the HQ experimental observable in the QGP. This approach provides new insights into non-equilibrium dynamics, offering a novel perspective on heavy-ion collision phenomenology and the role of fractional dynamics in understanding QCD matter at extreme conditions.

\subsection{Formalism}
\label{sec:formalism}

 The  FLE  to study the HQs momentum and position evolution in the QGP medium is \cite{li2012spectral,kobelev2000fractional,Prakash:2024irm,Prakash:2024rdz}, 
\begin{align}\label{Langevin_x_rel}
&^{C} D^\beta_{0+}x(t)=\frac{p(t)}{E(t)},
\\
&^{C} D^\alpha_{0+}p(t) = -\gamma p(t)+\xi(t),
\label{Langevin_p_rel}
\end{align}
$^{C}D_{0+}^{\alpha}$ denotes Caputo fractional derivative, $\alpha$ and $\beta$ are the fractional parameter, with $n-1 < \alpha \leq n$, and $n-1<\beta\leq n$, $n \in \mathbb{N}$ ($n$ is a natural number).
The Caputo fractional derivative \cite{10.1111/j.1365-246X.1967.tb02303.x} is defined as, 
\begin{align}\label{Caputo}
 ^{C} D^\nu_{0+}u(t) = \frac{1}{\Gamma({n-\nu})}\int_0^t \frac{u^{(n)}(s)}{(t-s)^{1+\nu-n}}ds,   
\end{align}
 where $u^{(n)}$ represents the $n^{th}$ derivative of $u$ and $\Gamma(\cdot)$ denotes the gamma function. In the FLE,  $E = \sqrt{p^2+ M^2}$ denotes the energy, and $p$ denotes the momentum of the HQs. The HQs experience two forces: a dissipative force, which contains the drag coefficient $\gamma$, and a stochastic force $\xi(t)$. The stochastic force, modelled as white Gaussian noise and its correlation  decays instantaneously such as,
\begin{align}\label{corr}
&\langle\xi(t)\xi(t')\rangle=2\mathcal{D}\delta(t-t'), \\
 &\langle\xi(t)\rangle=0,
\end{align}
 where $\mathcal{D}$ is the diffusion coefficient of the HQs. Here $\gamma$ is related to the $\mathcal{D}$ via the Fluctuation-Dissipation Theorem (FDT) as follows \cite{Moore:2004tg}:
\begin{align}\label{FDT}
\gamma = \frac{\mathcal{D}}{MT}.
\end{align}

\subsection{Results}

 \subsubsection{The Evolution of $\langle p^2(t)\rangle$ and $\langle x^2(t)\rangle$ of the Heavy Quarks}

\begin{figure*}[htp]
		\centering
        \includegraphics[scale = .26]{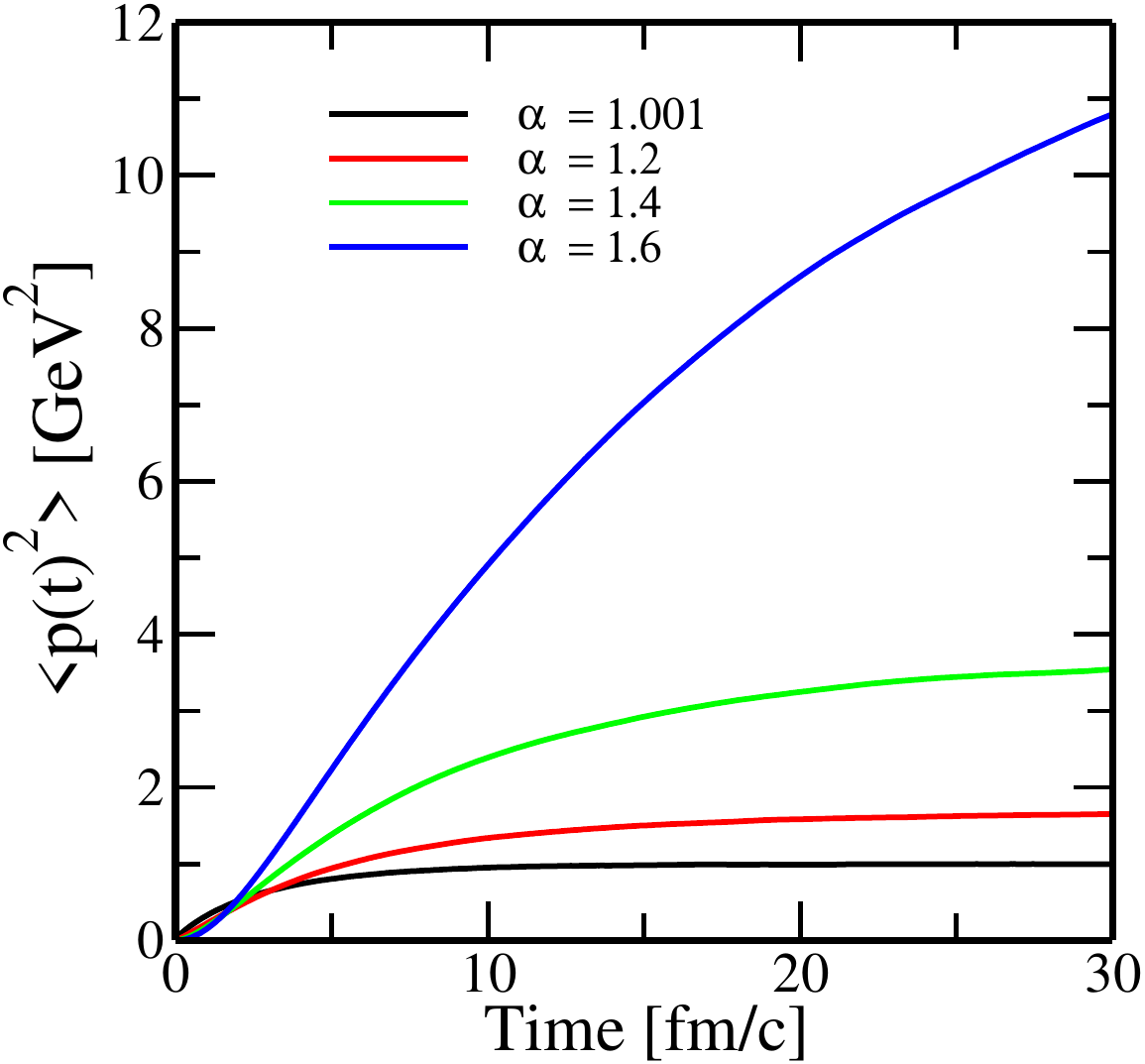}
         \hspace{10mm}
		\includegraphics[scale = .26]{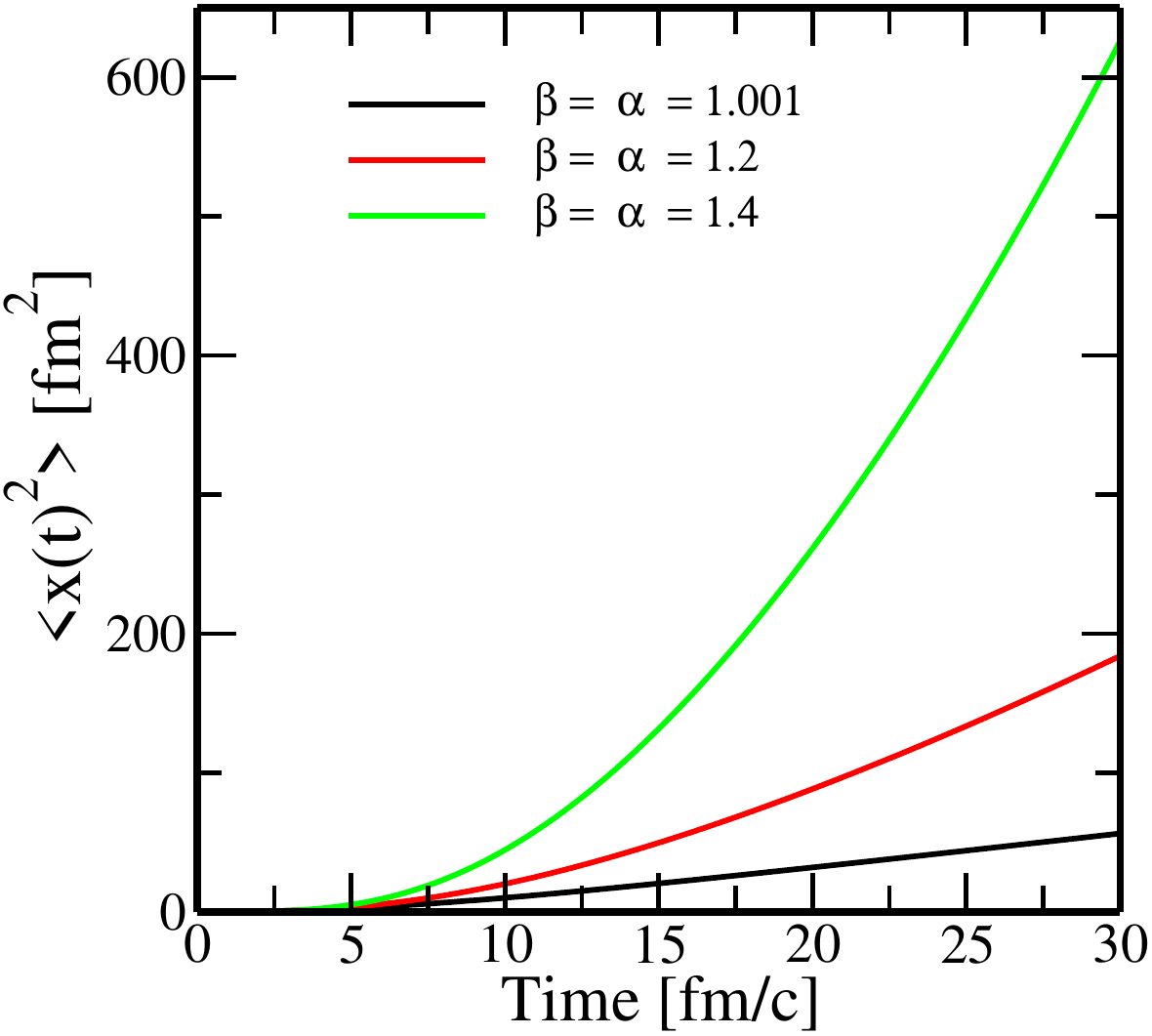}
        
		\caption{$\langle p^2(t) \rangle$  (left panel) and $\langle x^2(t) \rangle$  (right panel)  versus time, for the various values of the $\alpha$ and $\mathcal{D} = 0.1$ {GeV}$^2$/fm at $T = 250$ MeV. }
  	\label{p_2D}
	\end{figure*}

We have calculated the mean squared momentum ($\langle p^2(t)) \rangle$ and mean squared displacement ($\langle x^2(t) \rangle$) of HQs in the hot QCD matter. The definition of $\langle p^2(t) \rangle$ and $\langle x^2(t) \rangle$ used in our study is:
\begin{equation}
\langle p^2(t) \rangle = \langle p_x^2(t) + p_y^2(t) \rangle,
\end{equation}and,
\begin{align} \langle x^2(t) \rangle = X^2(t) + Y^2(t), \end{align}
where $p_x(t)$ and $p_y(t)$ represent the components of the HQ momentum along the $X$ and $Y$ axes, respectively.
We compute $\langle p^2(t) \rangle$ (left panel)  over time for different values of the fractional parameter $\alpha$ (specifically $\alpha = 1.001, 1.2, 1.4, 1.6$) for a medium $T = 250 $ MeV, $M$ = 1.3 GeV and $\mathcal{D} = 0.1$ GeV$^2$/fm, corresponding to charm quarks. The initial momentum is set to $p_x(t_0) = p_y(t_0) = 0$. From Fig.~\ref{p_2D} (left panel) \cite{Prakash:2024rdz}, it is observed that the value of $\alpha$ significantly influences the momentum evolution, with larger values of $\alpha$ indicating a superdiffusive process for the charm quark within the QCD medium. This effect is more noticeable at higher values of $\alpha$. As $\alpha \rightarrow 1$, the process slowly approaches normal diffusion, highlighting the transition from superdiffusive to standard diffusion. At later times, $\langle p^2(t)\rangle$ tends to approach $3MT$, which aligns with previous findings \cite{Moore:2004tg, Das:2013kea}.

In Fig.~\ref{p_2D} (right panel)  \cite{Prakash:2024rdz}, we plot the time evolution of the MSD, $\langle x^2(t) \rangle$, for three different values of $\alpha = \beta = 1.001, 1.2, 1.4$, while keeping the other parameters consistent with those used in Fig.~\ref{p_2D} (left panel). The initial conditions for the position are set to be $X(t_0) = Y(t_0) = 0$. It can be observed that for $\alpha = \beta = 1.2$ and 1.4, a distinct transition towards superdiffusion is seen. As $\alpha \rightarrow 1$ and $\beta \rightarrow 1$, the system tends to normal diffusion, where $\langle x^2(t) \rangle$ becomes proportional to time \cite{Moore:2004tg, Svetitsky:1987gq}. This analysis shows the complex behavior of $\langle p^2(t) \rangle$ and $\langle x^2(t) \rangle$ of the HQs in a QCD matter, highlighting the significance of incorporating anomalous diffusion processes to capture the complexities of HQ transport within the QGP.

\subsubsection{Nuclear modification factor}

\begin{figure*}
		\centering
        \includegraphics[scale = .23]{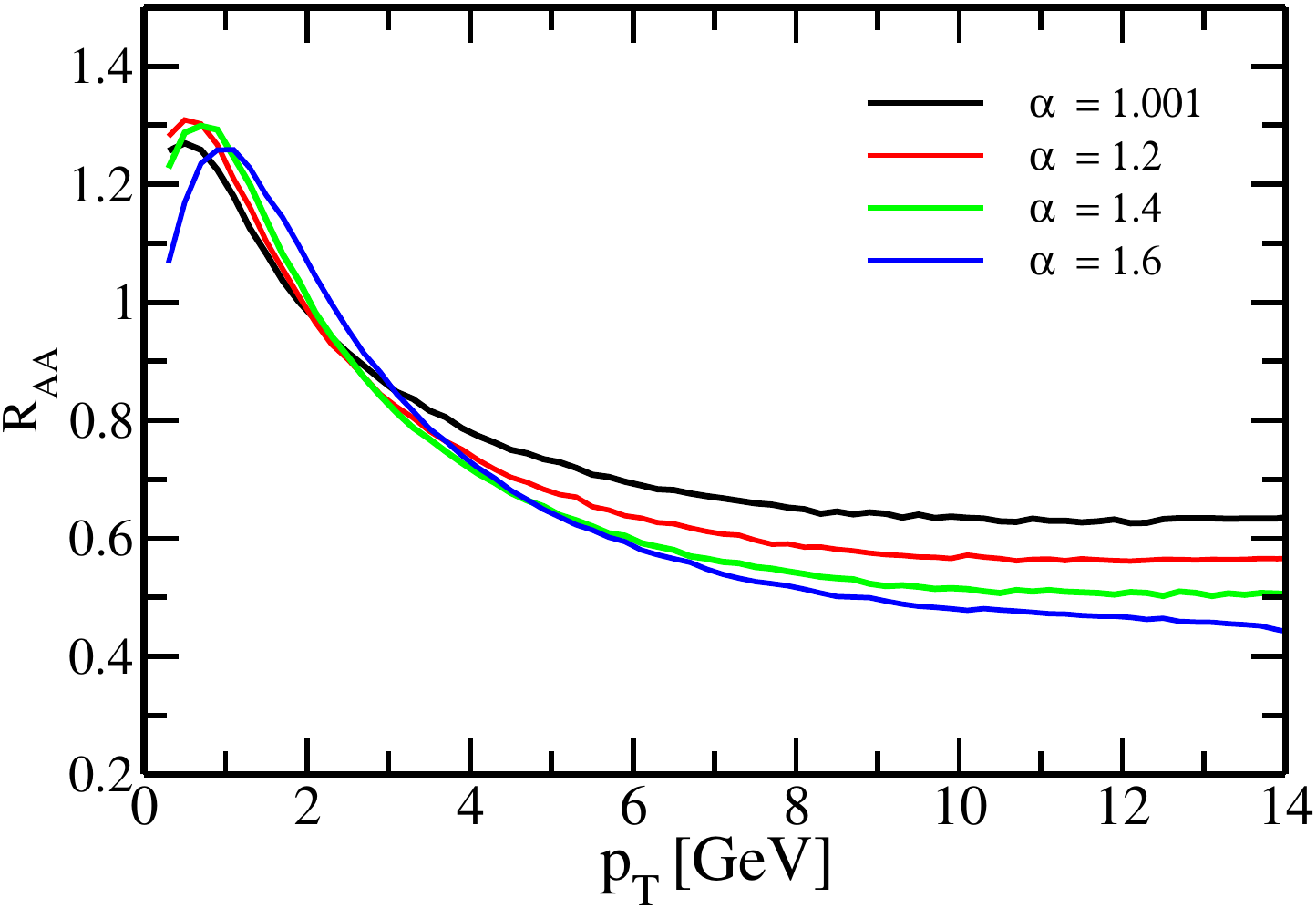}
         \hspace{10mm}
		\includegraphics[scale = .23]{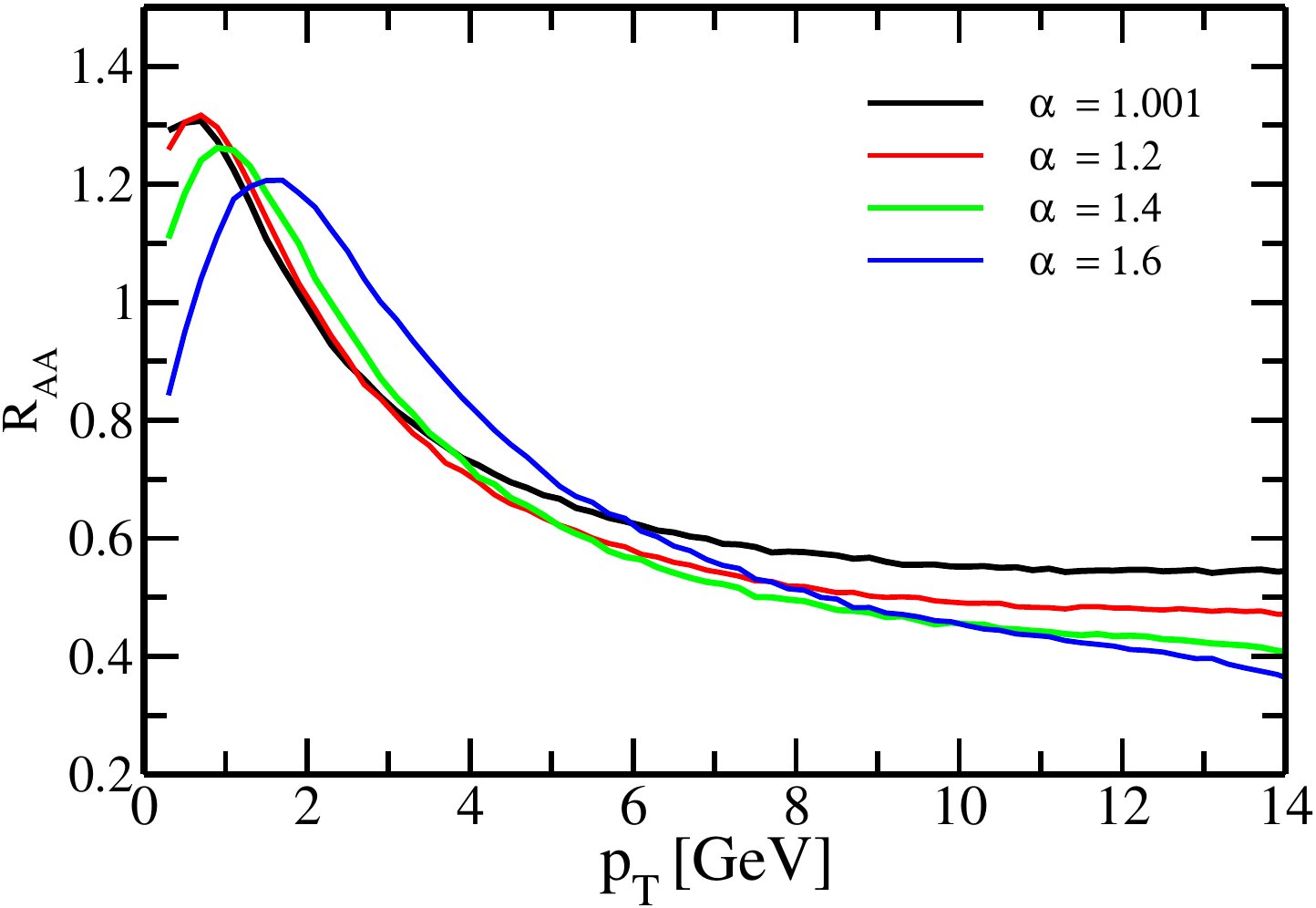}
        
		\caption{  $R_{AA}$ is shown as a function of $p_T$ at $\tau_f = 6 \, \text{fm/c}$ for two different temperatures: $T = 250 \, \text{MeV}$ (left panel) and $T = 350 \, \text{MeV}$ (right panel), with four different values of $\alpha$.}

		\label{raa}
	\end{figure*}

The definition of the $R_{AA}(p_T)$ of the HQs we used as fellow \cite{Moore:2004tg},

\begin{align}
 R_{AA}(p_T)=\frac{f_{\tau_f} (p_T )}{f_{\tau_0} (p_T)}.   \end{align}
The momentum spectrum,  $f_{\tau_f}(p)$, of the charm quark calculated for the time, $\tau_f$ = $6$ fm/c in our numerical results. To determine the interaction of the HQ  in the QGP medium consisting of massless quarks and gluons, we used perturbative Quantum Chromodynamics (pQCD) transport coefficients for elastic processes with the well-established diffusion coefficients~\cite{Svetitsky:1987gq}. In Fig.~\ref{raa}  \cite{Prakash:2024rdz}, the $R_{AA}$ is shown as a function of $p_T$, calculated using the FLE as defined in Eq.~\eqref{Langevin_p_rel} for various values of $\alpha$. The calculations are performed at two distinct temperatures: $T = 250 \, \text{MeV}$ (left panel) and $T = 350 \, \text{MeV}$ (right panel).

For $T = 250 \, \text{MeV}$, the drag force dominates across the entire $p_T$ range, with the effect of energy loss becoming more significant as $p_T$ increases, as seen in the left panel of Fig.~\ref{raa}. For $\alpha > 1$ (e.g., $\alpha = 1.2$, $\alpha = 1.4$, $\alpha = 1.6$), the transition from normal diffusion to superdiffusion is evident, showing more suppression in $R_{AA}$ at higher $p_T$. When $\alpha \rightarrow 1$ (black line), the behavior of $R_{AA}$ corresponds to normal diffusion, consistent with findings in the literature \cite{PhysRevC.93.014901,Das:2013kea,Das:2015ana}. At $T = 350 \, \text{MeV}$ (right panel), diffusion dominates the propagation of HQs, leading to the diffusion of low-momentum quarks to higher $p_T$. The higher temperature enhances the diffusion process, resulting in more pronounced suppression in $R_{AA}$. For $\alpha > 1$, $R_{AA}$ exhibits significant suppression at higher $p_T$, with $\alpha = 1.6$ showing the largest suppression as more particles tend to remain at low $p_T$ due to the superdiffusion process.

\subsection{Summary and outlook}
\label{sec:Summary_Conclusions}

We discussed anomalous diffusion using the FLE with Caputo fractional derivatives, focusing on superdiffusion in the HQ dynamics within the QGP medium. Our study highlighted the impacts of superdiffusion for various values of $\alpha$ and $\beta$, showing that as these parameters approach 1, the anomalous diffusion turns back to normal diffusion. Key quantities like $\langle p^2(t)\rangle$ and $\langle x^2(t)\rangle$ were calculated. Further, we extended the analysis to  $R_{AA}$, demonstrating that superdiffusion leads to increased suppression in $R_{AA}$ at high $p_T$. The potential impact of superdiffusion on other observables, such as $v_2$ and particle correlations, suggests intensified HQ-bulk interactions at later stages. Future work will include a refined study of $v_2$ with realistic initial conditions and a comparison with experimental data to validate these findings.



\section{A modular perspective to the jet suppression from  a small to large radius  in very high transverse momentum jets}

\author{Manaswini Priyadarshini, Om Shahi, Vaishnavi Desai, Prabhkar Palni}

\bigskip

\begin{abstract}
In this work, we extend the scope of the JETSCAPE framework to cover the jet radius parameter ($R$) dependence of the jet nuclear modification factor, ${R_{AA}}$, for broader area jet cones, going all the way up to $R$ = 1.0 . The primary focus of this work has been the in-depth analysis of the high-${p_{T}}$ inclusive jets and the quenching effects observed in the quark-gluon plasma formed in the Pb-Pb collisions at ${\sqrt{\rm s_{NN}}}$ = 5.02 TeV for the most-central (0-10\%) collisions. The nuclear modification factor is calculated for inclusive jets via coupling of the MATTER module (which simulates the high virtuality phase of the parton evolution) with the LBT module (which simulates the low virtuality phase of the parton evolution). These calculations are then compared with the experimental results in the high jet transverse momentum (${p_{T}}$). The predictions made by the JETSCAPE are consistent in the high ${p_{T}}$ range as well as for extreme jet cone sizes, with the deviations staying within 10-20\%. We also calculate the double ratio (${R^{\mathrm{R}}_{\mathrm{AA}}/R^{\mathrm{R=small}}_{\mathrm{AA}}}$) as a function of jet radius, where the observations are well described by the JETSCAPE framework which is based on the hydrodynamic multi-stage evolution of the parton shower.
\end{abstract}


\ccode{PACS numbers:}


\subsection{Introduction}
The extremely hot and dense conditions created at the start of the Big Bang led to what we now understand as the soup of deconfined state of the partons, the quark-gluon plasma (QGP)\cite{1,2,3}, which can be now created for very short instance of time in a heavy-ion collision. The high-${p_{T}}$ jets produced in these heavy ion collisions undergo strong yield suppression and medium modification which are together referred to as jet quenching phenomena\cite{4,5,6}. Jets modification has been studied in nucleus-nucleus collisions relative to proton-proton collisions to probe the properties of the QGP via constraints from model-to-data comparison\cite{7,8,9}. The measurement of the nuclear modification factor provides strong confirmation of the interaction of partons with the deconfined plasma, the respective medium modifications, and the eventual hydrodynamization with the medium. In this paper, we aim to advance the limits of the current event generators and various energy loss modules for a better description of jet quenching phenomena at very high transverse momenta and broader jet cones with the multi-stage evolution of the parton shower within the JETSCAPE framework\cite{10}.

\begin{figure} [htbp]
\centering
\includegraphics[width=4.3cm]{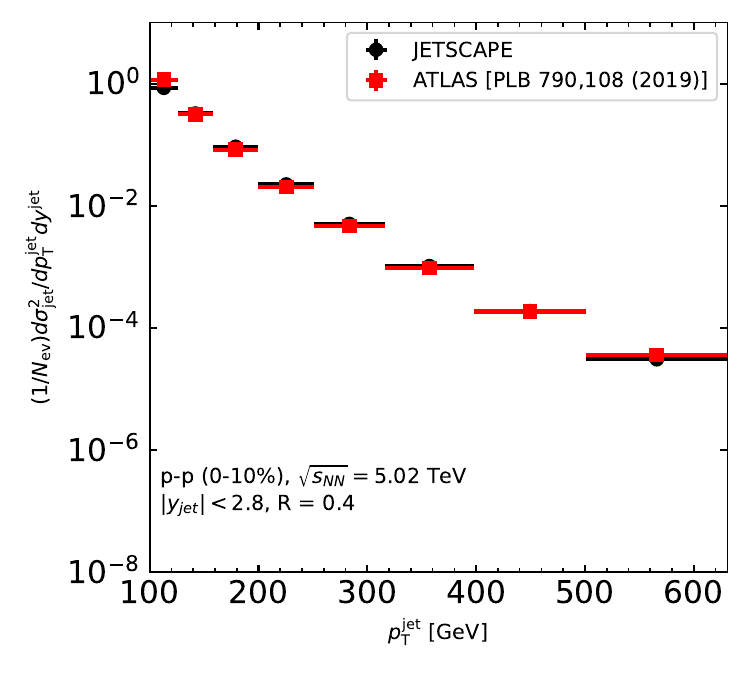}
\includegraphics[width=4.3cm,height= 4.3cm]{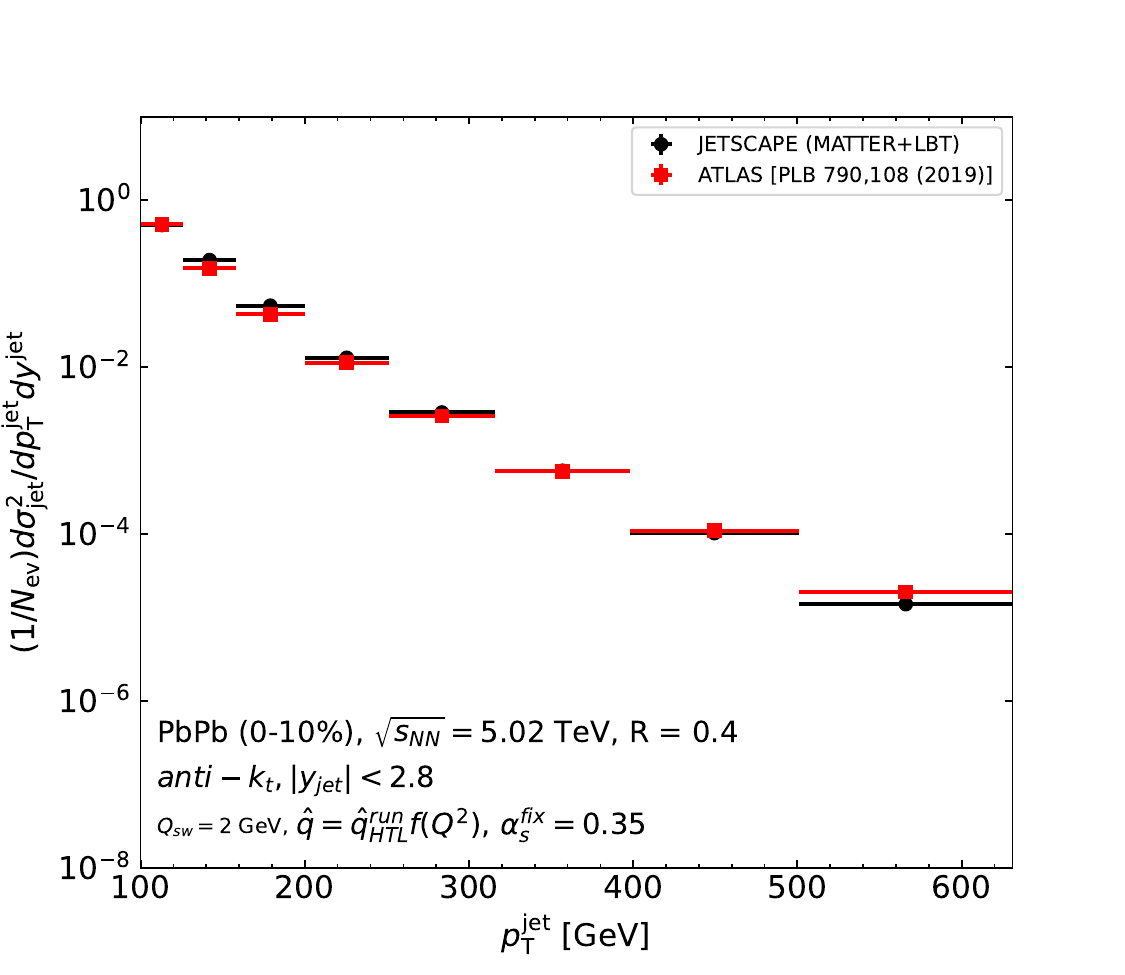}
\includegraphics[width=4.3cm,height=4cm]{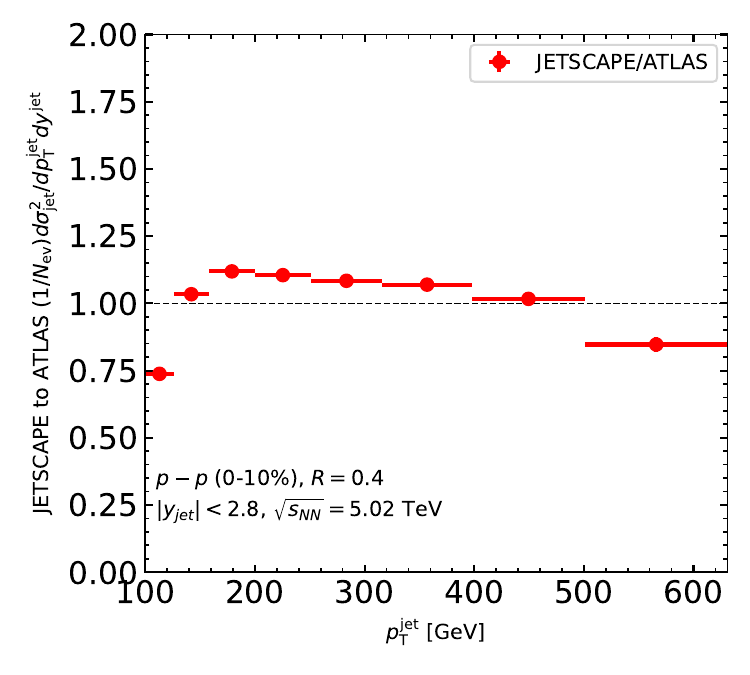}
\includegraphics[width=4.3cm]{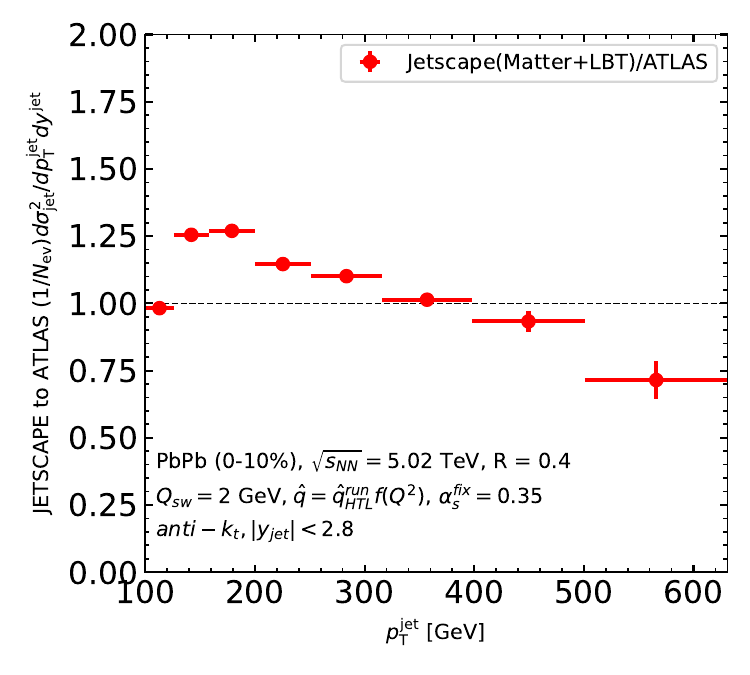}
\caption{Differential cross-section of inclusive jets for p+p and Pb+Pb collisions with cone size \\ R = 0.4, with a minimum track requirement of ${p^\mathrm{track}_\mathrm{T}>}$ 0.5 GeV. The bottom pannel shows the ratio of JETSCAPE to ATLAS data.}
\label{fig:jet-spectra}
\end{figure}

\subsection{Results}

This paper covers the collision energy: ${\sqrt{s_\mathrm{NN}}= 5.02}$ TeV for the most central (0-10\%) Pb-Pb collisions and includes a comparison with selected experimental data available from the ATLAS and the CMS collaborations for high-${p_{T}}$  jets. The energy loss depicted in this work is achieved by the coupling of MATTER with LBT module. The ratio of inclusive jet cross-section using the JETSCAPE to the ATLAS data\cite{11} is shown in the bottom pannel of Fig.~\ref{fig:jet-spectra}, which shows that the results are in the acceptable range.

\begin{figure} 
\centerline{\includegraphics[width=4.3cm]{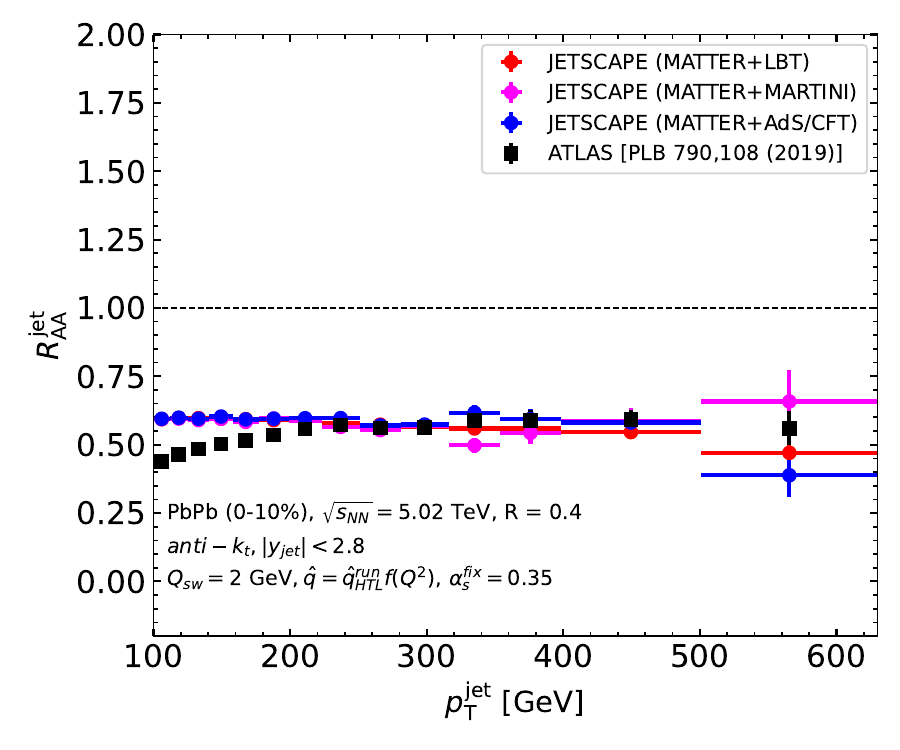}
\includegraphics[width=4.3cm]{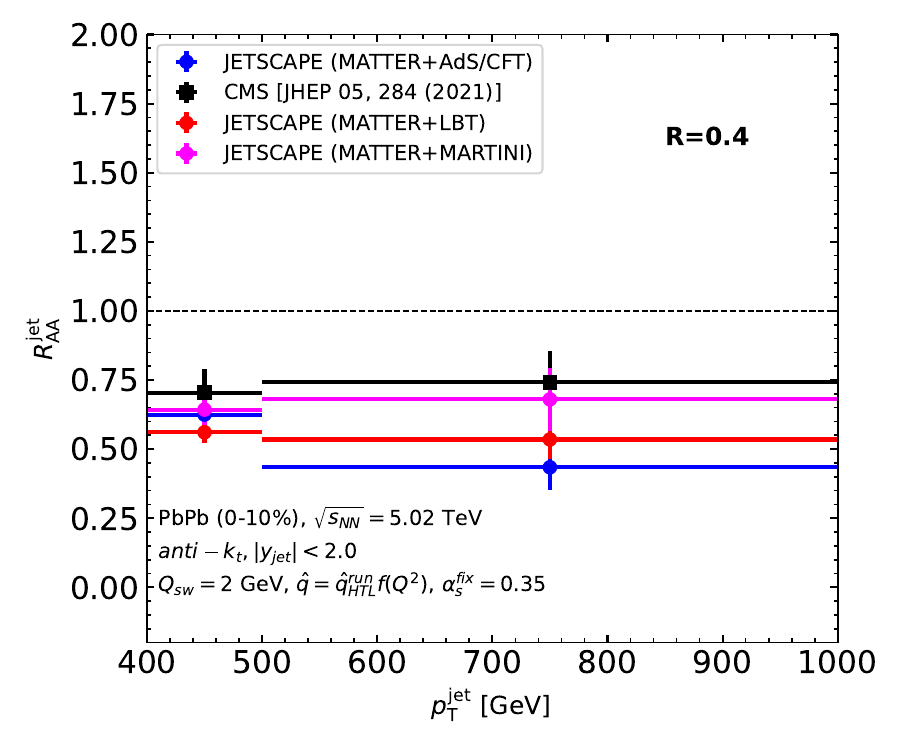}}
\caption{Jet-${R_{AA}}$ as a function of jet-${p_{T}}$ for inclusive jets in the most central (0-10\%)  Pb+Pb collisions at ${\sqrt{s_\mathrm{NN}}= 5.02\rm~TeV}$ for the jet cone radius $R$ = 0.4, are compared with the ATLAS and CMS data.}
\label{fig:RAA-ATLAS}
\end{figure}

 Simulations of Pb-Pb collisions are carried out using the MATTER module for both vacuum and medium showers, integrated with three distinct time-ordered parton shower models: LBT\cite{13}, MARTINI\cite{14} and AdS/CFT\cite{15}. The transition between different energy loss modules is performed independently for each parton, with the switching parameter $Q_0$ set at 2 GeV.  The inclusive jet-${R_{AA}}$ is calculated as the ratio of the Pb-Pb and p-p spectra, which is shown in Fig.~\ref{fig:RAA-ATLAS} in comparison to the ATLAS\cite{11} and CMS data\cite{12}. The (MATTER + LBT) shows significantly better alignment with the actual measurements, whereas the (MATTER + AdS/CFT) model shows more suppression at high ${p_{T}}$ region compared to the other configurations.

\subsubsection{Jet radius (${R}$) dependence of the ${R_{AA}}$}

\begin{figure} 
\centerline{{\includegraphics[width=4.3cm]{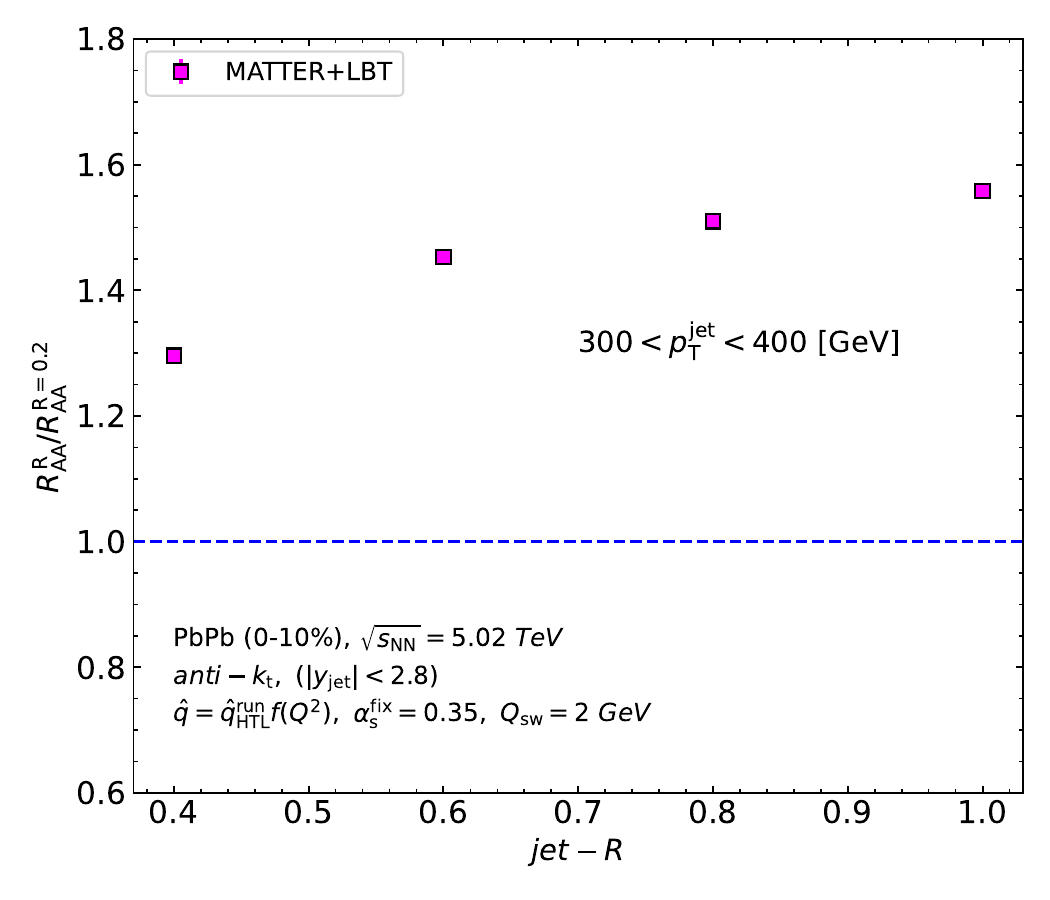}}
\includegraphics[width=4.3cm]{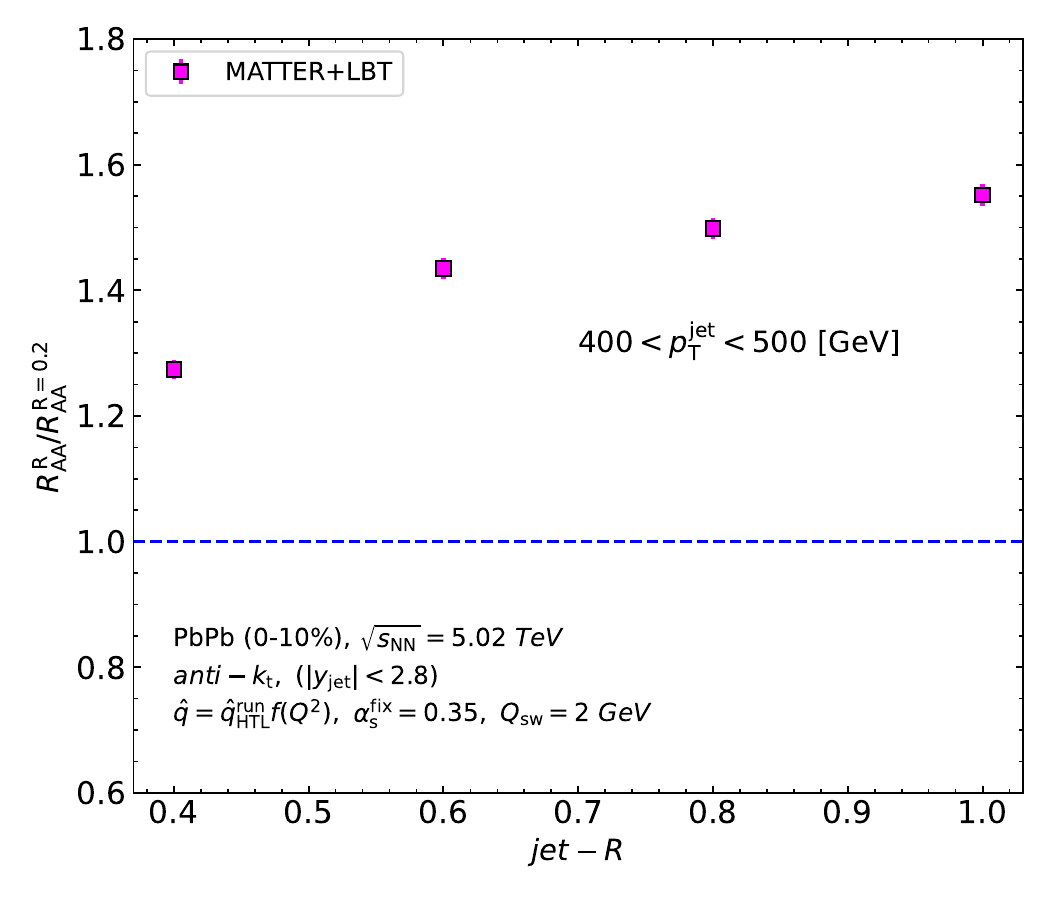}
\includegraphics[width=4.3cm]{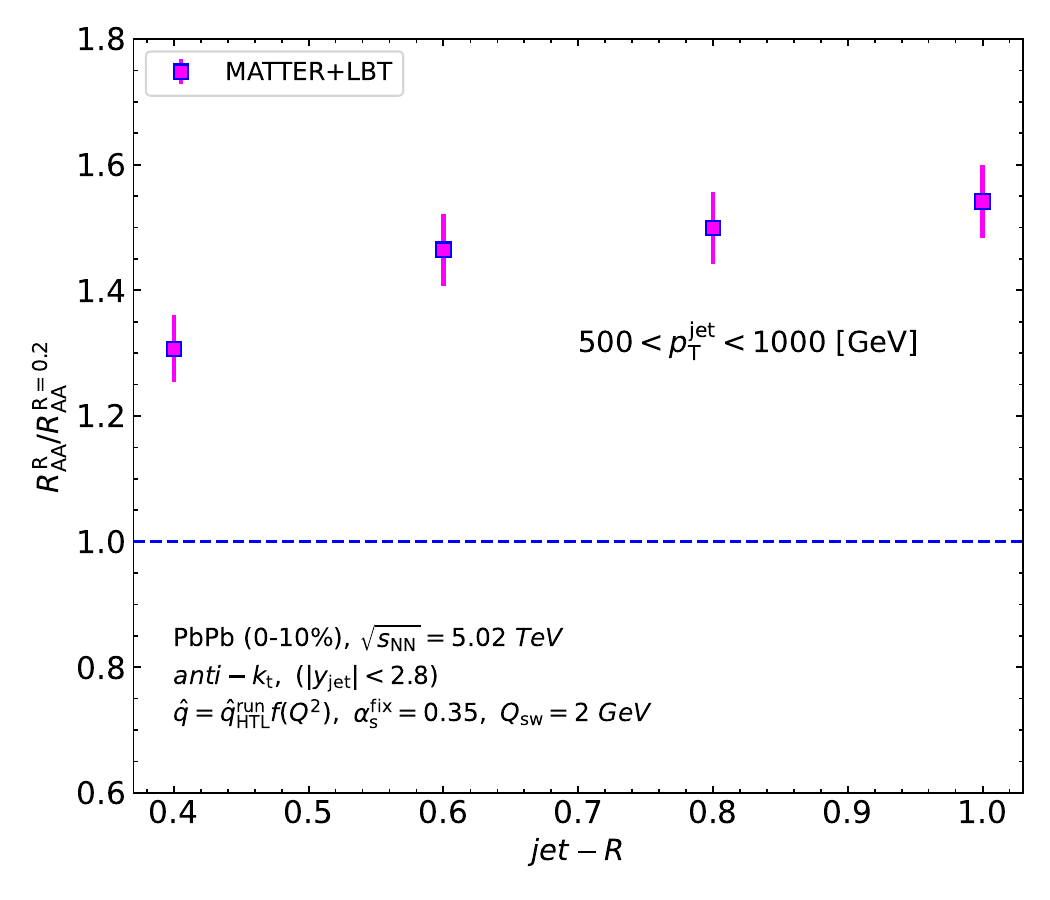}}
\caption{  The double ratio (${R^{\mathrm{R}}_{\mathrm{AA}}/R^{\mathrm{R=0.2}}_{\mathrm{AA}}}$) as a function of jet-R for inclusive jets in the most central(0-10\%) Pb+Pb collisions at ${\sqrt{s_\mathrm{NN}}= 5.02\rm~TeV}$ for different jet radii with a minimum track requirement of ${p^\mathrm{track}_\mathrm{T}>}$ 0.5 GeV.}
\label{fig:double-ratioR-300-400}
\end{figure}

We calculate the jet-${R_{\mathrm{AA}}}$ double ratio (${R^{\mathrm{R}}_{\mathrm{AA}}/R^{\mathrm{R=0.2}}_{\mathrm{AA}}}$) as a function of jet radius for the most central (0-10\%) Pb-Pb collisions over a range of jet-${p_{T}}$ from 300 GeV up to 1 TeV. The results are further sub-categorized by three jet-${p_{T}}$ intervals as shown in  Fig.~\ref{fig:double-ratioR-300-400}. We observe a consistent and identical trend across the entire ${p_{T}}$ range, indicating a marginal enhancement in the double ratio. This implies that with a larger jet radius, the jet retains a significant proportion of the extensively distributed momentum and energy deposited in the plasma.

\subsubsection{Conclusion}

In this paper, we present the jet-${R_{AA}}$ predictions from the JETSCAPE framework incorporating the (2+1)D MUSIC model for viscous hydrodynamic evolution and compare to the ATLAS and CMS data for Pb+Pb collisions at ${\sqrt{s_\mathrm{NN}}= 5.02}$ TeV. We observe more suppression in jet-${R_{AA}}$ in the low to intermediate  ${p_{T}}$ region, which provides a better description of the observed data. The predictions made by the JETSCAPE framework for the jet-${R_{\mathrm{AA}}}$ double ratio (${R^{\mathrm{R}}_{\mathrm{AA}}/R^{\mathrm{R=0.2}}_{\mathrm{AA}}}$) as a function of jet radius R, indicate that the large radius jets retains a larger fraction of the initial hard parton momentum as a result the energy lost by the jets is partially gained as the area of the jet cone area increases.








\section{Boosting Signal Detection in Rare Higgs Decay to  Z$\gamma$ at $\sqrt{s}$ = 13 Tev}

\author{Manisha Kumari, Amal Sarkar}

\bigskip

\begin{abstract}
Higgs Boson is a particle described in the Standard Model characterized by quantum numbers $I^\pi$ = $0^+$ and devoid of both color charge and electric charge. Higgs Boson fundamentally forms the cosmos as it interacts with other particles to impart mass. Analysis for the Higgs boson decay to a photon and a Z boson which further decays into a pair of leptons ($\mu^\pm$ or $e^\pm$) is performed using collision data from p-p interactions produced by the event generator PYTHIA8 at collision energy of 13 TeV. Reconstruction of the Higgs mass by applying the selection criteria based on specific kinematic variables is done. A critical aspect of this analysis is to focus on the study of the angular correlation between the $P_{Z}$ vs $\theta_{\ell^+\ell^-}$ and $P_{H}$ vs $\theta_{Z\gamma}$ to enhance the signal intensity and calculate the signal-to-background ratio to effectively differentiate them. Acceptance and correction studies have also been done to incorporate into the cross-section calculation.    
\end{abstract}
\subsection{Introduction}
In 2012 CERN confirmed the existence of the Higgs boson, validating the predictions made by the standard model \cite{chatrchyan2012observation}. The search was conducted in five decay modes, $H \rightarrow \gamma\gamma$, $ZZ$, $W^+ W^-$, $\tau^+ \tau^-$, and $bb$. An observed excess in events above the expected background confirmed the Higgs boson’s existence with a 5$\sigma$ significance showing the most notable excess in the $\gamma\gamma$ channel followed by the next highest event count in the ZZ channel. The angular distributions of leptons in the bosonic decay channels were used to determine the particle's spin-parity. Which was found to align with the standard model's predicted value of \( J^P = 0^+ \).   
\begin{figure}[h!]
\centering
\includegraphics[width=9cm, height = 3cm]{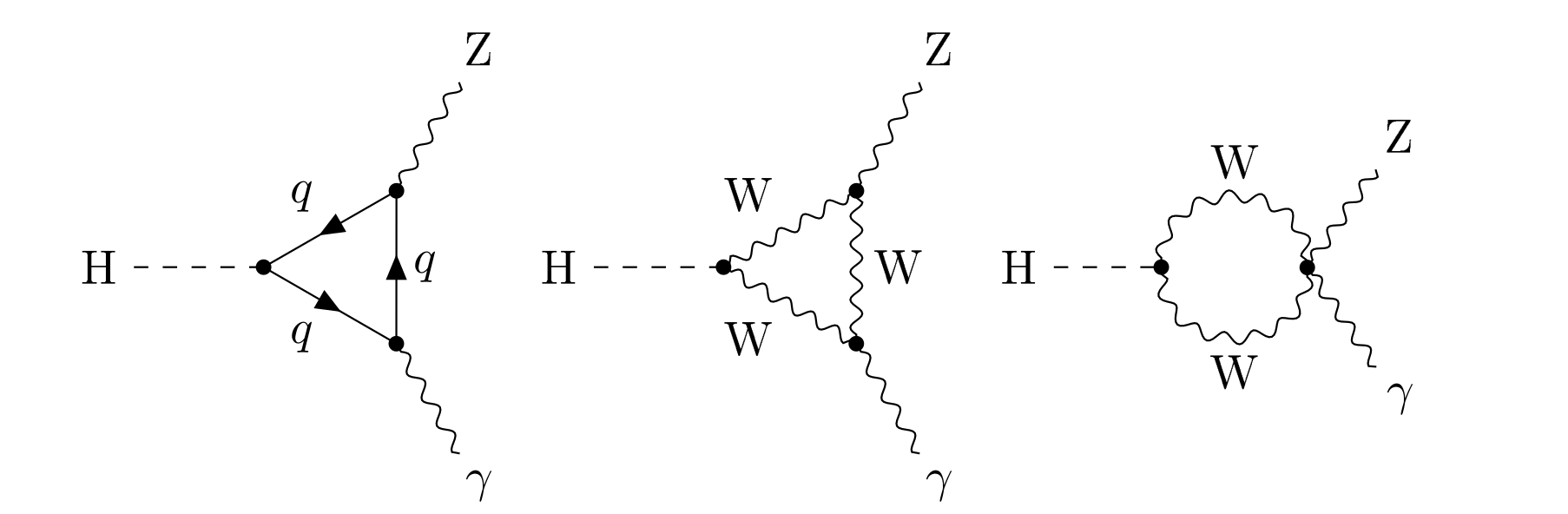}
\caption{Different potential Feynman diagrams for the decay process $H \rightarrow Z\gamma$.}
\label{fig:Feymann Diagrams}
\end{figure}
Recently, the CMS and ATLAS experiments reported results supporting the existence of Z$\gamma$ decay of Higgs\cite{giljanovic2023search}. In the Standard Model the Z$\gamma$ channel has a branching ratio of $\beta(H \rightarrow Z\gamma) = (1.57 \pm 0.09) \times 10^{-3}$. Since the Feynman diagram of Z$\gamma$ channel is similar to that of the $\gamma$$\gamma$ channel, investigating Z$\gamma$ is necessary. Loop diagrams in this process are particularly sensitive to BSM physics through loop corrections. The modifications in the couplings of the Higgs to $Z\gamma$ can impact the decay rate for $H \to Z\gamma$. The coupling modifiers $\kappa_{z\gamma}$ can be expressed in terms of $\kappa_i$. The Standard Model predicts a value of one for all  $\kappa$  parameters. However recent studies have measured $\kappa_{Z\gamma} = 1.65^{+0.34}_{-0.37}$ indicating a uncertainty 40\% \cite{cms2022portrait}. This can affect both the branching fraction of the $H \to Z\gamma$ decay and the production cross-section in associated processes. 
The ratio $\beta(H\rightarrow Z\gamma) / \beta(H\rightarrow \gamma\gamma) = 0.69\pm0.04$ serves as a probe for physics beyond Standard Model including supersymmetry and extended Higgs sector models and could imply that particles involved in the loop structure may not be those which are predicted by the standard model. This study centers on examining the angular correlation between the decay products and the momentum of the parent particle i,e 
 $P_Z$  versus $\theta_{\ell^+\ell^-}$ and  $P_H$ versus $\theta_{Z\gamma}$ . The aim is to use these correlations to enhance Higgs signal intensity by selecting optimal pairs.
 \subsection{Results and Discussion}
\label{sec:Results and Discussion}
{\bf{$\ell^{+}\ell^{-}\gamma$ Analysis:}} The study examines the angular distribution between $\theta_{\ell^+\ell^-}$ and  $\theta_{Z\gamma}$ vs $P_{Z}$  and  $P_{H }$ respectively concentrating on a single Higgs channel.
\begin{figure}[h!]
    \centering
    \begin{minipage}[b]{0.45\textwidth}
        \centering
        \includegraphics[width=1\textwidth]{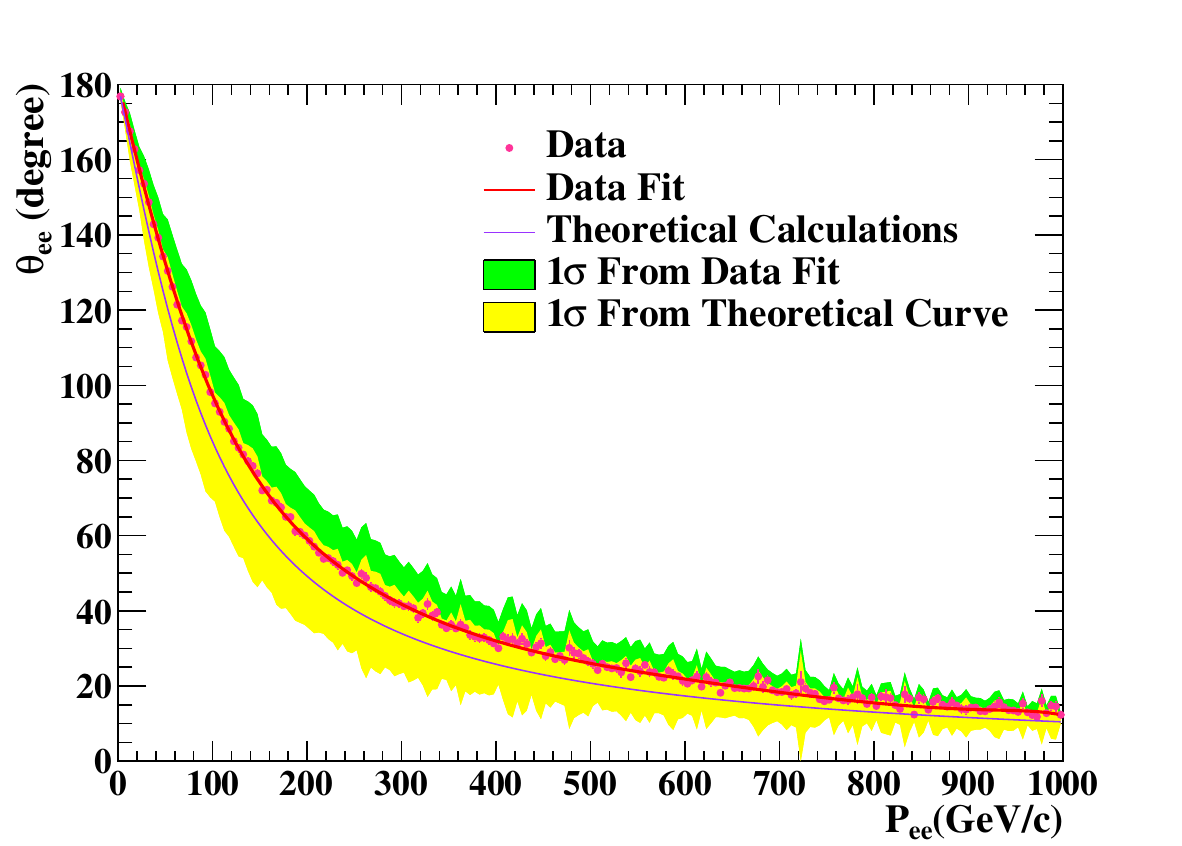}
    \end{minipage}%
    \hspace{0pt}  
    \begin{minipage}[b]{0.45\textwidth}
        \centering
        \includegraphics[width=1\textwidth]{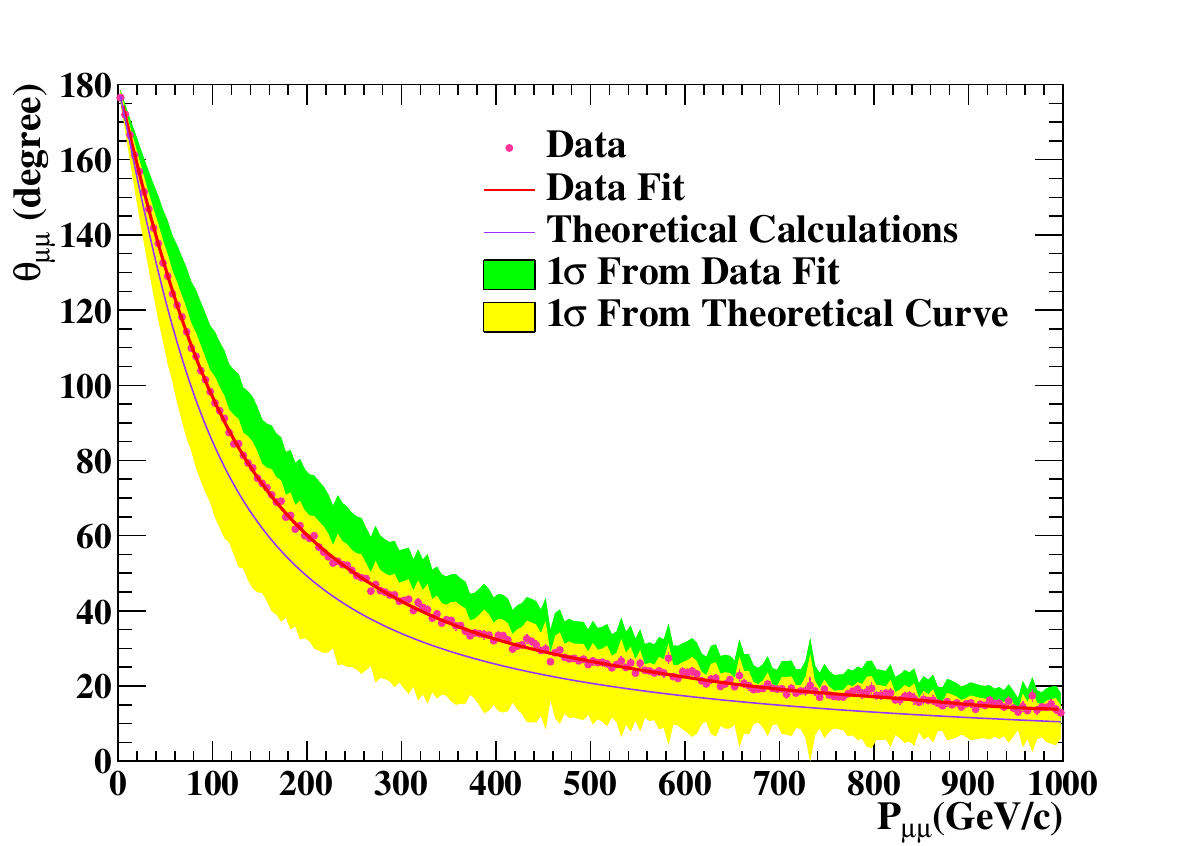}
     
    \end{minipage}%
    \vspace{0cm}  
    \begin{minipage}[b]{0.45\textwidth}
        \centering
        \includegraphics[width=1\textwidth]{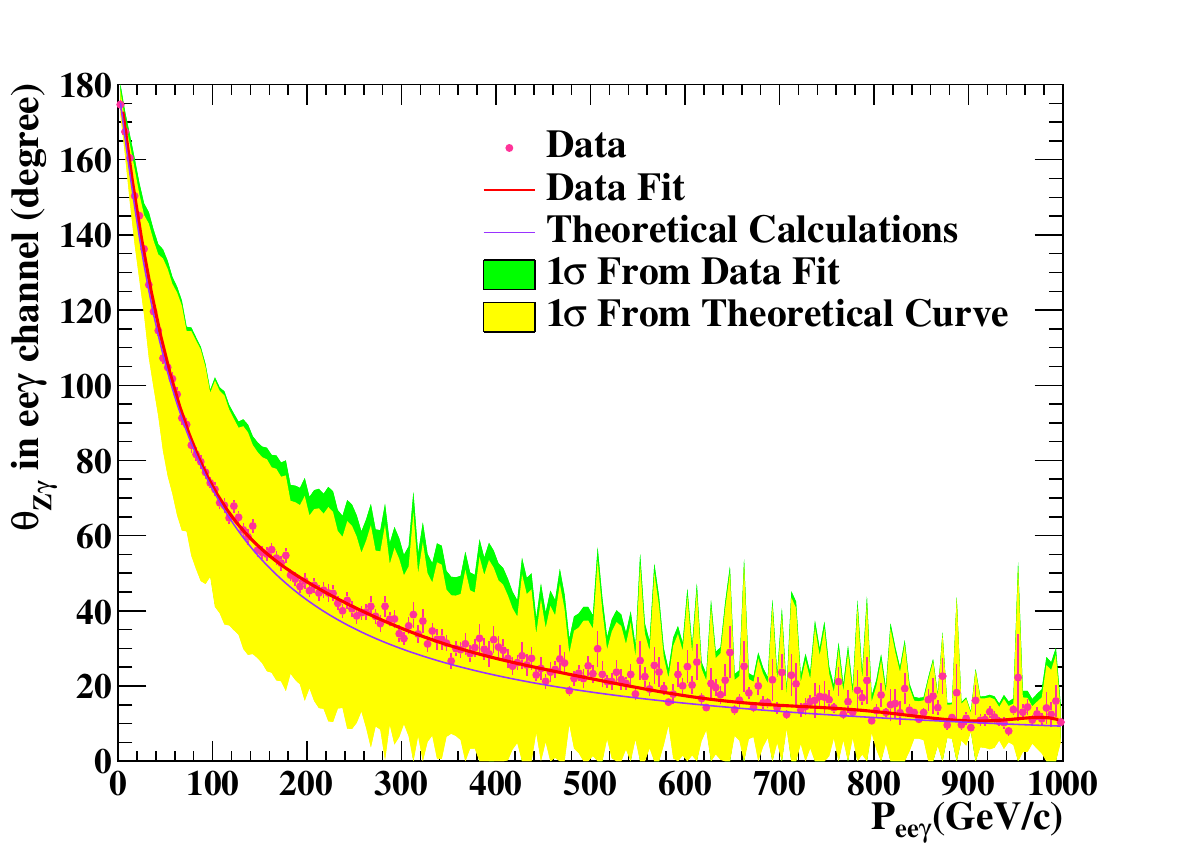}
     
    \end{minipage}%
    \hspace{0pt} 
    \begin{minipage}[b]{0.45\textwidth}
        \centering
        \includegraphics[width=1\textwidth]{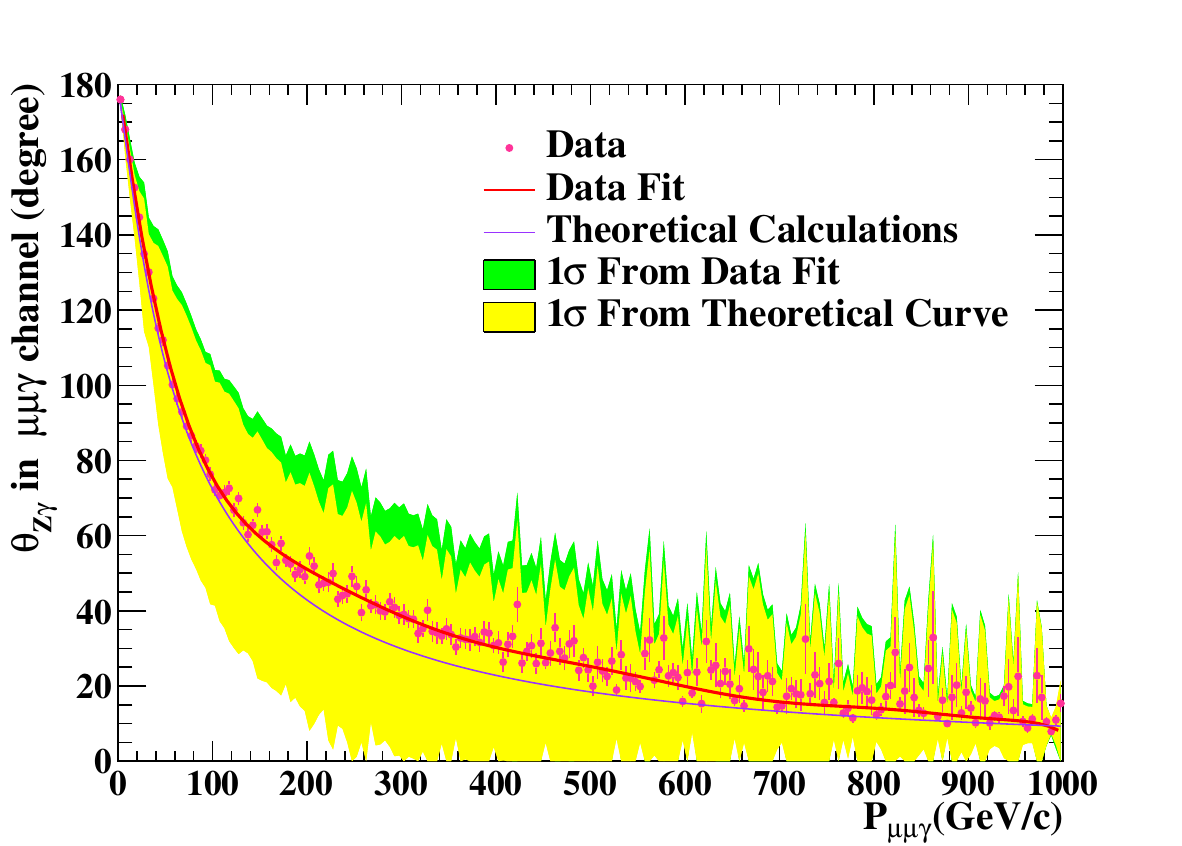}
   \end{minipage}%
    \caption{Upper left Plot for the angle $\theta_{ee}$ as a function of $P_{ee}$, Upper right Plot for the angle $\theta_{\mu\mu}$ as a function of $P_{\mu\mu}$, lower left Plot for the angle $\theta_{Z\gamma}$ as a function of $P_{ee\gamma}$, lower right Plot for the angle $\theta_{Z\gamma}$ as a function of $P_{\mu\mu\gamma}$.}
    \label{fig:angular_correlation}
\end{figure}
From the projection of each x bin of $P_{Z}$ and  $P_{H}$ along the y axis, parameters like mean (degree) and standard deviation (degree) have been extracted for all distributions shown in Figure~\ref{fig:angular_correlation}. Pairs ($\ell^+\ell^-$ or $Z\gamma$) satisfying the condition within one standard deviation (1$\sigma$) from both the theoretical value and extracted mean are taken into consideration for further process. This strategy have been applied while considering all possible decay channels of Higgs.
\\
 Invariant mass $m_{\ell^+\ell^-}$ is computed from the four momenta of the $\ell^+\ell^-$ for both muons and electrons.
Further  $m_{\ell^+\ell^-\gamma}$ is computed from the reconstructed momenta for $Z\gamma$.
\begin{figure}[h!]
    \centering
    \begin{minipage}[b]{0.45\textwidth}
        \centering
     
        \includegraphics[width=1\textwidth]{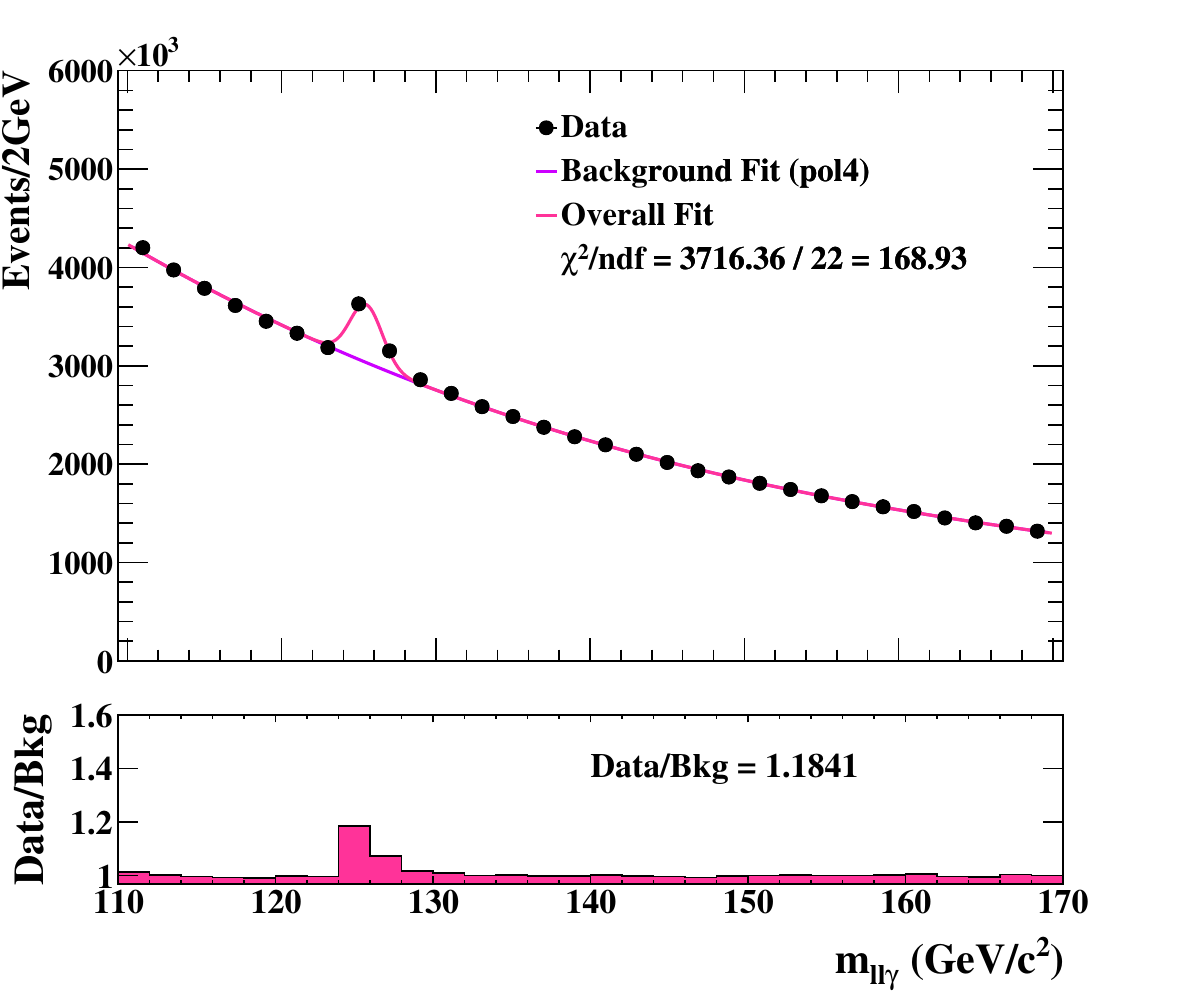}
        
    \end{minipage}%
    \hspace{0pt} 
    \begin{minipage}[b]{0.45\textwidth}
        \centering
      
        \includegraphics[width=1\textwidth]{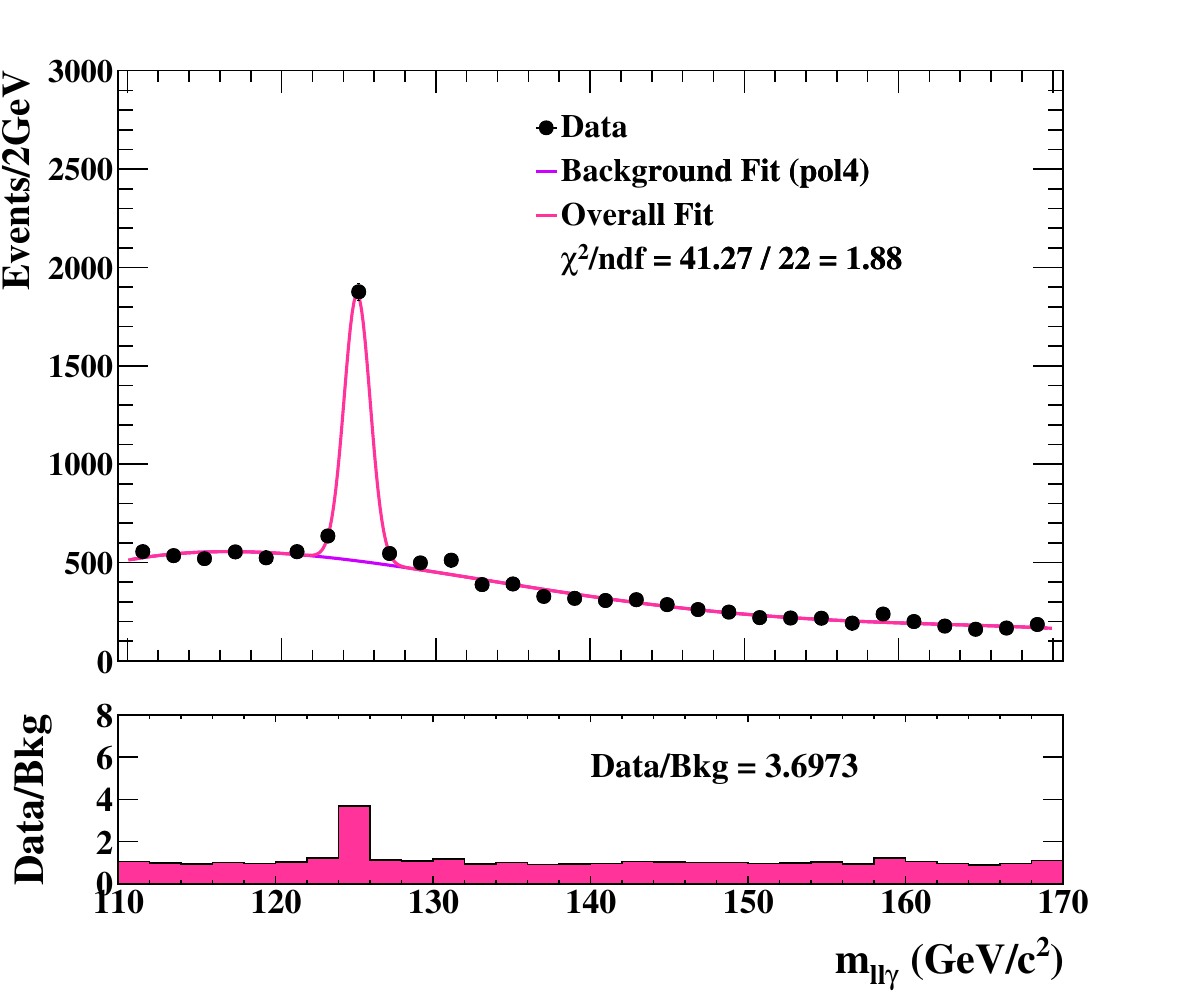}
       
    \end{minipage}%
    \vspace{0cm}  
    \begin{minipage}[b]{0.45\textwidth}
        \centering
    
        \includegraphics[width=1\textwidth]{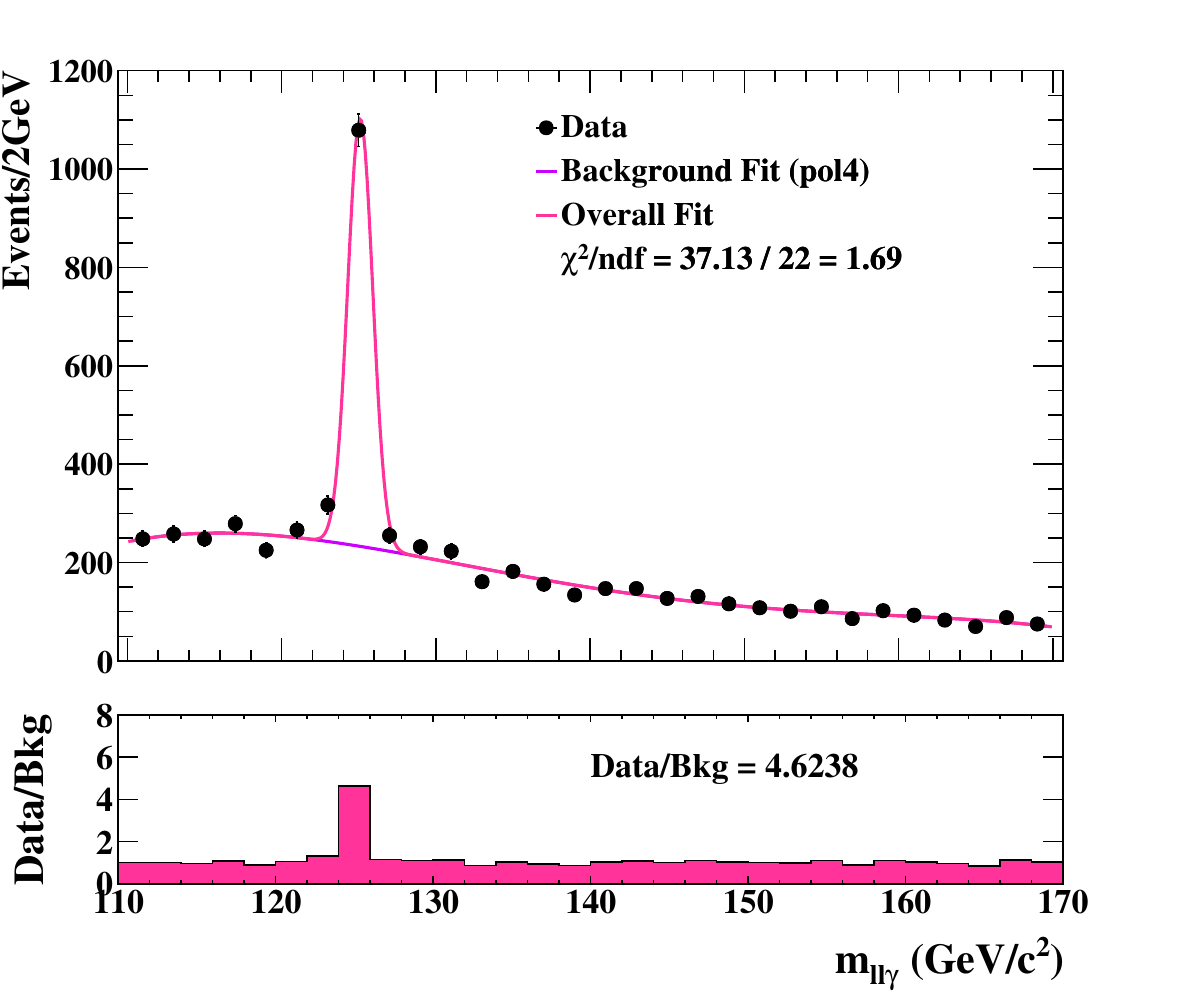}
       
    \end{minipage}%
    \hspace{0pt} 
    \begin{minipage}[b]{0.45\textwidth}
        \centering
      
        \includegraphics[width=1\textwidth]{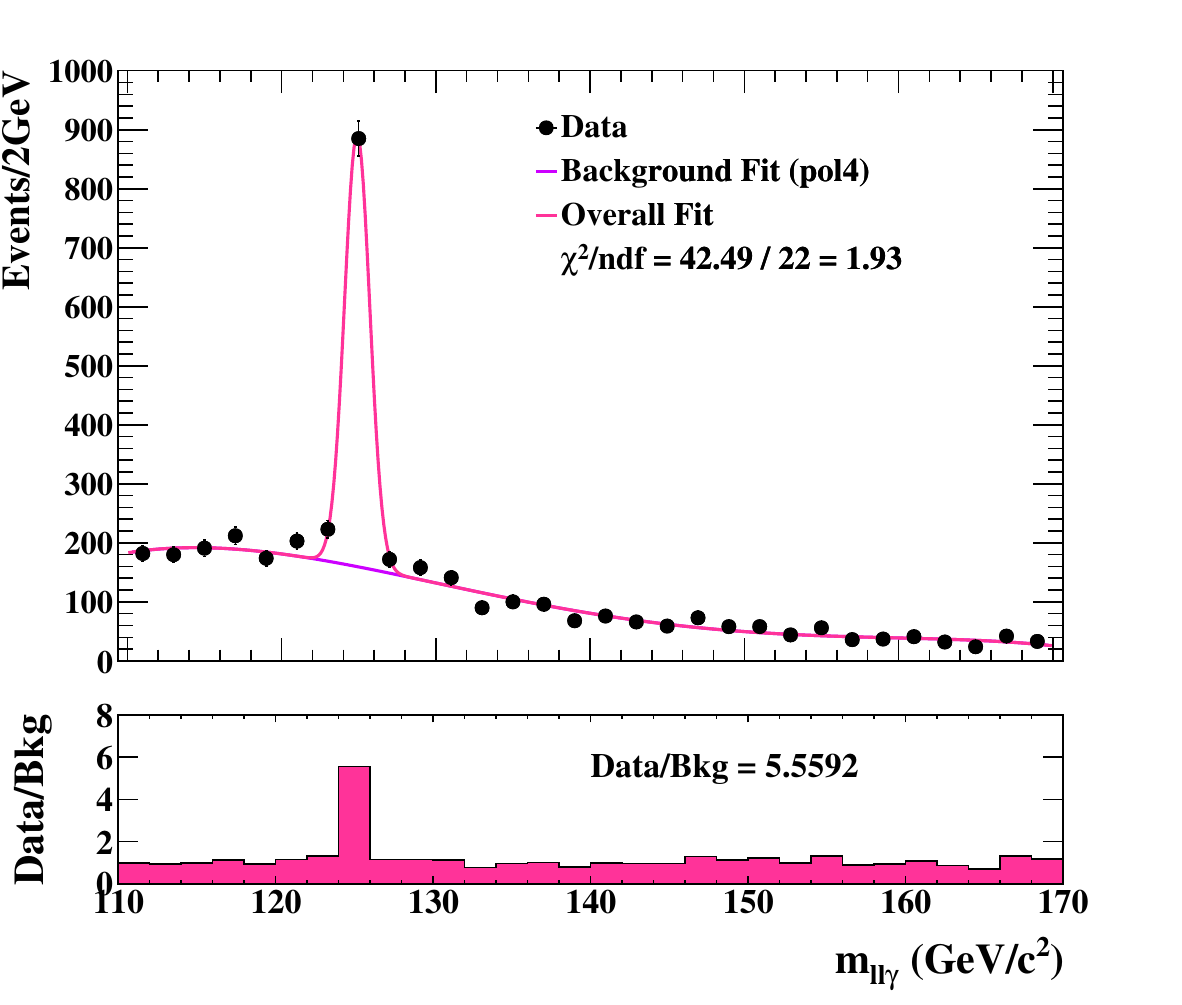}       
    \end{minipage}%
    \caption{Results are presented in four panels show $m_{\ell^+\ell^-\gamma}$ reconstruction plots(top) and Data/Bkg plots(bottom) in four analysis phases. Top left and right panels: Phase one and two analysis respectively as described in section \ref{sec:Results and Discussion}. Lower left panel: Introduces angular correlation between the $\ell\ell$ and $P_{Z}$ . Lower right panel: Incorporates correlation between Z$\gamma$ and $P_{H}$.}
\label{fig:Results}
\end{figure}   
 Z boson pole mass is taken to be 91.2GeV. The definition of $m_{ll\gamma}$ is:
\begin{equation}
    m_{\ell^+\ell^-\gamma}^{2} = \left( \sqrt{p_{\ell^+\ell^-}^{2} + m_{\ell^+\ell^-}^{2}} + \sqrt{p_\gamma^{2} + m_{\gamma}^{2}} \right)^{2} - \left| \vec{p}_{\ell^+\ell^-} + \vec{p}_\gamma \right|^{2}
\end{equation}
where $p_{\ell^+\ell^-}^{T}$ and $m_{\ell^+\ell^-}$ are the di-lepton transverse momentum and invariant mass, respectively. In phase one, events with two leptons and a photon where \( Z \rightarrow e^{+}e^{-} \) or \( \mu^{+}\mu^{-} \) were selected and all possible combinations taken into consideration. During phase two, selection criteria required \( p_T > 25/15 \ \text{GeV} \) for the highest-\( p_T \) and next-highest-\( p_T \) electron and \( p_T > 20/10 \ \text{GeV} \) for the highest-\( p_T \) and next-highest-\( p_T \) muon. Additionally, the photon needed \( p_T > 15 \, \text{GeV} \) and the di-lepton pair had to satisfy \( m_{\ell\ell} > 50 \, \text{GeV} \). In events with a higher number of lepton pairs the invariant mass of the di-lepton pair must be near the mass of the Z boson.
In phases three and four Angular correlation successfully helped in reducing the background and increasing signal-to-background ratio as compared to applying only kinematics selection in phase one and phase two.
\\
{\bf{Acceptance $\times $ Efficiency Calculations:}} In the analysis, PYTHIA8 model data have been used. But real detectors have some phase-space and kinematics limitations. To make analysis more realistic we should included acceptance and efficiency factors in cross-section studies.
\begin{equation}
\boldsymbol{\textbf{Acceptance} \times \textbf{Efficiency}} = \frac{\boldsymbol{\textbf{N}_{\textbf{reconstructed in Fiducial Space}}}}{\boldsymbol{\textbf{N}_{\textbf{Total Phase Space}}}}
\end{equation}
Acceptance is about how much of the relevant events our detector can see given its design. Efficiency refers to the proportion of events that fall within the selection criteria.\cite{santos2018combined}.
\begin{figure}[ht!]
    \centering
    \begin{minipage}[b]{0.44\textwidth}
        \centering
        \includegraphics[width=1\textwidth, height=0.2\textheight]{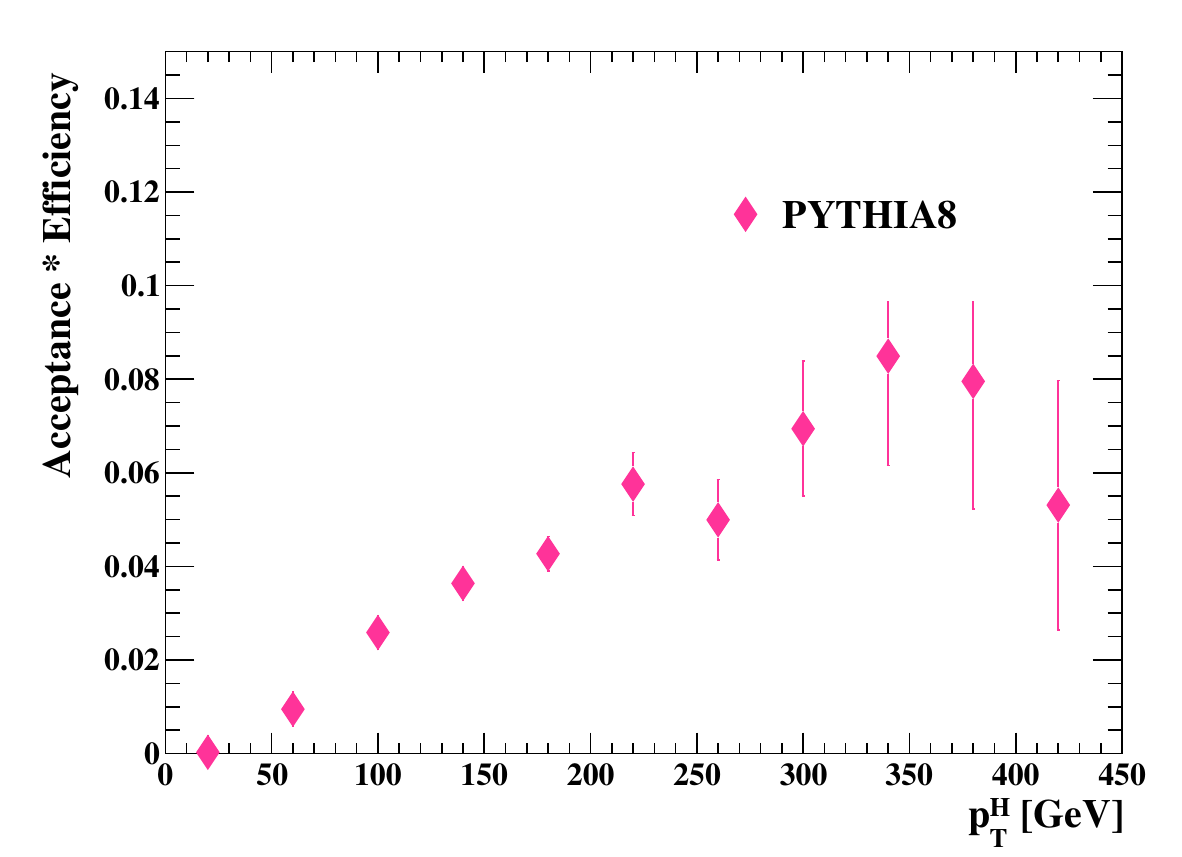} 
    \end{minipage}
    \hspace{0pt}
    \begin{minipage}[b]{0.44\textwidth}
        \centering
        \includegraphics[width=1\textwidth, height=0.2\textheight]{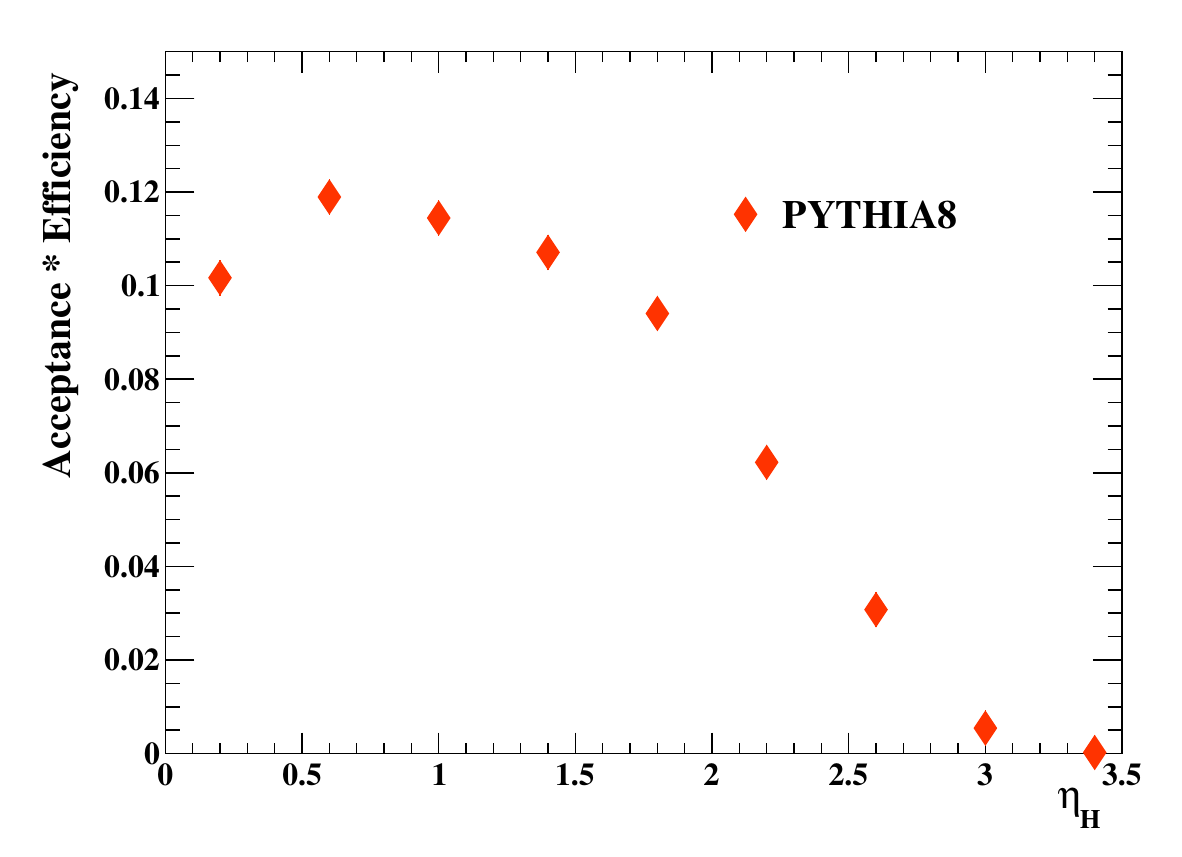}
    \end{minipage}
    \caption{Left plot: Acceptance $\times$ Efficiency as a function of $P^H_{t}$; Right plot: Acceptance $\times$ Efficiency as a function of $\eta_{H}$}
\end{figure}
Product of these two factors that have been calculated and presented. 
In fiducial Space $\left| \eta \right|$ have been taken as $ 0<\left| \eta \right|< 4.5$.
\subsection{Summary and conclusion}
In conclusion, 
pp collision data at an energy of $\sqrt{s} = 13 \text{TeV}$ generated using PYTHIA8, have been used for $H\rightarrow Z\gamma$ analysis. The invariant mass \( m_{\ell^+\ell^-} \)  and \( m_{\ell^+\ell^-\gamma} \) derived from the reconstructed Lorentz vector of the  Z(\(\ell^+\ell^-\)) and \(Z\gamma\) respectively by applying kinematic cuts in all four phases. The $1^{st}$ angular correlation in phase three, and $2^{nd}$ angular correlation in phase four both up to 1$\sigma$ were applied which enhanced the signal-to-background ratio up to several orders of magnitude.

\section{Impact of magnetic field on the properties of heavy quarkonia}

\author{Rishabh Sharma, Vineet Kumar Agotiya}

\bigskip

\begin{abstract}
The N-dimensional radial Schrödinger equation has been analytically solved using the Nikiforov-Uvarov (NU) method by incorporating the medium-modified Cornell potential and quasi-particle Debye mass under the influence of a strong magnetic field. It is found that the variation of binding energy and mass spectra decrease as the magnetic field increases, leading to early quarkonium dissociation. We also found that a higher dimensionality factor leads to higher initial binding energy values, but binding energy still decreases with temperature. The results obtained are consistent with recent theoretical studies.
\end{abstract}



\subsection{Introduction}
A significant challenge in non-relativistic quantum mechanics is obtaining exact and analytical solutions for the Schrödinger equation using specific potentials that are of particular interest in physics. In recent years, the potentials~\cite{Ikhdair, Inyang} which are used in analytical methods to find solutions in N- dimensions are Hulthen Hellmann potential, coulomb potential, Pseudo harmonic potential, Mie type potential, Hua potential, Screened Kratzer potential, global potential, and Cornell Potential with some modifications. Several methods~~\cite{Ikhdair} to solve the Schrodinger equation have been designed from time to time such as asymptotic iteration method (AIM), Laplace transformation method, supersymmetry quantum mechanics method (SUSYQM), power series method, analytical exact iteration Method (AEIM) has been used to get solution of Schrodinger equation in N-dimensions.\\ 
The NU method is an elegant and analytical method based on 2nd order differential equation.
From the last Decades the studies mainly focused on the effect of  magnetic field on the quark gluon using different phenomenology. Here in this manuscript we have studied the effect of variable magnetic field on the binding energy and mass spectra of ground state of heavy quarkonium.
\subsection{Methodology}
NU method~\cite{Rishabh_2024} is an appropriate mathematical tool to get the solution of 2nd order differential equation. By using coordinate transformation into hypergeometric type generalized equation can be written as,
\begin{equation}
	\label{1}
	\psi^{\prime\prime}\left(s\right)+\frac{\bar{\tau}\left(s\right)}{\sigma\left(s\right)}\psi^\prime\left(s\right)+\frac{\bar{\sigma}\left(s\right)}{\sigma^2\left(s\right)}\psi\left(s\right)=0.                 
\end{equation}
Where $\sigma(s)$ and $\bar{\sigma}(s)$ are the polynomials of second degree and $\bar\tau(s)$ is polynomial of first degree.
According to quantum mechanics the two particle system can be described by Schrodinger equation.The Schrodinger equation for two interacting quarks in N-dimensional space \cite{Shady} can be written as:
\begin{equation}
	\label{2}
	\frac{d^2R}{dr^2}+2\mu\left[E-V\left(r\right)-\frac{\left[L+\left(\frac{N-1}{2}\right)\right]^2-\frac{1}{4}}{2\mu r^2}\right]R\left(r\right)=0.     
\end{equation}
 Where V(r) is the medium modified Cornell potential~\cite{Solanki} which is represented as:
\begin{equation}
	\label{3}
	V\left(r\right)=\ -\left[\left(\frac{2\sigma}{m_D^2}-\alpha\right)\frac{m_D^3}{6}\right]r^2+\left[\left(\frac{2\sigma}{m_D^2}-\alpha\right)\frac{m_D^2}{2}\right]r+0r^0-\frac{\alpha}{r}.
\end{equation}
Where $\sigma$ is string term and $\alpha$ is two loop running coupling constant. Quasi particle Debye mass ($m_D$) in the presence of strong magnetic field (eB) for hot QCD medium is given as: \cite{Lal}  
\begin{equation}
	\label{4}
	m_D^2\left(T,eB\right)=4\pi\alpha\left(T^2+\frac{3eB}{2\pi^2}\right).	
\end{equation}
The value of potential from eq. (\ref{3}) putting in eq. (\ref{2}) we get the equation look like
  
\begin{equation}
	\begin{aligned}
		\label{5}
		\frac{d^2R}{dx^2}+\frac{2x}{x^2}\frac{dR}{dx}+\frac{2\mu}{x^4}\left[E-\frac{a_1}{x^2}-\frac{a_2}{x}+a_4x-\left(\frac{\left[L+\left(\frac{N-1}{2}\right)\right]^2-\frac{1}{4}}{2\mu}\right)x^2\right]R(x)=0.
	\end{aligned}    
\end{equation}
Now to solve the eq. \ref{5} with NU method\cite{Rishabh_2024} and we get the binding energy (B.E.) and Mass Spectra (M.S.) which is represented as:
\begin{equation}
	\label{7}
	B.E=\frac{a_2}{\delta}(2-m_D )-\frac{2\mu\left[\frac{a_2}{\delta^2}\left(1-\frac{2m_D}{3}\right)+\alpha\right]^2}{\left[(2n+1)+\sqrt{1+4\left[\left(L+\frac{N-2}{2}\right)^2-\frac{1}{4}\right]}\right]^2}.
\end{equation}

\begin{equation}
	\label{8}
	M. S.=2m_Q+\frac{a_2}{\delta}(2-m_D )-\frac{2\mu\left[\frac{a_2}{\delta^2}\left(1-\frac{2m_D}{3}\right)+\alpha\right]^2}{\left[(2n+1)+\sqrt{1+4\left[\left(L+\frac{N-2}{2}\right)^2-\frac{1}{4}\right]}\right]^2}. 
\end{equation}

\begin{figure*}[ht]
	\centering
	\begin{minipage}[t]{0.24\textwidth}
		\centering
		\includegraphics[height=3cm, width=3cm]{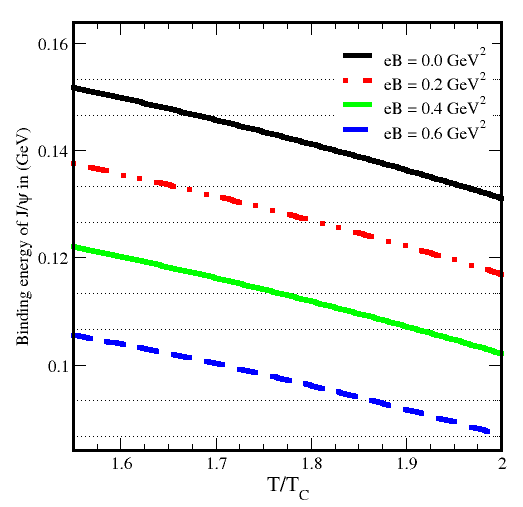}
		\label{fig:Bpsi}
	\end{minipage}
	\hfill
	\begin{minipage}[t]{0.24\textwidth}
		\centering
		\includegraphics[height=3cm, width=3cm]{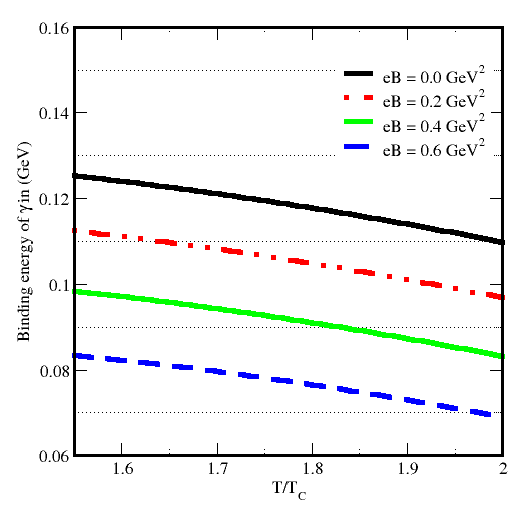}
		\label{fig:Bupsilon}
	\end{minipage}
	\hfill
	\begin{minipage}[t]{0.24\textwidth}
		\centering
		\includegraphics[height=3cm, width=3cm]{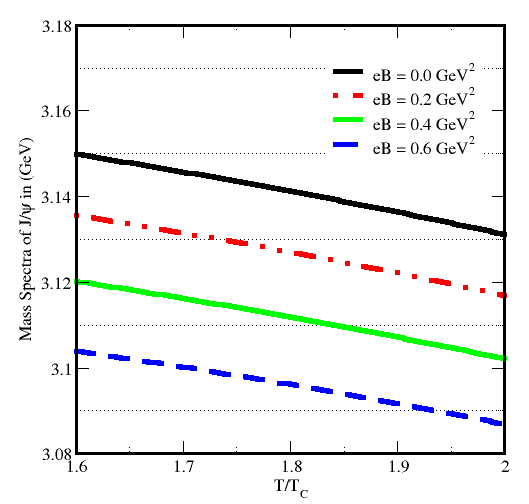}
		\label{fig:mpsi}
	\end{minipage}
	\hfill
	\begin{minipage}[t]{0.24\textwidth}
		\centering
		\includegraphics[height=3cm, width=3cm]{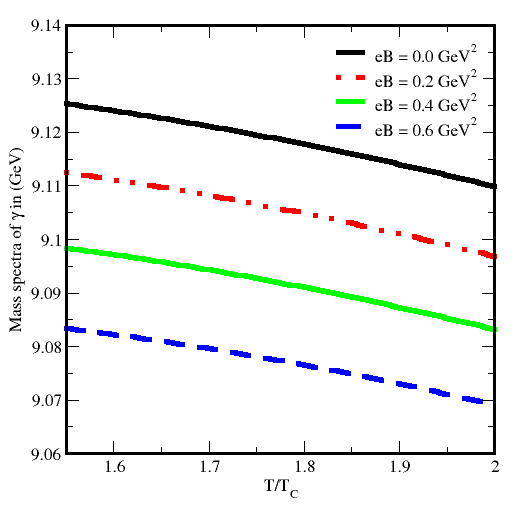}
		\label{fig:mupsilon}
	\end{minipage}
	
	\caption{Variation of binding energy (B.E.) of $J/\psi$ and $\Upsilon$ in the (a) and (b), and mass spectra (M.S.) for $J/\psi$ and $\Upsilon$ in (c) and (d), respectively, with $T/T_C$ at different values of magnetic fields at $N_f=4$.}
	\label{fig.group}
\end{figure*}

\subsection{Results and Conclusion}

The present study focuses on investigating the impact of magnetic fields on the binding energy (B.E.) and mass spectra (M.S.) of heavy quarkonium using the NU method, incorporating a medium-modified Cornell potential. Our results indicate that the binding energy of heavy quarkonium decreases in the presence of a magnetic field. Fig.~\ref{fig.group} illustrates the variations in binding energy for $J/\Psi$ and $\Upsilon$ in panels (a) and (b), and the variations in mass spectra for $J/\Psi$ and $\Upsilon$  in panels (c) and (d), as a function of $T/T_C$ for different values of magnetic field (eB = 0, 0.2, 0.4, 0.6 $GeV^2$) at fixed values of temperature (T= 0.3 GeV) and dimensionality (N=4). From Fig.~\ref{fig.group}, we conclude that both the binding energy and mass spectra decrease as the magnetic field strength increases. In future work, we plan to extend this research to explore additional thermodynamic properties.



\section{$c {\bar c}$ suppression in the pre-equilibrium stage}
	
\author{Mohammad Yousuf Jamal, Pooja, Partha Pratim Bhaduri, Marco Ruggieri, Santosh K. Das}

\bigskip

\begin{abstract}

This study explores the dissociation dynamics of \( c\bar{c} \) pairs in the pre-equilibrium, gluon-rich stage of high-energy nuclear collisions. The attractive force between $c$ and $\bar{c}$ pairs is modelled using Cornell potential. Employing the Wong equations, we simulate the interactions of \( c\bar{c} \) pairs with the evolving Glasma fields, where the dominant color field interaction drives the pairs apart, leading to dissociation. The finite probability of dissociation reveals the role of QCD dynamics in suppressing \( c\bar{c} \) states in the pre-equilibrium phase. Since, there is a possibility that the pair generate in the octet state. Hence, we also investigate the results considering a perturbative form of octet repulsive potential. The results offer insights for experimental interpretations.\\

\end{abstract}


\ccode{PACS numbers:}


\subsection{Introduction}
\label{sec:intro}
Relativistic heavy-ion collision (HIC) experiments, such as those conducted at RHIC and LHC, provide a crucial platform for exploring the pre-equilibrium phase, known as the Glasma. This phase occurs immediately after high-energy collisions of nuclei like Au-Au or Pb-Pb, where the system transforms into a dense state dominated by gluons and quarks. Intense gluon fields generated in this process lead to the formation of color flux tubes, which evolve into the Glasma—a highly non-equilibrium phase characterized by strong color fields and high energy densities. The Glasma plays a pivotal role in the early evolution of the system before it transitions into the Quark-Gluon Plasma (QGP).

Heavy quarks, such as charm and beauty quarks, are produced at very early stages in these collisions and traverse through multiple phases of the system, making them key probes for understanding both the Glasma and QGP \cite{ Singh:2023zxu, Agotiya:2016bqr, Jamal:2018mog, Jamal:2020rvh, Nilima:2024nvd}. Their interactions with the medium, particularly during the Glasma phase, impact diffusion and dissociation processes. Recent studies suggest that heavy quarks experience significant diffusion in the Glasma, which could have observable experimental effects. Moreover, the pair of charm and anticharm ($c\bar{c}$)—are of particular interest in this context. As this state interacts with the Glasma, they may dissociate, affecting the formation of mesons, such as D mesons, in the later stages of the collision.  This may play a significant role in setting the initial conditions that govern the yields and properties of heavy mesons.

\subsection{Formalism}
\label{sec:formalism}

The evolution of the Glasma fields is described within the McLerran-Venugopalan (MV) model, which is based on the color-glass condensate (CGC) effective theory \cite{	Lappi:2007ku, 	Das:2016cwd}. This model treats colliding nuclei as Lorentz-contracted sheets of colored glass, where fast-moving partons act as sources for slow gluon fields. The color charge densities of the colliding nuclei, denoted by \( \rho_A \) and \( \rho_B \), follow Gaussian statistics, with the saturation momentum \( Q_s \) being the key energy scale for the Glasma.

The Glasma fields are computed by solving Poisson's equation for the gauge potentials, which are used to calculate the Wilson lines. These, in turn, yield the gauge fields of the colliding nuclei. At the moment of the collision, the Glasma gauge potential is expressed as a sum of the gauge fields from the two nuclei:
\ba
    A_i = \alpha_i^{(A)} + \alpha_i^{(B)}, \quad A_z = 0,
\ea
where \( i = x, y \) and \( z \) is the beam direction. The longitudinal electric and magnetic fields in the Glasma at the initial time are given by the commutators of the gauge fields:
\ba
    E^z = -ig \sum_{i=x,y} [\alpha_i^{(B)}, \alpha_i^{(A)}], \quad B^z = -ig \left( [\alpha_x^{(B)}, \alpha_y^{(A)}] + [\alpha_x^{(A)}, \alpha_y^{(B)}] \right).
\ea
The classical Yang-Mills equations then govern the time evolution of these fields \cite{Das:2024vac, Ruggieri:2018rzi}. Charm quarks are produced in the early stages of the collision and propagate through the evolving Glasma. Their dynamics are governed by the Wong equations, which describe the quarks' motion under the influence of both the non-Abelian Lorentz force and a potential interaction between the charm and anticharm \cite{Pooja:2024rnn}:
\ba
    \frac{dx^i}{dt} = \frac{p^i}{E}, \quad \frac{dp^i}{dt} = g Q_a F_a^{i\nu} \frac{p_\nu}{E} - \frac{\partial V}{\partial x_i}.
\ea
The interaction potential is modelled as:
\ba
    V(r) = -\frac{3 \alpha_s}{4r} + \sigma r,
    \label{eq:prima_1}
\ea
where \( r \) is the ${c {\bar c}}$ separation, \( \alpha_s \) is the strong coupling constant, and \( \sigma \) is the string tension. The first term represents the perturbative short-range force, while the second term accounts for the non-perturbative long-range interaction.

\subsection{Results and Discussions}
\label{sec:results}

The transverse momentum distribution of  charm quark is initialized within the Fixed Order + Next-to-Leading Log (FONLL) QCD result that reproduces the D-mesons and B-mesons spectra in \( pp \) collisions after fragmentation~\cite{FONLL, Cacciari:2012ny}:
\begin{equation}
	\left.\frac{dN}{d^2 p_T}\right|_\mathrm{prompt} = \frac{x_0}{(x_1 + x_3{p_T^{x_1})}^{x_2}},
	\label{eq:FFNLO}
\end{equation}
the parameters that we use in the calculations are \( x_0  \), \( x_1\), \( x_2 \) and \( x_3 \) obtained via the slope of the spectrum that has been calibrated to a collision at \( \sqrt{s}=5.02 \) TeV.
The color charges are initialized on a \( 3 \)-dimensional sphere with radius one; \( Q_a Q_a \) corresponds to the total squared color charge, which is conserved in the evolution. Within this initial setup, we monitor the evolution of the \( c \bar{c} \)  pairs within the background Glasma and observe their separation and dissociation percentage under various scenarios. Our criterion for determining whether a pair dissociates or survives is based on the final distance between the evolving pairs in the Glasma. If the separation between the \( c \bar{c} \) pair after evolution time \( \tau \) is greater than or equal to a specified cutoff distance \( r_{c} \), we classify it as a dissociated pair; otherwise, it is considered as survived.

\begin{figure}[t!]
	\centering
	\includegraphics[width=5.3cm]{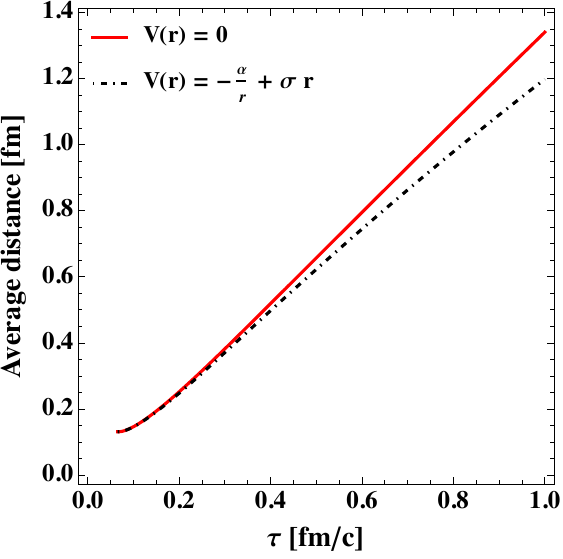}
	\hspace{3mm}
		\includegraphics[width=5.3cm]{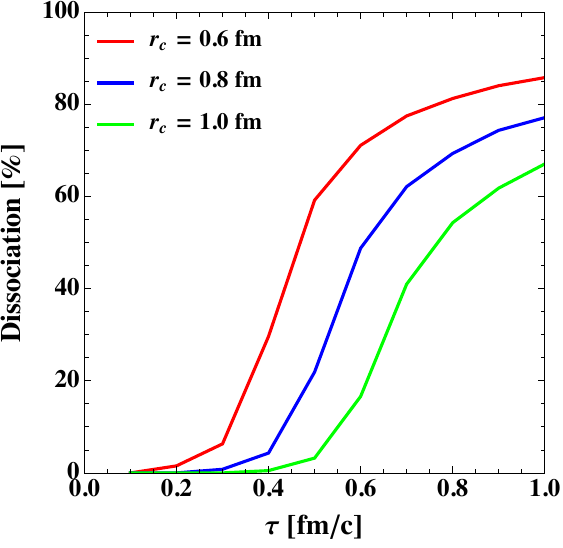}
	\caption{(Left Panel) Average distance between the \( c{\bar{c}} \) pair  versus proper time (\( \tau \)), with and without  potential~\eqref{eq:prima_1}. 
(Right Panel) Dissociation percentage of \( c{\bar{c}} \)  versus \( \tau \) at different values of cutoff, \( r_c \).
	}
	\label{Fig:separation}
\end{figure}	

In Fig.~\ref{Fig:separation}(left panel), we plot the average distance between quarks and antiquarks in \( c\bar{c} \)  pairs versus proper time ($\tau$), with (dot-dashed lines) and without (solid lines) the confining potential~\eqref{eq:prima_1}. It is interesting to note that \( V(r) \) slightly slows down the coordinates broadening of the \( c\bar{c} \) pair, indicating a strong attractive force among \( c \) and \( \bar{c} \); however, the effect of the strong gluon fields in which the charm quark diffuse is stronger, pushing the \( c \) and \( \bar{c} \) apart. In Fig.~\ref{Fig:separation}(right panel), we plot the dissociation probability of \( c\bar{c} \) pairs versus $\tau$ at specific values of \( r_c  = 0.6, 0.8, 1.0\) fm. It is observed that with the increase in \(\tau\), dissociation increases; however, as \( r_c \)  increases, the dissociation decreases. 

{ Furthermore, we can not neglect the possibility of $c\bar{c}$ to produce in the color octet state. Hence, we revised the analysis considering a perturbative form of octet repulsive potential. It is found that up to \( \tau \approx 0.5 \) fm/c (the expected lifetime of Glasma), the difference between results using singlet and octet potentials is under 10\%. This suggests that during the pre-equilibrium, gluon-dominated stage, fluctuations from color-singlet to color-octet states, while possible, likely do not significantly impact the dissociation rate predictions. However, a more detailed study is needed, which will be addressed in future work.}

\subsection{Summary and Conclusions}
\label{sec:Summary_Conclusions}
We investigated the evolution and the possibility of dissociation of \( c \bar{c} \) pairs within the Glasma medium, considering the Cornell potential. Despite this potential being attractive, the separation between \( q \bar{q} \) pairs increases due to the dominant influence of strong Glasma fields, leading to dissociation. The dissociation percentage varies from 0\% for \( \tau = 0.6 \) fm/c and \( r_c = 1 \) fm to approximately 80\% for \( \tau = 1.0 \) fm/c and \( r_c = 0.6 \) fm, illustrating the sensitivity to these parameters.  This study contributes to understanding the complex dynamics of \( c \bar{c} \) pairs in the Glasma medium, shedding light on their dissociation mechanisms. Future investigations could refine the initialization process to better estimate the final results, such as the amount of Charmonia and D mesons entering the QGP medium produced in HICs \cite{Das:2015ana, Das:2017dsh, Das:2015aga, Song:2019cqz}. Different approaches could also be employed to understand this complex phenomenon \cite{Das:2013kea,  Prakash:2023zeu,  Prakash:2024rdz, Jamal:2024qzy, Pooja:2023gqt, Ruggieri:2022kxv}.
Exploring dissociation in an expanding medium and integrating this study with QGP analyses, including recombination possibilities, may enhance our understanding of experimental results \cite{Das:2016llg, Das:2016cwd, Ghosh:2014oia, Plumari:2019hzp, Das:2022lqh, Das:2015ana}.


\section{Probing Hard Jet Substructure Using a Multi-Stage Approach}

\author{Gurleen Kaur, Prabhakar Palni}

\bigskip

\begin{abstract}
This study investigates hard jet substructure observables in p-p and Pb-Pb collisions at \( \sqrt{s} = 5.02 \) TeV using the JETSCAPE framework. The MATTER module simulates high virtuality phase with reduced medium interactions, while LBT, MARTINI, and AdS/CFT modules model jet energy loss at low virtuality. We analyze jet splitting momentum fraction \( z_g \), jet splitting radius \( \theta_g \), mass drop parameter \( \mu_g \), and groomed jet mass \( \rho \) in high \( p_T \) ranges (60-80 and 160-180 GeV) for 0-10\% centrality collisions, with and without coherence effects, to identify QGP-induced modifications. Results are compared with LHC experimental data.

\end{abstract}




\subsection{Introduction}

The Big Bang theory proposes that Quark-Gluon Plasma (QGP), a state of deconfined partons, briefly existed before matter formed \cite{G1}. Evidence of QGP was found in ultra-relativistic heavy-ion collisions at CERN's LHC and Brookhaven's RHIC, marking a major breakthrough in nuclear physics.

Hard interactions in hadronic collisions produce off-shell partons that become real through gluon radiation, forming jets \cite{G2}. In nuclear collisions, partons interact with QGP, modifying jet rates and structure, a phenomenon called jet quenching. With advanced detectors and high-statistics data from RHIC and LHC, jet substructure modifications are now studied, revealing insights into jet-medium interactions \cite{G3}.

Initially, highly virtual partons interact minimally with the medium, leading to Vacuum-like Emissions \cite{G4}. As partons split repeatedly, their virtuality decreases, increasing medium interaction and energy loss through scattering.

Coherence effects\cite{G4} play a significant role in jet modification in high-energy nuclear collisions. The medium can resolve jet partonic fragments only if their separation exceeds the jet resolution scale (R); otherwise, the fragments act coherently as a single emitter. These effects are captured by the jet quenching strength \cite{G5} \(\hat{q}\) defined as $\hat{q}$ = $\hat{q}_{HTL}f(Q^2)$, where \( f(Q^2) \) represents the modulation factor, given as:
\begin{equation}
f(Q^2) =
\begin{cases} 
\frac{1+10ln^2(Q_{SW}^2) + 100ln^4(Q_{SW}^2)}{1+10ln^2(Q^2) + 100ln^4(Q^2)} & \text{if } Q > Q_{SW} \\
1 & \text{if } Q < Q_{SW}
\end{cases}. \label{two}
\end{equation}

JETSCAPE (Jet Energy Loss Tomography with a Statistically and Computationally Advanced Program Envelope) is used for simulations. This paper provides a study of jet substructure observables using JETSCAPE. The multi-stage approach within this framework explicitly considers the virtuality-dependent jet evolution. 

\subsection{Simulations of A-A collisions: An overview}

To investigate medium-induced modifications in jet substructure, event-by-event space-time profiles of the QGP medium in A-A collisions are calculated using the MUSIC \cite{G6} module, based on initial conditions from TRENTO \cite{G7}. Hard scattering is simulated with PYTHIA 8 \cite{G8}, and the produced partons undergo multi-stage in-medium evolution within the JETSCAPE framework. The high virtuality stage is modeled using MATTER\cite{G9}, while LBT \cite{G5} handles the low virtuality stage, with bi-directional switching between modules based on a virtuality parameter \( Q_{SW} = 2 \, \text{GeV} \). If the parton virtuality drops below the cutoff \( Q^2_{\text{min}} = 1 \, \text{GeV}^2 \), hadronization occurs via the Colorless Hadronization module using the Lund string model \cite{G10} in PYTHIA 8.

\subsection{Groomed jet observables}

Jets are groomed using the Soft Drop (SD) grooming algorithm \cite{G11} from the FastJet package \cite{G12} with FastJet Contrib \cite{G13}. It iteratively checks for the SD condition:
\begin{equation}
\frac{p_{T,\text{subleading}}}{p_{T,\text{leading}} + p_{T,\text{subleading}}} \geq z_{\text{cut}} \left(\frac{\delta r}{R}\right)^{\beta} \label{four}
\end{equation}
where subleading prong is the one with lower $p_T$. $\delta r$ is the distance between the prongs in $\eta$-$\phi$ plane, and \( R \) is the jet cone radius. \( z_{cut} \) and \( \beta \) are the parameters controlling the grooming procedure. 

In this study, four groomed observables, namely jet splitting momentum fraction ($z_g$), jet splitting radius (scaled) ($\theta_g$), mass drop parameter ($\mu_g$), and groomed jet mass ($\rho$), are explored. $z_g$ is presented by the left side of Eq.~(\ref{four}), $\theta_g$ = $\sqrt{(\eta_1 - \eta_2)^2 + (\phi_1 - \phi_2)^2}/{R}$, $\mu_g$ = $max(m_1, m_2)/{M}$, and $\rho$ = $M_g/{p_T^{jet}}$. `1' and `2' in subscripts denote the prongs of the jets satisfying SD condition. 

The general form of the distribution of these observables is given as:
\begin{equation}
    \frac{1}{N_{jet}}\frac{dN_{SD,jet}}{d(observable)} = \frac{1}{\sigma_{jet}}\frac{d\sigma_{SD,jet}}{d(observable)}
    \label{thirteen}
\end{equation}
where \( N_{jet} \) is the number of inclusive jets, \( N_{SD,jet} \) is the number of jets satisfying the Soft Drop condition, and \( \sigma_{jet} \) and \( \sigma_{SD,jet} \) are the corresponding cross-sections.

\subsection{Results and Discussions}

This section presents JETSCAPE (JS) results for jet substructure observables in Pb-Pb collisions at $\sqrt{s_{NN}} = 5.02$ TeV for two $p_T$ ranges (60–80 GeV, 160–180 GeV) in 0–10\% centrality, with and without coherence effects. Parameters are $R=0.2$, $\eta<0.7$, $z_{\text{cut}}=0.2$, $\beta=0$ for the low $p_T$ range, and $R=0.4$, $\eta<1.3$, $z_{\text{cut}}=0.1$, $\beta=0$, $\Delta R_{12}>0.1$ for the high $p_T$ range.

\begin{figure}[th]
    \centering
    \begin{minipage}{0.5\textwidth}
        \centering
        \includegraphics[width=5.7cm, height=4.2cm]{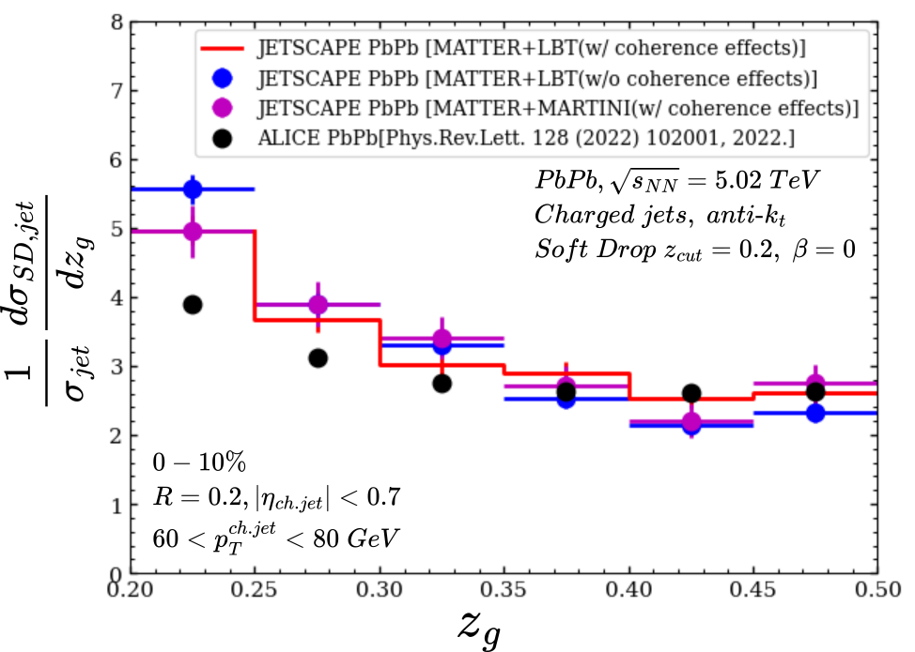}
    \end{minipage}\hfill
    \begin{minipage}{0.5\textwidth}
        \centering
        \includegraphics[width=5.7cm, height=4.1cm]{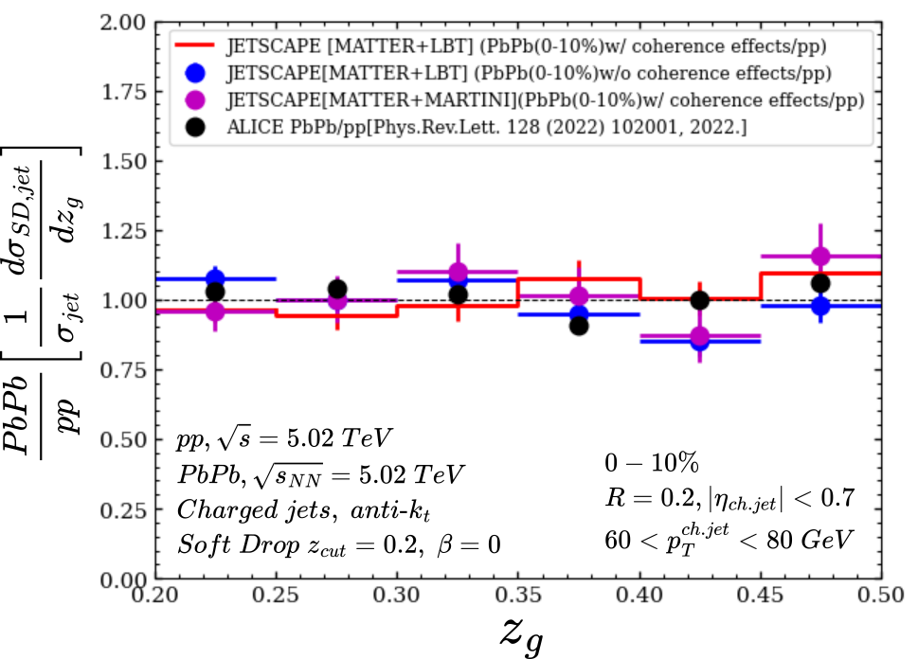}
    \end{minipage}
    \caption{Analysis of \( z_g \) in the 60--80 GeV range, comparing the energy loss models MATTER+LBT and MATTER+MARTINI with ALICE experimental data \cite{G14}. The left panel shows the \( z_g \) distribution for Pb-Pb collisions, while the right panel presents the ratio of \( z_g \) distributions between Pb-Pb and p-p collisions.}
    \label{fig:1}
\end{figure}

In Fig.~\ref{fig:1}, JS with coherence effects (MATTER+LBT, MATTER+MARTINI) slightly overestimates $z_g$ distribution at low $z_g$ but matches experimental data at mid-to-high $z_g$. Results without coherence effects show a more pronounced overestimation at low $z_g$. The ratio plot (right panel) shows the ratio near one, suggesting minimal QGP impact on $z_g$.

\begin{figure}[th]
    \centering
    \begin{minipage}{0.5\textwidth}
        \centering
        \includegraphics[width=5.6cm, height=4.1cm]{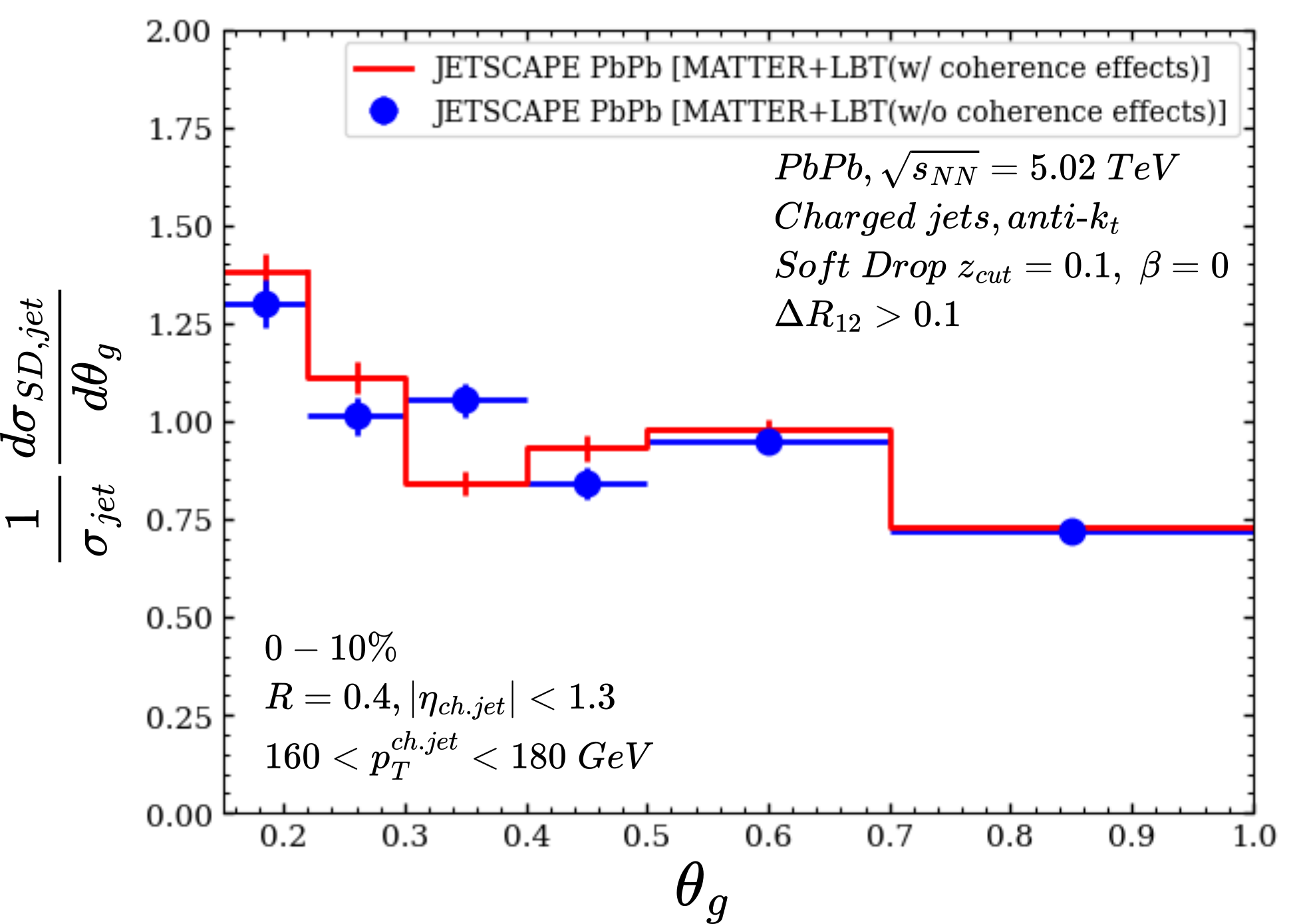}
    \end{minipage}\hfill
    \begin{minipage}{0.5\textwidth}
        \centering
        \includegraphics[width=5.6cm, height=4.1cm]{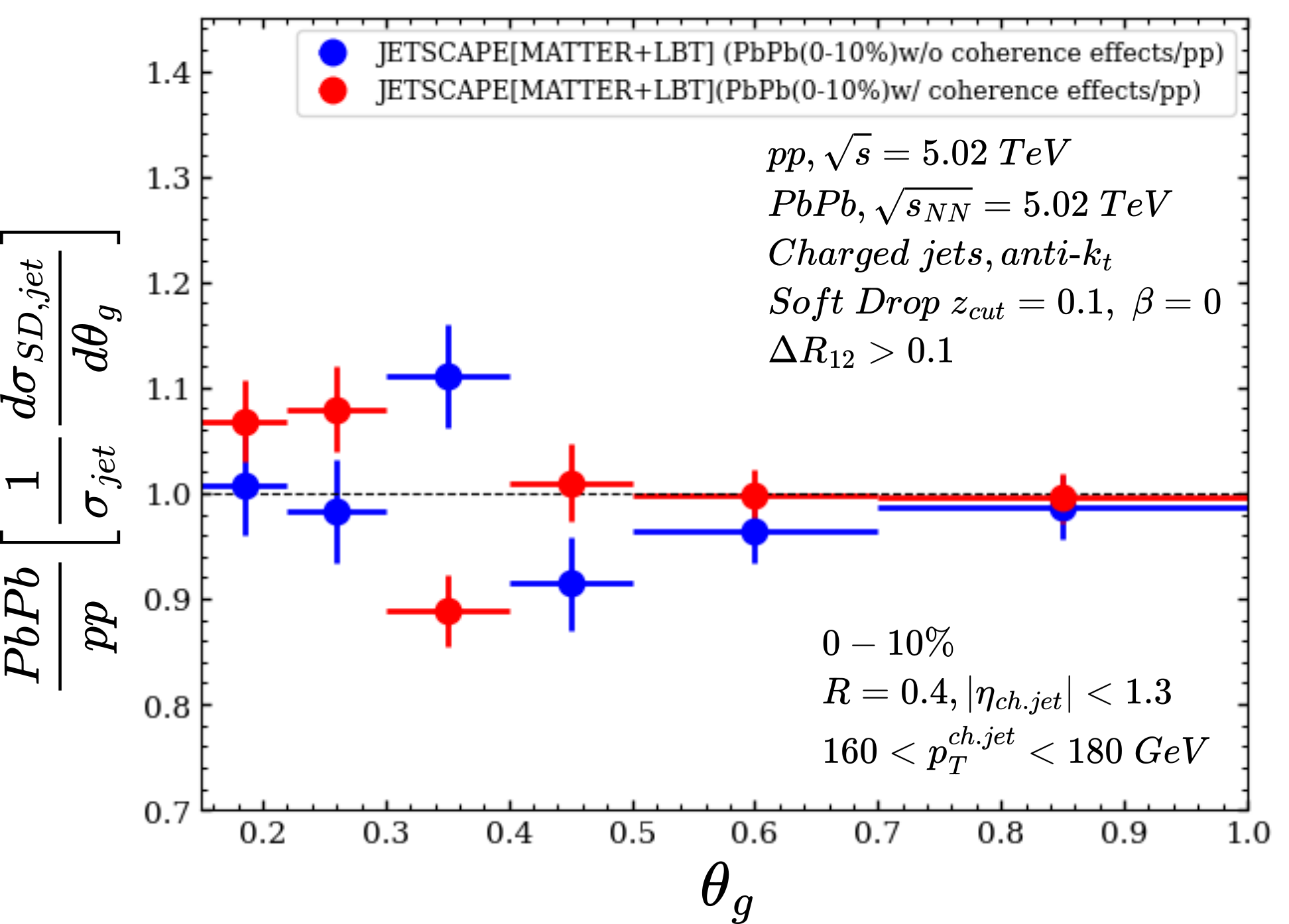}
    \end{minipage}
    \caption{Results for \(\theta_g\) using MATTER+LBT in the 160--180 GeV range, with and without coherence effects, as simulated by JS. Displayed on the left is the \(\theta_g\) distribution for Pb-Pb collisions; on the right, the ratio of the \(\theta_g\) distribution for Pb-Pb to p-p collisions.}
    \label{fig:2}
\end{figure}

Fig.~\ref{fig:2} (left) shows more jets with closely spaced prongs in the $\eta$-$\phi$ plane than those separated by $R$, with fluctuations in the intermediate $\theta_g$ region. The right panel indicates that coherence effects enhance the Pb-Pb $\theta_g$ distribution at low $\theta_g$ but have minimal impact at higher $\theta_g$. Without coherence, Pb-Pb shows suppression at low $\theta_g$. This highlights the significant role of coherence effects in parton-QGP interactions and experimental observations.

\begin{figure}[th]
    \centering
    \begin{minipage}{0.5\textwidth}
        \centering
        \includegraphics[width=5.6cm, height=4.1cm]{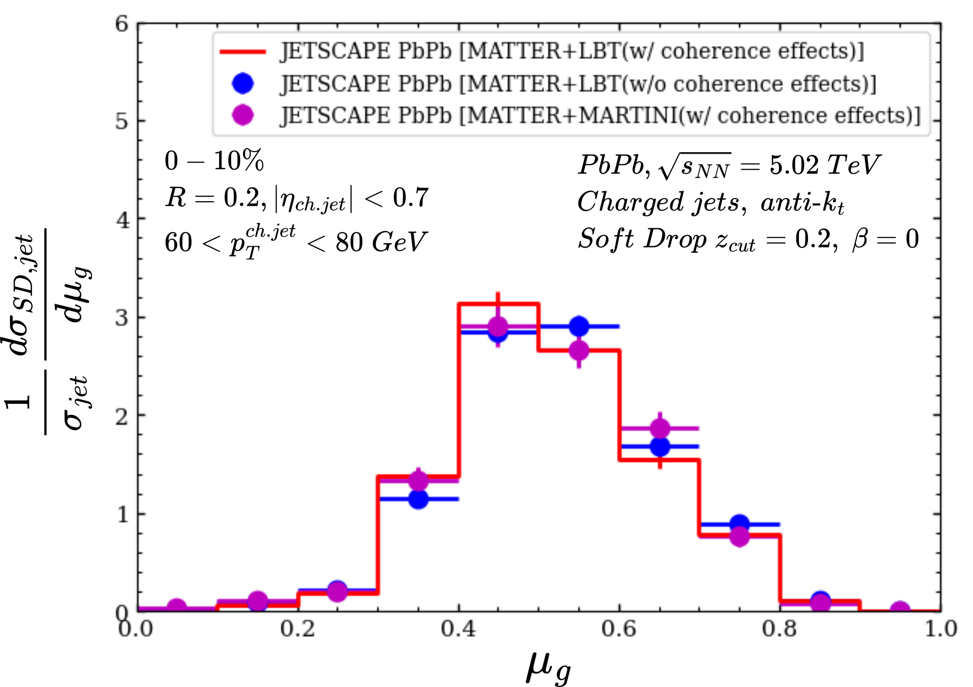}
    \end{minipage}\hfill
    \begin{minipage}{0.5\textwidth}
        \centering
        \includegraphics[width=5.6cm, height=4.1cm]{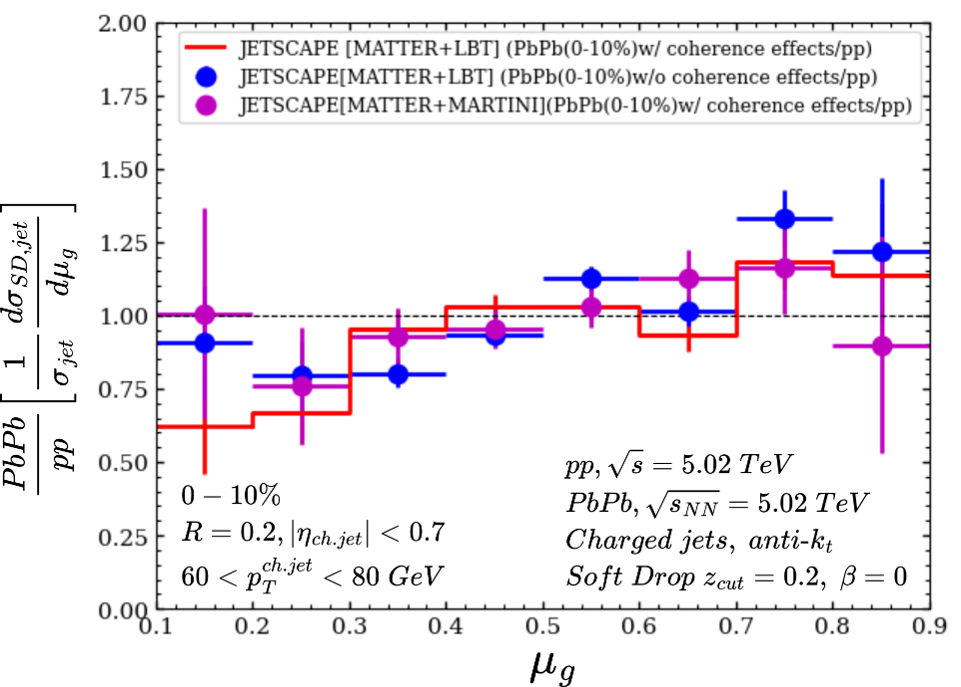}
    \end{minipage}
    \caption{Depiction of \( \mu_g \) distributions in the 60--80 GeV range, generated from JS simulations using MATTER+MARTINI (with coherence effects) and MATTER+LBT (both with and without coherence effects). The left panel illustrates the \( \mu_g \) distribution for Pb-Pb collisions, while the right panel shows the corresponding Pb-Pb to p-p ratio.}
    \label{fig:3}
\end{figure}

The left panel of Fig.~\ref{fig:3} shows $\mu_g$ distributions for the two combinations of energy loss models. Fewer jets have a minimal or full contribution from the heavier subjet, compared to those where it contributes around 50\%. The right panel shows that MATTER+LBT with coherence effects leads to suppression at low $\mu_g$, minimal modification at intermediate $\mu_g$, and enhancement at high $\mu_g$, while other results show fluctuations. These findings suggest QGP alters the mass distribution within jets, increasing the heavier subjet’s contribution.

\begin{figure}[th]
    \centering
    \begin{minipage}{0.5\textwidth}
        \centering
        \includegraphics[width=5.5cm, height=4.1cm]{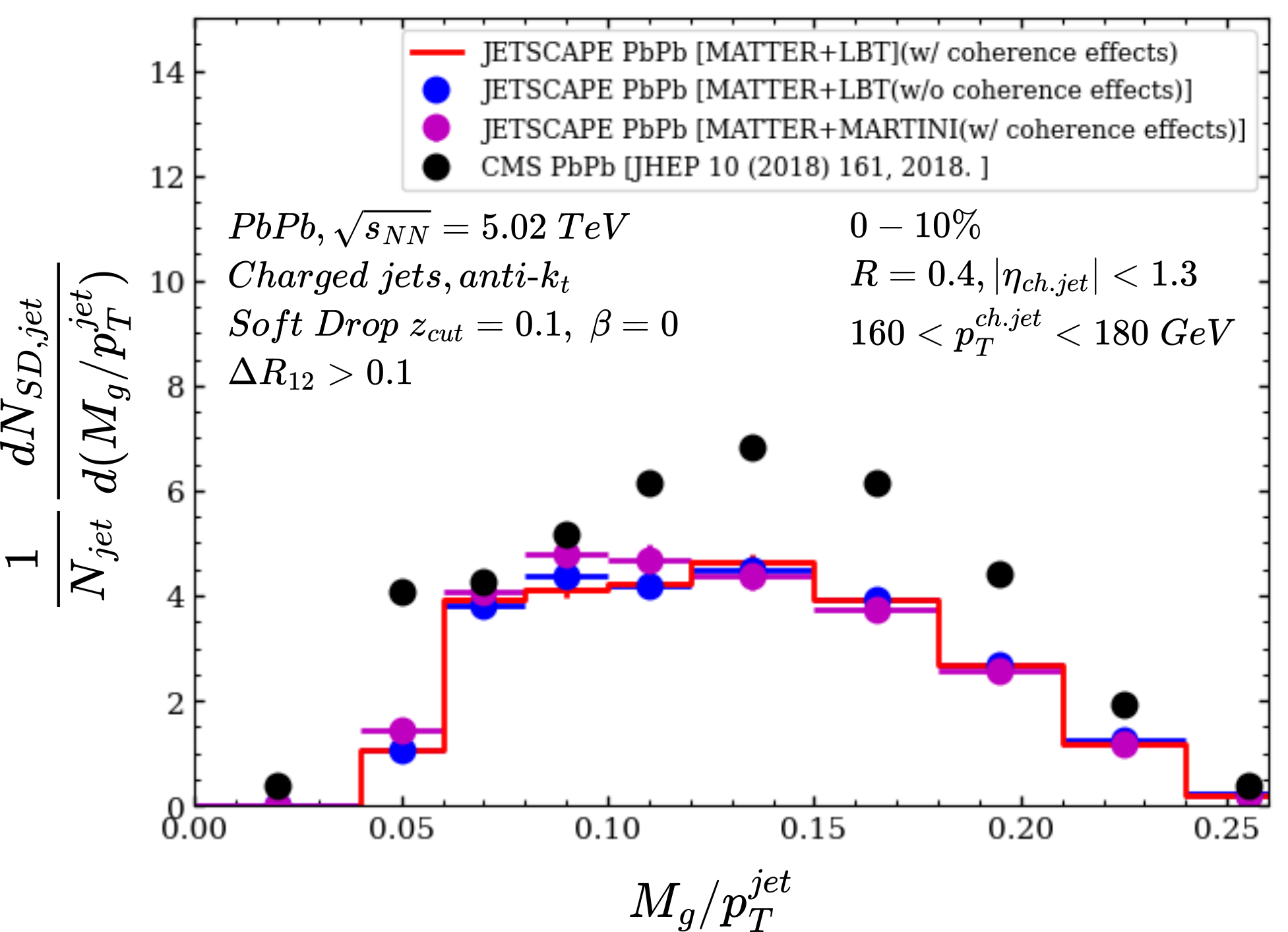}
    \end{minipage}\hfill
    \begin{minipage}{0.5\textwidth}
        \centering
        \includegraphics[width=5.5cm, height=4.1cm]{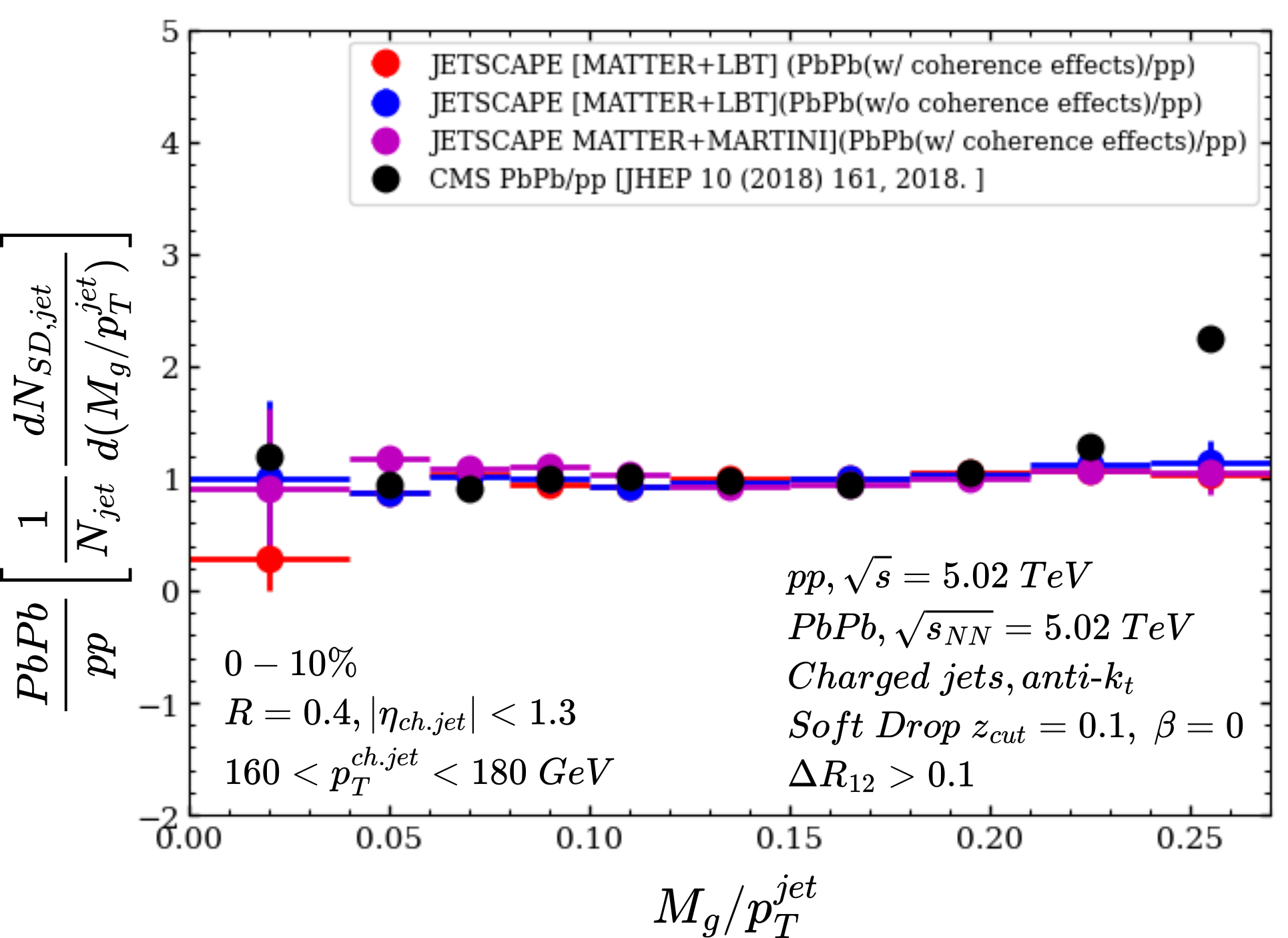}
    \end{minipage}
    \caption{Results for \( \rho \) from JS in the 160--180 GeV range, compared to CMS data \cite{G15}. The left panel shows the \( \rho \) distribution for Pb-Pb collisions, and the right panel displays the ratio of \( \rho \) distributions between Pb-Pb and p-p collisions.}
    \label{fig:4}
\end{figure}

The left panel of Fig.~\ref{fig:4} shows that the Pb-Pb $\rho$ distribution is suppressed at low and high values, and enhanced at intermediate values. JS underestimates the distribution but follows the trend of the experimental data. The right panel shows the ratio is near one, deviating only at the highest $\rho$. Thus, QGP does not significantly modify the groomed mass of the core of the jets in Pb-Pb compared to p-p collisions.


\section{Enhancing Large Radius Boosted
Top-Quark Tagging with Machine Learning}

\author{Devanshu Sharma, Prabhakar Palni}

\bigskip

\begin{abstract}
We investigate the performance of Residual Neural Network (ResNet) as jet taggers on
large radius boosted top quark jets reconstructed from optimized jet input objects in
simulated proton-proton collisions at $\sqrt s$ = 13 TeV. Events with boosted top quarks
are of great interest for both precision measurements of Standard Model processes and
searches for new physics beyond the Standard Model. For this reason, identifying large
radius jets that result from boosted top decays is a crucial task in data analysis, also
known as boosted top tagging. In this work, we use the simulated dataset provided
by the ATLAS collaboration at the Large Hadron Collider to train the models. The
performance of models trained on raw and processed jet images are compared.
\end{abstract}




\subsection{ Introduction }

The top quark, the heaviest particle in the Standard Model, has been a key focus of particle physics since its discovery in 1995. With a mass of 173 GeV/c², it decays before hadronization, producing a distinct signature involving a W boson and a bottom quark. In high-energy collisions at the LHC, boosted top quarks with high transverse momentum decay into collimated, large-radius jets.

Identifying these jets, known as boosted top quark tagging, is crucial for studying Standard Model processes and exploring new physics. Traditional methods have advanced with machine learning, particularly using Residual Neural Networks (ResNet)\cite{resnet} for jet image analysis. This work examines how improving jet image processing can enhance boosted top quark tagging performance in simulated data, leading to more accurate event identification.

\subsection{ Methodology }


\subsubsection{ Jet Images }

Jet images are a way to represent the complex structure of a particle jet as a two-dimensional image. The jets are converted into “jet images” by binning each constituent’s $\eta$ and $\phi$ coordinates into 64 bins, equally spaced in the range [-2, 2]. Each “pixel” in this grid reflects the energy of particles in that region.\cite{jet-images} 

\subsubsection{ Image preprocessing }

The image transformations explored in this study were:

\begin{itemlist}
    \item Resizing: Images were resized by binning the $\eta$ and $\phi$ coordinates between [-1, 1].\\

    \item Gaussian Filtering: A Gaussian filter was applied to the resized images to spread the information from each pixel across neighboring pixels. 
    This process smooths out the sparse regions, effectively filling in gaps and reducing the impact of sparsity.
\end{itemlist}

\begin{figure}[th]
\centering
    \includegraphics[width=0.4\linewidth]{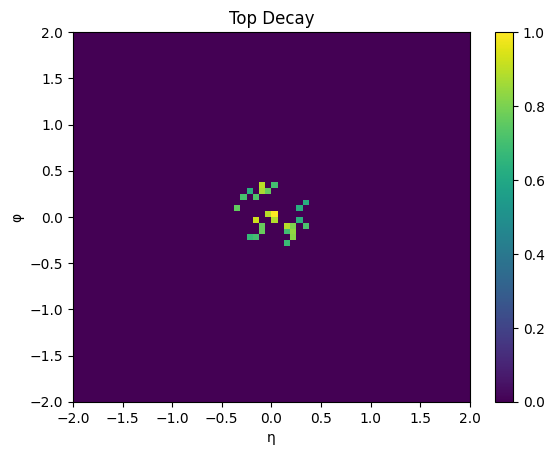}
    \includegraphics[width=0.4\linewidth]{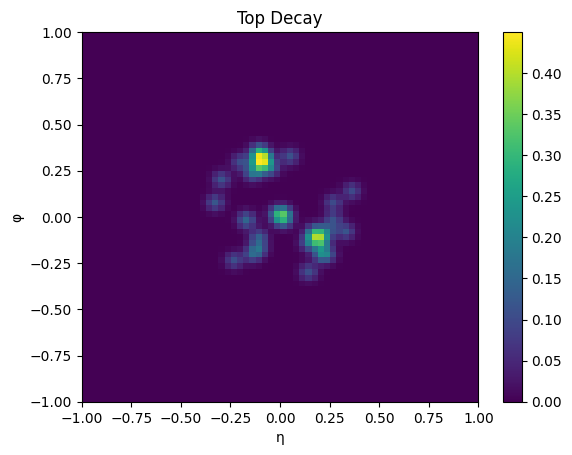}
    \caption{Example jet images before(left) and after(right) preprocessing}
\end{figure}

\subsubsection{ Model Training and Evaluation }

An image classification model, ResNet50\footnote{The optimal model hyperparamters and training specifications are obtained from ATLAS PUB Note ATL-PHYS-PUB-2022-039.\cite{main}}, is trained on the data. \\

Model evaluation is conducted using accuracy and the area under the receiver operating characteristic (ROC) curve (AUC). Accuracy measures the proportion of correctly classified jets, while AUC reflects the model's ability to distinguish between signal and background across different classification thresholds.

\subsection{ Results }

The results demonstrate that models trained on processed jet images significantly outperform those trained on raw images, particularly in terms of accuracy. Preprocessing steps improve the model's ability to recognize top quark jets’ distinct substructure by reducing variations in scale and orientation.

\begin{figure}[th]
\centerline{\includegraphics[width=0.5\linewidth]{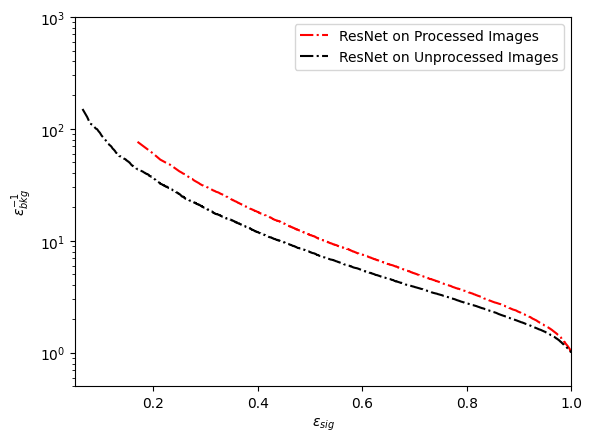}}
\caption{ROC curves comparing the performance of ResNet using processed versus unprocessed
jet images. The inverse of efficiency in background QCD jets $(\epsilon^{-1}_{\text{bkg}})$ versus the efficiency in top quark jets $(\epsilon_{\text{sig}})$ for the
top quark tagger studied. The plot highlights the impact of image preprocessing on model efficiency.}
\end{figure}

Processed images enable the ResNet model to more accurately identify the three-pronged decay pattern of boosted top quarks, a feature often obscured in raw images due to noise and inconsistencies.
\\
These findings highlight the value of preprocessing in optimizing machine learning models for boosted top quark tagging and suggest that further refinements may continue to improve model performance in high-energy physics analyses.

\begin{table}[pt]
\tbl{Comparison of model performance using processed and unprocessed jet images for
ResNet.}
{\begin{tabular}{@{}cccc@{}} \toprule
Binning Range & Filter & Accuracy Score &
Area Under Curve \\
\colrule
$[-2, 2]$\hphantom{00} & \hphantom{0}None & \hphantom{0}70.90\% & 0.795 \\
$[-1, 1]$\hphantom{00} & \hphantom{0}Gaussian Filter ($\sigma = 1$) & \hphantom{0}76.12\% & 0.837 \\
\botrule
\end{tabular}}

\end{table}

\subsection{ Conclusion }

In this study, we show that preprocessing jet images can significantly improve the performance of machine learning models for large-radius boosted top quark tagging. By enhancing jet image clarity through normalization and alignment, models like ResNet more accurately capture the distinct substructure of boosted top quarks, resulting in higher tagging accuracy and improved background rejection. These results highlight the importance of preprocessing in optimizing model performance and suggest that further exploration of preprocessing techniques may lead to even greater advancements in top quark tagging.

In future work, other filters can be tested to determine the optimal configuration, potentially further enhancing tagging performance.


\section{Heavy Flavor Production at the Large Hadron Collider: A Machine Learning Approach}

\author{Raghunath Sahoo}


\bigskip

\begin{abstract}
Charmonia suppression has been considered as a smoking gun signature of quark-gluon plasma. However,
the Large Hadron Collider has observed a lower degree of suppression as compared to the Relativistic Heavy Ion Collider energies, due to regeneration effects in heavy-ion collisions. Though proton collisions are considered 
to be the baseline measurements to characterize a hot and dense medium formation in heavy-ion collisions,
LHC proton collisions with its new physics of heavy-ion-like QGP signatures have created new challenges.
To understand this, the inclusive charmonia production at the forward rapidities in the dimuon channel is compared with the corresponding measurements in the dielectron channel at the midrapidity as a function of final state 
charged particle multiplicity. None of the theoretical models quantitatively reproduce the
experimental findings leaving out a lot of room for theory. To circumvent this and find a reasonable understanding,
we use machine learning tools to separate prompt and nonprompt charmonia and open charm mesons using the
decay daughter track properties and the decay topologies of the mother particles. Using PYTHIA8 data, we 
train the machine learning models and successfully separate prompt and nonprompt charm hadrons from the
inclusive sample to study various directions of their production dynamics. This study enables a domain of using
machine learning techniques, which can be used in the experimental analysis to better understand 
charm hadron production and build possible theoretical understanding.

\end{abstract}


\ccode{PACS numbers:}


\subsection{Introduction}
In the heavy flavor sector, charmonia (J/$\psi$), the bound state of a charm and anti-charm quark 
($\rm c\bar{c}$) plays an
important role, the suppression of which in heavy-ion collisions at ultra-relativistic energies is considered a 
signature of the deconfined primordial matter, called quark-gluon plasma (QGP). This, on the other hand, leads 
to the enhancement of open charm mesons like $D^0$. A complementary study taking both J/$\psi$ and $D^0$
can give a full picture of probing the produced QCD medium using charmonia. While making such a study in
heavy-ion collisions to probe the hot QCD medium, usually the minimum-bias proton-proton (pp) collisions are 
taken as a baseline measurement. It should also be noted here that the degree of charmonia suppression at the LHC energies is observed to be lower compared to RHIC energies, because of the availability of higher energy
phase space at the LHC leading to charmonia regeneration effects\cite{ALICE:2015jrl}.  At the TeV LHC energies, several partonic collisions may occur
in a single pp collision, which affects the total multiplicity through the production of light quarks and gluons.
These high-multiplicity pp events are of special importance at the LHC energies, where one observes
several heavy-ion-like signatures, which include collective flow pattern\cite{CMS:2010ifv}, enhancement of strangeness\cite{strange,strange1} etc. For a general review of such QGP-like signatures in proton collisions, please see Ref.\cite{Sahoo:2021aoy}. On the contrary, the absence of evidence of jet quenching in such
high-multiplicity pp events makes them illusive. These observations add to the earlier conjecture that high-energy 
pp collisions could produce statistical systems capable of showing hydrodynamics behavior\cite{Fermi,Hagedorn,Landau}. Such statistical
systems with partonic quanta, which are locally thermalized and behave like strongly interacting hot fluid of QGP
are to be confronted with experimental tests. However, high-multiplicity proton-antiproton collisions at the 
Tevatron energies ($\sim \sqrt{s} = $ 1.8 TeV) didn't support the formation of QGP in such collisions\cite{Tevatron}.

In this report, we present the forward rapidity inclusive J/$\psi$ production as a function of the final state charged
particle multiplicity measured with ALICE\cite{ALICE:2021zkd}, which is not quantitatively explained by the existing theoretical models. Further, we employ Machine Learning (ML) techniques to separate prompt and nonprompt J/$\psi$ and
$D^0$ meson in pp collisions at TeV energies to study various production dynamics of charmonia and open
charms using PYTHIA8 event generator using track-level properties and
decay topology\cite{Prasad:2023zdd,Goswami:2024xrx}.

\subsubsection{Charmonia measurement in proton collisions}

\begin{figure*}[ht!]
\begin{center}
\includegraphics[scale = 0.28]{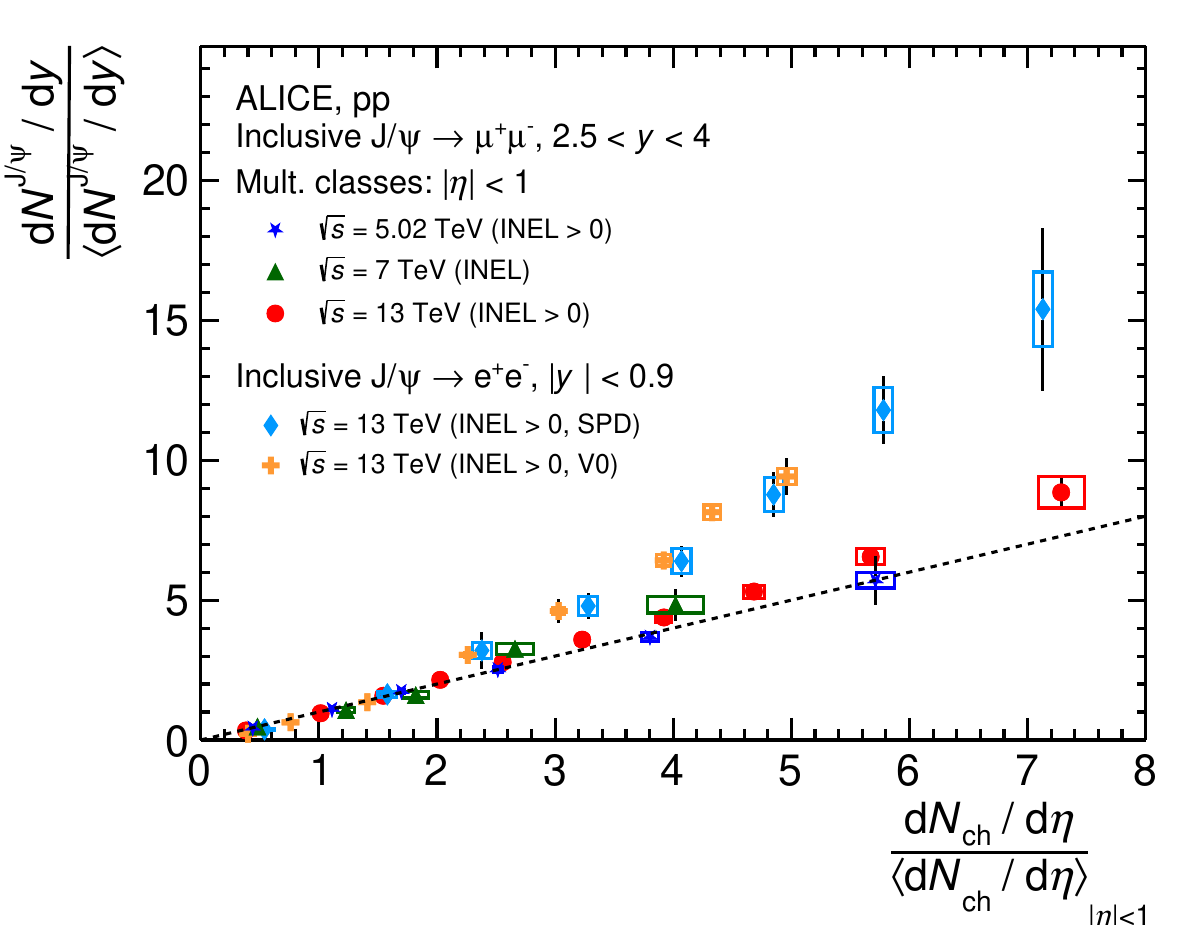}
\caption{(Colour Online) Forward rapidity relative J/$\psi$ yields in pp collisions at $\sqrt{s}$ = 5.02, 7 and 13 TeV compared to $\sqrt{s}$ = 13 TeV measurement at midrapidity \cite{ALICE:2021zkd}.}
\label{fig1}
\end{center}
\end{figure*}

\begin{figure*}[ht!]
\begin{center}
\includegraphics[scale = 0.38]{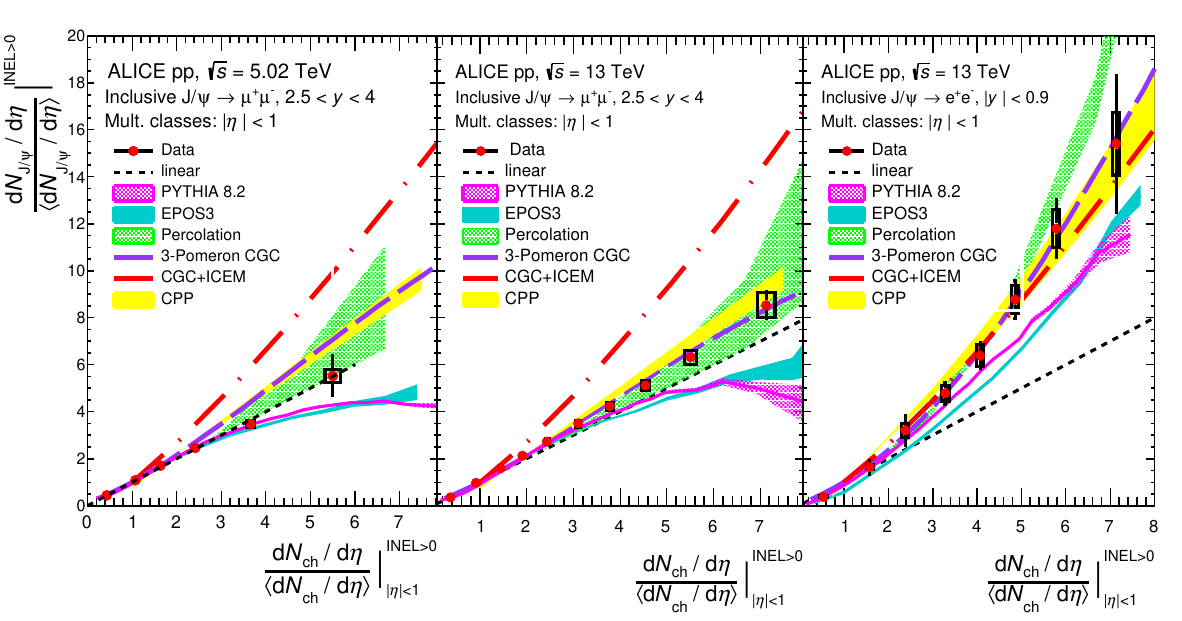}
\caption{(Colour Online) Forward rapidity relative J/$\psi$ yields in pp collisions at $\sqrt{s}$ = 5.02 and 13 TeV compared various theoretical model predictions as discussed in the text\cite{ALICE:2021zkd}.}
\label{fig2}
\end{center}
\end{figure*}

ALICE has measured the forward rapidity $(2.5 < y < 4)$ J/$\psi$ in pp collisions at $\sqrt {s} = 5.02, 7$ and 13 TeV
in the dimuon channel using a muon spectrometer. Whereas the proxy of centrality (or impact parameter) is the
final state charged particle multiplicity, which is measured at the midrapidity ($|\eta| < 1$) using a Silicon Pixel Detector (SPD) to avoid possible
autocorrelation bias in the measurement. The details of the measurement procedure can be found in Ref.\cite{ALICE:2021zkd}.

The self-normalized yield of J/$\psi$ is defined as the ratio of the yield in a given multiplicity window to the average yield across all the measured multiplicity bins, i.e. $(dN_{J/\psi}/dy)$/$<dN_{J/\psi}/dy>$.  This is shown in Fig.\ref{fig1} as a function of the corresponding self-normalized charged particle yield. For all the discussed energies 
the relative J/$\psi$ yield shows a nearly linear rise with the midrapidity charged particle relative multiplicity. 
A similarity in the forward rapidity J/$\psi$ production across various energies suggests in a given multiplicity window, J/$\psi$ 
production is more or less independent of collision energy. These forward rapidity measurements are compared with
the corresponding measurements at midrapidity in the dielectron channel for pp collisions at $\sqrt{s}$ = 13 TeV, where the multiplicity estimators are taken both at midrapidity and forward rapidity, $-3.7 < \eta < -1.7 $ and $2.8 < \eta < 5.1$  (V0 detector). Although the results are consistent with experimental uncertainties showing the absence
of possible auto-correlation bias arising from multiplicity selection bias, there is a deviation from the linear behavior
of J/$\psi$ production in the midrapidity dielectron channel as compared with the forward rapidity dimuon channel. 
To have a better understanding of J/$\psi$ production, these results are compared with available theoretical models
like Coherent Particle Production (CPP), CGC with ICEM (improved color evaporation model), 3-Pomeron CGC,
Percolation, EPOS3, and PYTHIA8\cite{ALICE:2021zkd}. Both EPOS and PYTHIA8 describe the forward rapidity
J/$\psi$ yield at low-multiplicity while underestimating the high-multiplicity data. The CPP model with a phenomenological parametrization for mean multiplicities of light hadrons and J/$\psi$, shows a very good agreement with the high-multiplicity measurements both for $\sqrt{s}$ = 5.02 and 13 TeV. CGC+ICEM employs the
NRQCD framework to describe J/$\psi$ hadronization. While this model describes the midrapidity dielectron
channel results, it predicts a faster-than-linear increase of J/$\psi$ yield with multiplicity for pp collisions. The 3-gluon fusion model seems to describe the multiplicity-dependent J/$\psi$ yield both at midrapidity and forward rapidity
for $\sqrt{s}$ = 13 TeV, whereas it fails to describe the high-multiplicity region for $\sqrt{s}$ = 5.02 TeV. Percolation
model with larger uncertainties however seems to describe the forward rapidity multiplicity-dependent J/$\psi$
production both for $\sqrt{s}$ = 5.02 and 13 TeV. None of the theoretical models seem to describe the multiplicity-dependent inclusive J/$\psi$ production in pp collisions across rapidity, multiplicity, and production channels. Let us
now move to use ML techniques to separate prompt and nonprompt  J/$\psi$  and $D^0$ for a better understanding
of the underlying production mechanisms.
 
\subsubsection{Machine Learning tools heavy flavor studies}

\begin{figure*}[ht!]
\begin{center}
\includegraphics[scale = 0.4]{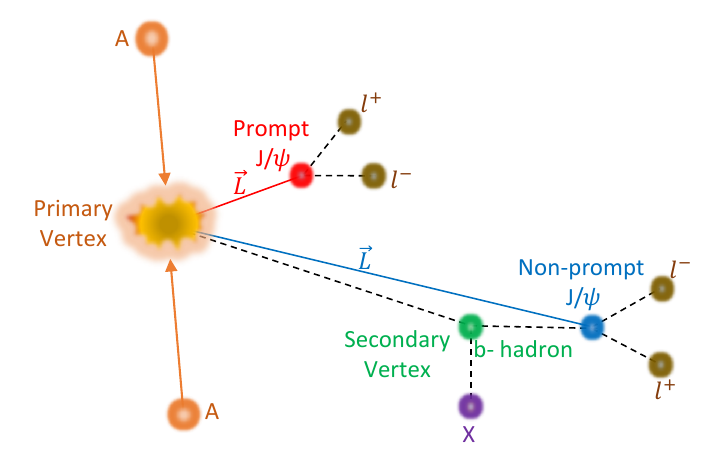}
\includegraphics[scale = 0.3]{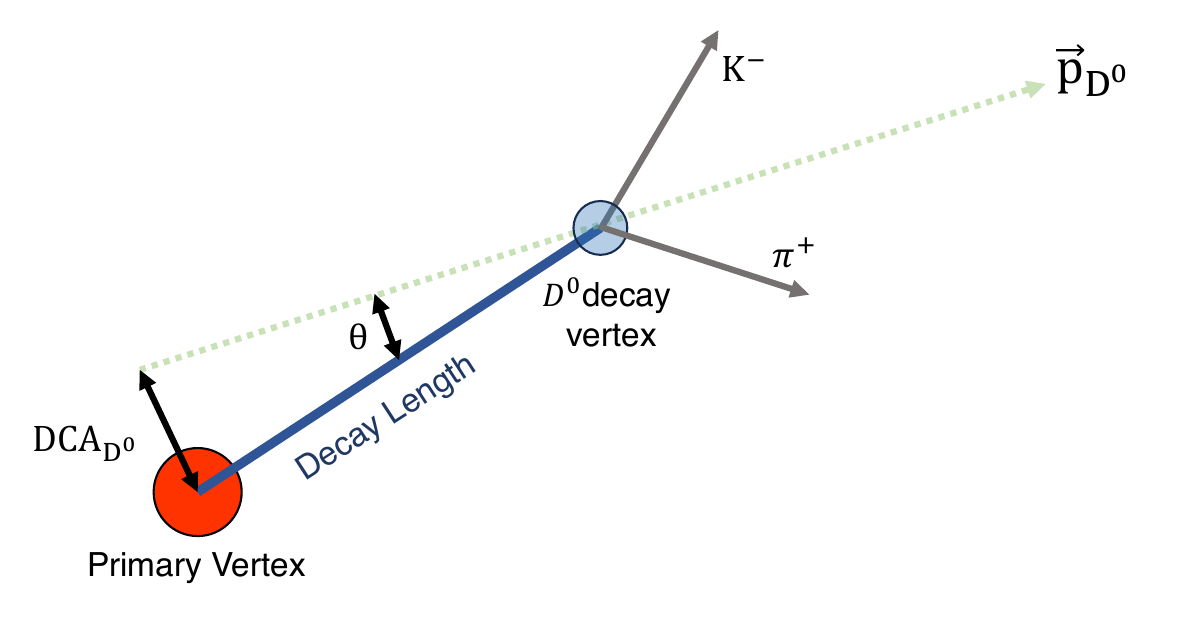}
\caption{(Colour Online) (Left) Decay topology of prompt and nonprompt J/$\psi$\cite{Prasad:2023zdd}, and (Right) for $D^0$ \cite{Goswami:2024xrx}.}
\label{fig3}
\end{center}
\end{figure*}

\begin{figure*}[ht!]
\begin{center}
\includegraphics[scale = 0.15]{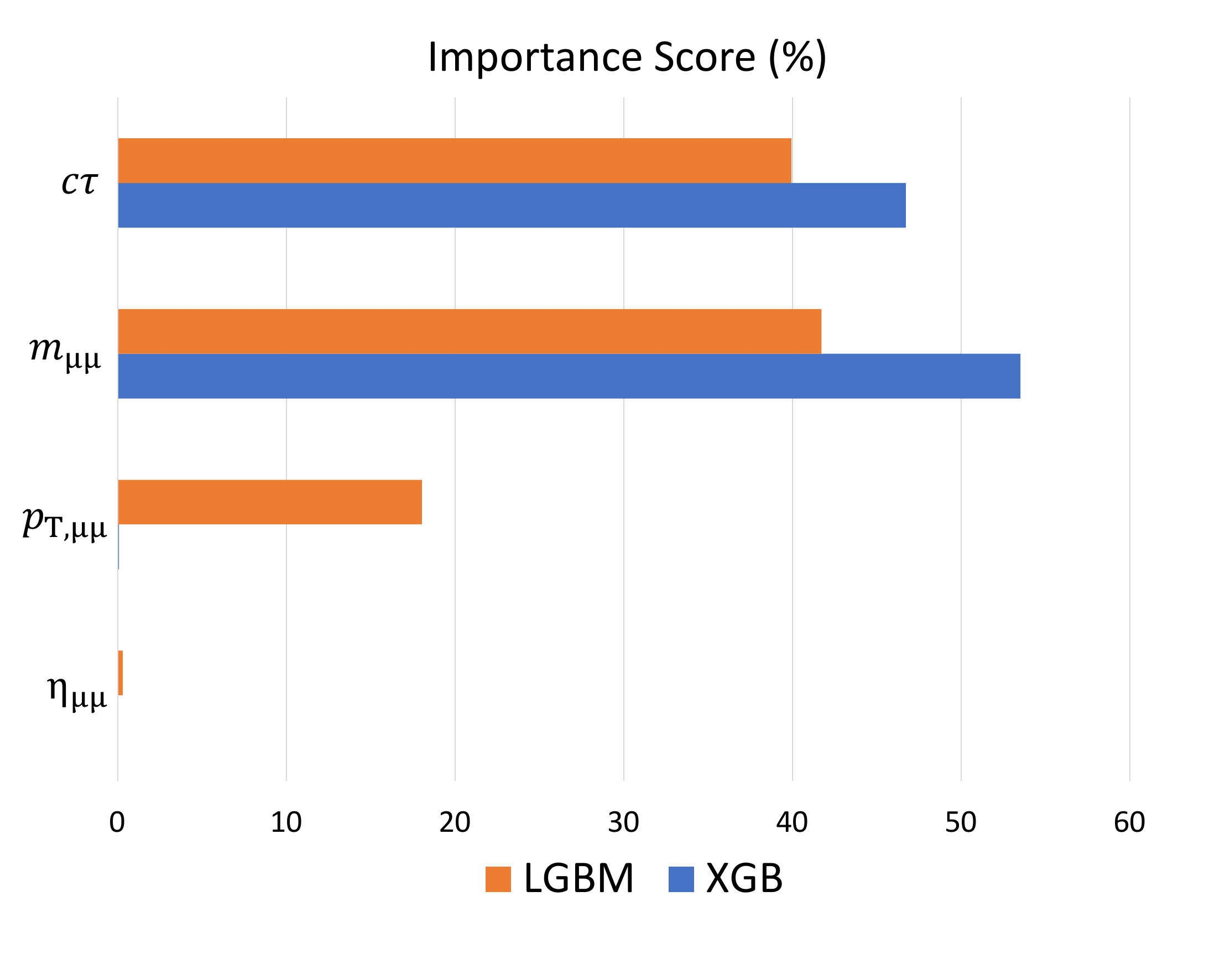}
\includegraphics[scale = 0.18]{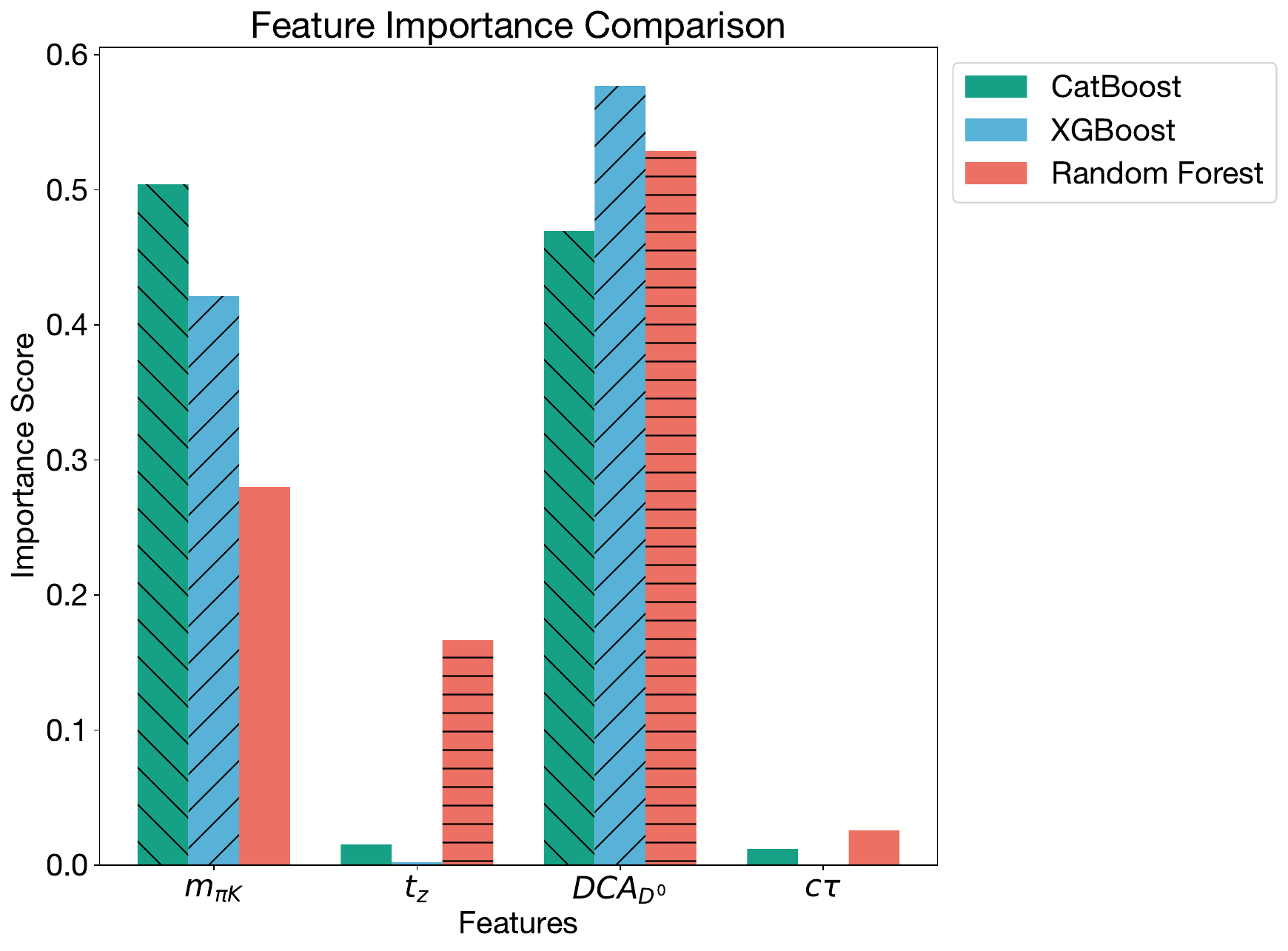}
\caption{(Colour Online) (Left) Comparison of importance scores (\%) for the input variables used for prompt and nonprompt J/$\psi$ separation\cite{Prasad:2023zdd}, and (Right) for separating prompt and nonprompt $D^0$ \cite{Goswami:2024xrx}.}
\label{fig4}
\end{center}
\end{figure*}

The inclusive production of charmonia consists of J/$\psi$ produced directly in the hadronic/nuclear collisions and
those produced via the feed down from directly produced higher charmonium states like $\psi(2S)$ and $\chi_c$.
These are called prompt J/$\psi$. Whereas those produced from the weak decay of beauty hadrons are classified as nonprompt J/$\psi$. The prompt J/$\psi$ is produced close to the interaction vertex, whereas the nonprompt
J/$\psi$ decay is associated with a secondary vertex, as shown in the decay topology (Fig.\ref{fig3}).  This classification can help
in the indirect estimation of the nuclear modification factor in the beauty sector in addition to the spin polarization 
measurements\cite{Prasad:2023zdd}. In the experimental data analysis, this separation method relies on statistical
methods with template fitting. However, the discussed technique uses track-level properties of the daughter
particles with the decay topology of  J/$\psi$.

We use ML models that use gradient-boosting-decision-trees-based classifications like XGBoost and LightGBM with simulated data of properly tuned PYTHIA8 for pp collisions at $\sqrt{s}$ = 13 TeV for training and prediction. Details of the methodology can be found in Ref.\cite{Prasad:2023zdd}. We estimate the pseudoproper decay length ($c\tau$), dimuon invariant mass, transverse momentum, and pseudorapidity as the
input variables for the topological separation of J/$\psi$. The $D^{0}$ meson uses the invariant mass, pseudo-proper time ($t_{z}$), distance of closest approach and $c\tau$. Model parameters such as the loss function, learning rate, number of trees, and maximum depth are 
tuned for each model. The optimized parameters are selected through a grid search method. As shown in Fig. \ref{fig4}, the importance score shows that both $c\tau$ and dimuon invariant mass play a significant role in the training and prediction compared to other parameters to identify prompt and nonprompt production of $\rm{J}/\psi$. Similarly, for the separation of prompt and nonprompt $D^0$, as shown in
Fig. \ref{fig4}, both the invariant mass and the distance of the closest approach are important parameters in the
ML techniques used.

\begin{figure*}[ht!]
\begin{center}
\includegraphics[scale = 0.25]{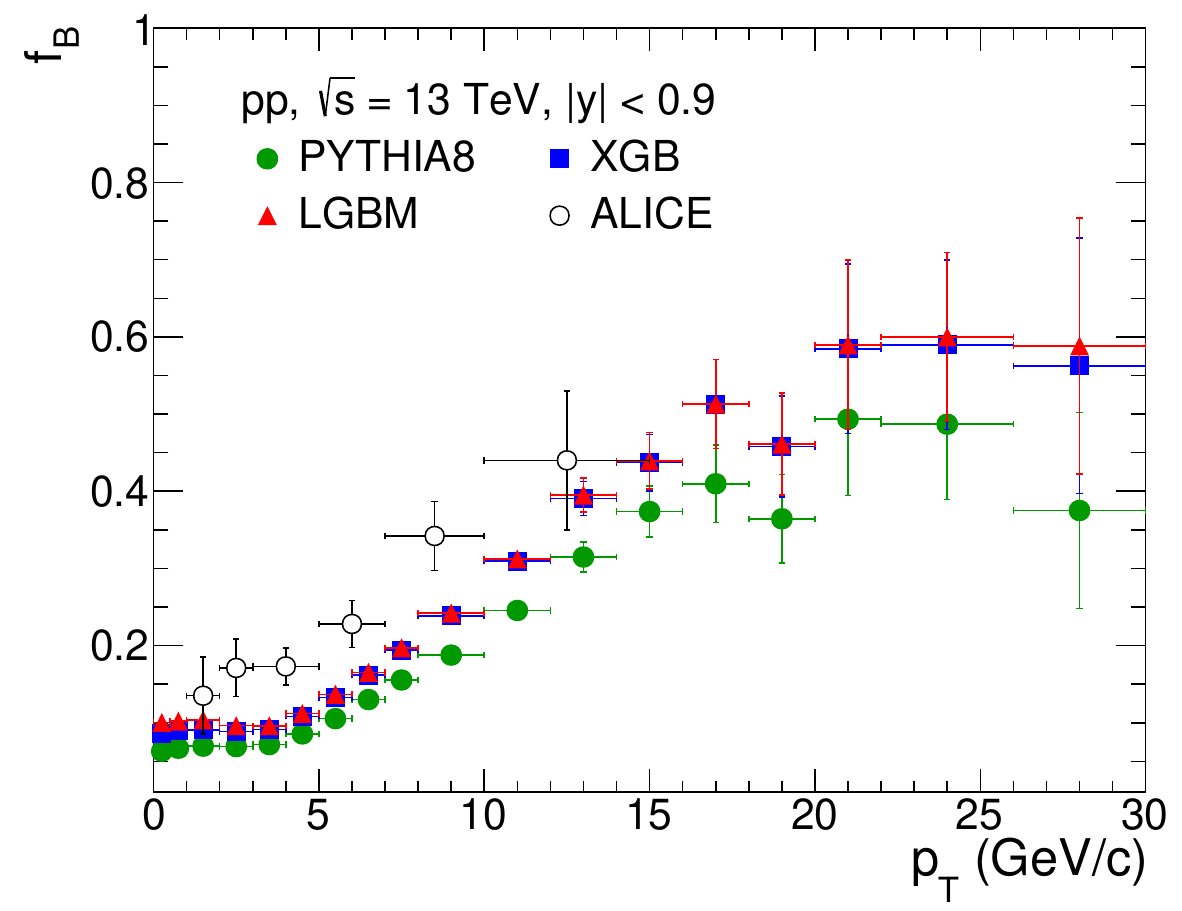}
\includegraphics[scale = 0.25]{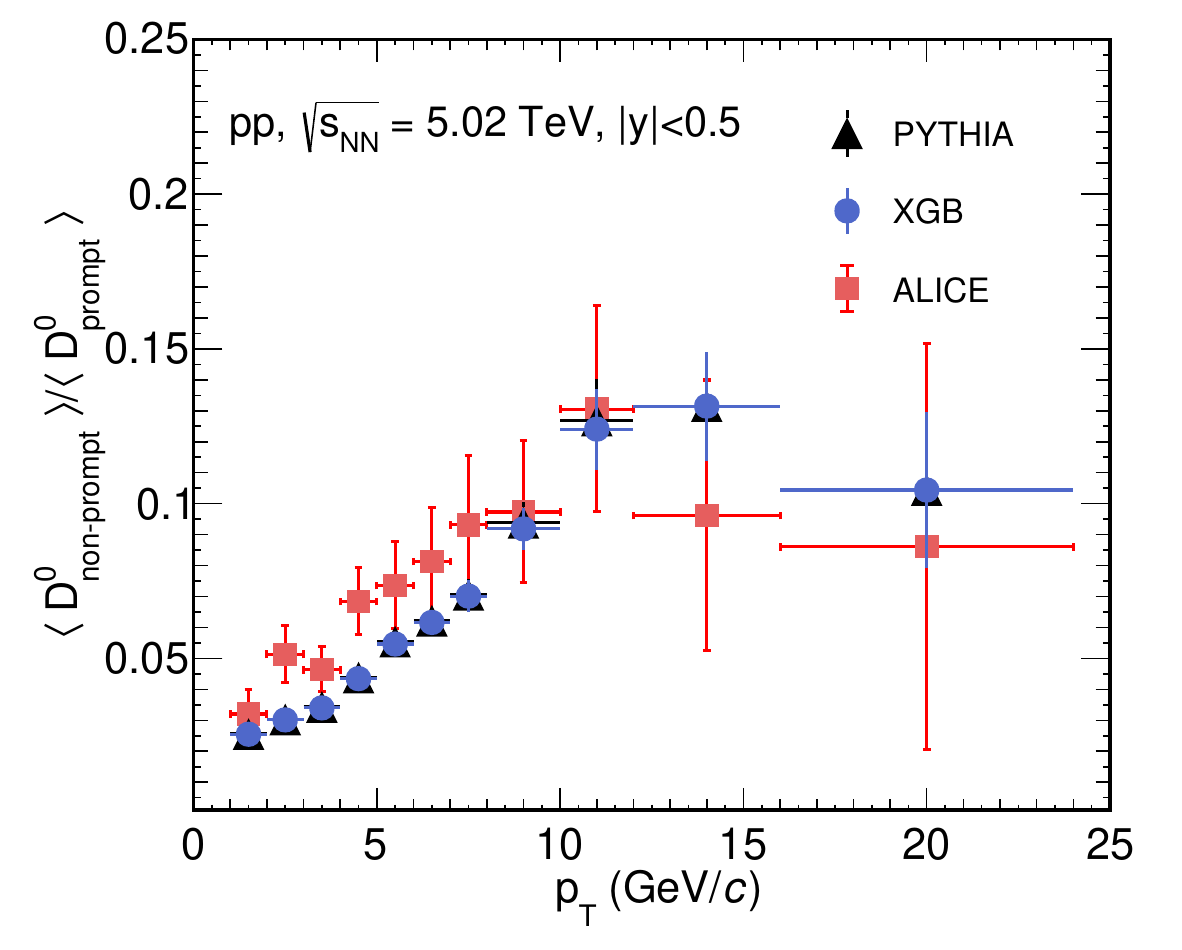}
\caption{(Colour Online) (Left) J/$\psi$ fraction from b-hadron decay ($f_B$) as a function of $p_{\rm T}$ at midrapidity for minimum-bias pp collisions at $\sqrt{s}$ = 13 TeV using PYTHIA8, predictions from XGB and LGBM
and a comparison with ALICE data is shown\cite{Prasad:2023zdd}. (Right) Nonprompt to prompt $D^0$ ratio for minimum-bias pp collisions at $\sqrt{s}$ = 5.02 TeV using PYTHIA8. Predictions from XGB and a comparison with ALICE data are also shown\cite{Goswami:2024xrx}.}
\label{fig5}
\end{center}
\end{figure*}

In Fig. \ref{fig5}, we show the J/$\psi$ fraction from b-hadron decay ($f_B$) as a function of $p_{\rm T}$ at midrapidity for minimum-bias pp collisions at $\sqrt{s}$ = 13 TeV using PYTHIA8, along with the predictions from XGB and LGBM,
and a comparison with ALICE data\cite{Prasad:2023zdd}. Here the trend of $f_B$ is similar to the experimental 
results and the ML models seem to do a good job even estimating $f_B$ in finer bins of $p_{\rm T}$ using ML.
Fig. \ref{fig5} (right) shows the ratio of nonprompt to prompt $D^0$ at midrapidity ($|y| < 0.5$) for minimumbias pp collisions at $\sqrt{s}$ = 5.02 TeV using PYTHIA8. Predictions from XGB and a comparison with ALICE data are also shown\cite{Goswami:2024xrx}. This ratio quantifies the fraction of $D^0$ coming from beauty hadron decays
in comparison to the direct production from the charm sector. There is a linear increase in this ratio with $p_{\rm T}$
upto $p_{\rm T} \simeq 12$ GeV/c. XGB seems to agree quantitatively with the estimates of
PYTHIA8. Details of the associated results for both charmonia and open charm meson can be found in the
Refs. \cite{Prasad:2023zdd,Goswami:2024xrx}. 

\subsection{Summary}
The experimental measurement of inclusive J/$\psi$ at forward rapidity in the dimuon channel is shown for TeV pp collisions as a function of final state charged particle multiplicity. These measurements are compared to those obtained in the dielectron channel. A linear trend is seen in the former case, whereas an increase
higher than linear is observed in the latter case. None of the discussed theoretical models seem to explain the data quantitatively in the discussed region of charged particle multiplicity. Further, with the need to separate prompt from
non-prompt charmonia and open charms, we use ML techniques, which are found to do an excellent job taking the 
track-level properties and decay topologies. This method, augmented to the mainstream of data analysis, will be a boon, given that experimental methods use statistical template fitting or additional detectors for secondary vertexing is a need for such separation. For a general review of charm and beauty as next-generation measurements to study QCD plasma created in TeV hadronic and nuclear collisions, please see Ref.\cite{Das:2021igk}.


%
\newpage
\section*{Acknowledgments}
This write-up compiles the contributions on hard probes presented at the "Hot QCD Matter 2024 Conference," held from July 1-3, 2024, at IIT Mandi, India. Vyshakh B R, Rishi Sharma acknowledge the support of the Department of Atomic Energy, Government of
India, under Project Identification No. RTI 4002. Nihar Ranjan Sahoo expresses gratitude to Shandong University (Qingdao), China; the National Institute of Science Education and Research, India; and Texas A\&M University, USA, for their support. The author also thanks the organizers of the 2nd edition of Hot QCD Matter for the invitation and for hosting a wonderful conference on the beautiful IIT Mandi campus in the serene Himalayan mountains. Saumen Datta thanks Dibyendu Bala for discussion.
Saumen Datta's research is supported by the Department of Atomic Energy, Government of India, under Project Identification No.\ RTI 4002. Aritra Bandyopadhyay's work is supported by the Alexander von Humboldt foundation. Debarshi Dey acknowledges financial support provided by Indian Institute of Technology Bombay for attending the conference Hot QCD matter 2024. Pooja acknowledges IIT Goa and MHRD for funding this work. Santosh Kumar Das and Partha Pratim Bhaduri acknowledges the support from DAE-BRNS, India, Project No. 57/14/02/2021-BRNS. Jai Prakash acknowledge support from DAE-BRNS (Grant No. 57/14/02/2021-BRNS) and from SERB (Project Code: RD/0122-SERBF30-001). Manaswini Priyadarshini, Om Shahi, Vaishnavi Desai and Prabhkar Palni express our gratitude to the JETSCAPE Collaboration for making the state-of-the-art framework publicly available for extensive research use.  The authors would also like to thank IIT Mandi Param-Himalaya computing facility, and seed money grant support. Mohammad Yousuf Jamal would like to acknowledge the SERB-NPDF (National postdoctoral fellow) fellowship with File No. PDF/2022/001551. Gurleen Kaur and Prabhakar Palni would like to acknowledge JETSCAPE collaboration, its simulation framework, School of Physical Sciences IIT Mandi, and Param Himalaya computing facility. Prabhakar Palni would like to acknowledge the support from the SERB Seminar/Symposia Scheme (File no. SSY/2024/000612) and the IIT Mandi SRIC seed grant support (Ref. No. IITM/SG/2024/01-2348). Raghunath Sahoo gratefully acknowledges the DAE-DST, Government of India funding under the mega-science project “Indian participation in
the ALICE experiment at CERN” bearing Project No. SR/MF/PS-02/2021-IITI(E-37123)
under which these research works have been carried out.


\end{document}